\documentclass[12pt]{article}
\usepackage{epsfig,amsfonts,amssymb}
\usepackage{cite}
\input epsf.sty
\usepackage{cite}

\def\ZZZ{{\hbox{ Z\kern-1.6mm Z}}}
\def\RRR{{\hbox{ R\kern-2.4mm R}}}
\def\CCC{{\hbox{ C\kern-2.0mm C}}}
\def\zzz{{\hbox{z\kern-1mm z}}}

\newcommand{\nn}{\nonumber \\}

\newcommand{\qeq}{{\hbox{=\kern-2.3mm ? \kern.5mm }}}
\renewcommand{\qeq}{=}

\newcommand{\eps}{\epsilon}
\newcommand{\vareps}{\varepsilon}

\newcommand{\vp}{\varphi}
\newcommand{\ve}{\varepsilon}

\newcommand{\DD}{{\cal D}}

\newcommand{\II}{{\cal I}}
\newcommand{\AAA}{{\cal A}}

\newcommand{\KK}{{\cal K}}

\newcommand{\MM}{{\cal M}}

\newcommand{\OO}{{\cal O}}

\newcommand{\LL}{{\cal L}}

\newcommand{\wt}{\widetilde}
\newcommand{\wh}{\widehat}

\newcommand{\NN}{{\cal N}}

\newcommand{\SSS}{{\cal S}}

\newcommand{\be}{\begin{equation}}
\newcommand{\ee}{\end{equation}}
\newcommand{\ben}{\begin{eqnarray}\displaystyle}
\newcommand{\een}{\end{eqnarray}}

\newcommand{\refb}[1]{(\ref{#1})}
\newcommand{\p}{\partial}
\newcommand{\sectiono}[1]{\section{#1}\setcounter{equation}{0}}

\newcommand{\zet}{\zeta}

\newcommand{\Lamb}{\Lambda}

\def\one{{\hbox{ 1\kern-.8mm l}}}
\def\zero{{\hbox{ 0\kern-1.5mm 0}}}

\begin{document}

\baselineskip 24pt

\begin{center}
{\Large \bf  Logarithmic Corrections to $\NN=4$ and
$\NN=8$
Black Hole Entropy: A One Loop Test of Quantum Gravity}

\end{center}

\vskip .6cm
\medskip

\vspace*{4.0ex}

\baselineskip=18pt

\centerline{\large \rm Shamik Banerjee$^a$, 
Rajesh Kumar Gupta$^b$,
Ipsita Mandal$^c$ and Ashoke Sen$^c$}

\vspace*{4.0ex}

\centerline{\large \it $^a$Dept. of Physics,
Stanford University,
Stanford, CA 94305-4060,
USA}
\centerline{\large \it 
$^b$ Institute of Theoretical Physics, Utrecht University} 
\centerline{\large \it
Leuvenlaan 4, 3584 CE, Utrecht, The Netherlands}
\centerline{\large \it $^c$Harish-Chandra Research Institute}
\centerline{\large \it  Chhatnag Road, Jhusi,
Allahabad 211019, India}

\vspace*{1.0ex}
\centerline{\small E-mail: bshamik@stanford.edu,
R.K.Gupta@uu.nl, ipsita@mri.ernet.in, sen@mri.ernet.in}

\vspace*{5.0ex}

\renewcommand{\check}{\bar }

\centerline{\bf Abstract} \bigskip

We compute logarithmic corrections to the entropy of
supersymmetric extremal black holes in $\NN=4$ and
$\NN=8$ supersymmetric string theories and find results
in perfect agreement with the microscopic results. In
particular these logarithmic corrections vanish for quarter
BPS 
black holes in $\NN=4$ supersymmetric theories, but
has a finite coefficient for 1/8 BPS black holes
in the $\NN=8$ supersymmetric
theory. On the macroscopic side these computations
require evaluating the one loop determinant of massless
fields around the near horizon geometry, and include, in
particular, contributions from dynamical four dimensional
gravitons propagating in the loop. Thus our analysis provides
a test of one loop quantum gravity corrections to the
black hole entropy, or equivalently of the $AdS_2/CFT_1$
correspondence. We also extend our analysis to $\NN=2$
supersymmetric STU model and make a prediction for the
logarithmic correction to the black hole entropy in that
theory.

\vfill \eject

\baselineskip=18pt

\tableofcontents

\sectiono{Introduction} \label{s1}

Wald's formula gives a generalization of the 
Bekenstein-Hawking entropy
in a classical theory of gravity with higher derivative 
terms, possibly coupled to other matter 
fields\cite{9307038,9312023,9403028,9502009}.
In the extremal limit the near horizon geometry contains an
$AdS_2$ factor, and Wald's formula 
leads to a simple algebraic 
procedure for determining the near horizon field
configurations and the entropy\cite{0506177,0508042}. 
A proposal for computing quantum
corrections to this formula was suggested in
\cite{0809.3304}. In this formulation, 
called the quantum entropy function
formalism,
the degeneracy associated with the black hole
horizon is given by the string theory
partition function $Z_{AdS_2}$
in the near horizon geometry of the black hole.  
Such a partition function is divergent
due to the infinite volume of $AdS_2$, but the rules of 
$AdS_2/CFT_1$ correspondence gives a precise procedure
for removing this divergence.
While in the classical limit this prescription gives us
back the exponential of the Wald entropy, it can in principle be
used to systematically calculate the quantum corrections to the
entropy of an extremal black hole. 

Given this prescription one would like to test this by comparing
with some microscopic results. For $\NN=8$ supersymmetric
string theories obtained by compactifying type II 
string theory on $T^6$ and a class of $\NN=4$ 
supersymmetric string theories obtained by compactifying
type II string theory on $K3\times T^2$ and its 
various orbifolds, the exact formula for the microscopic
index is 
known\cite{9607026,0412287,0505094,0506249,
0508174,0510147,
0602254,0603066,0605210,0607155,0609109,0612011,
0708.1270,0802.0544,0802.1556,0803.2692}.
Furthermore it has been argued in
\cite{0903.1477,1009.3226} that for extremal supersymmetric
black holes preserving four supercharges the
black hole entropy also gives the index, and hence can be
directly compared with the microscopic index. 
Thus these
theories provide us with an ideal ground for testing
the macroscopic formula for the black hole entropy.

The microscopic formula for the index in these theories
shows that in the
limit in which all the components of the charge are large,
the logarithm of the index is given 
by\cite{0412287,0510147,0609109}
\be \label{en4}
S_{micro} = \pi \sqrt{\Delta} + \OO(1) \quad \hbox{for}
\quad \NN=4, \ee
and\cite{0908.0039}
\be \label{en8}
S_{micro} = 
\pi \sqrt{\Delta} - 2 \ln\Delta + \OO(1) \quad \hbox{for}
\quad \NN=8\, ,
\ee
where in both theories
$\Delta$ is the unique quartic combination of the
charges which is invariant under all the continuous
duality transformations.
Using the equality between index and degeneracy for a
black hole, eqs.\refb{en4}, \refb{en8} should be equal to
the entropies of the corresponding black holes.
Now for a classical black hole solution carrying these charges,
the radius of curvature $a$ of the near horizon $AdS_2\times S^2$
geometry is related to $\Delta$ via
\be \label{edelta}
\sqrt\Delta = a^2 / G_N\, , 
\ee
where $G_N$ is the four dimensional Newton's constant. Thus
the leading contribution $\pi\sqrt{\Delta}=4\pi a^2/4G_N$  
is the
classical Bekenstein-Hawking entropy\cite{9507090,9512031}.
This leads to a natural question: can we reproduce
the logarithmic corrections from the 
macroscopic side?\footnote{Earlier approaches to computing
logarithmic corrections to black hole entropy can be found
in \cite{9407001,9408068,9412161,9604118,
9709064,0002040,0005017,
0104010,0112041,0406044,0409024,0805.2220,
0808.3688,0809.1508,0911.4379,1003.1083,1008.4314}.}

This question was partially analyzed in \cite{1005.3044}
where it was argued that such corrections, if present, must
arise from one loop quantum correction to $Z_{AdS_2}$
due to massless fields of the supergravity theory. The stringy modes, massive Kaluza-Klein
excitations along compact directions
and/or higher derivative corrections to the effective action
play no role in this analysis and can be safely ignored.
Furthermore \cite{1005.3044} computed the contribution
to $Z_{AdS_2}$ due to the massless matter multiplets
in $\NN=4$ supersymmetric string theories, and found that
the net contribution vanishes, in agreement with the fact
that the result \refb{en4} is independent of the number of
matter multiplets in the theory.  In this paper we complete
the computation by including the contribution from the
gravity multiplet of $\NN=4$ supersymmetric theories
and also extend the analysis to $\NN=8$ supersymmetric
string theories. In both cases our results are in perfect agreement
with the microscopic results \refb{en4} and \refb{en8}. Since
the computation on the macroscopic side involves one loop
determinant of dynamical
gravitons propagating on $AdS_2\times S^2$, our results can
be taken as a non-trivial confirmation of quantum gravity
contribution to the black hole entropy, or equivalently of
$AdS_2/CFT_1$ correspondence\cite{0809.3304} on 
which the prescription for
computing quantum corrections to the entropy is 
based.\footnote{The analysis 
of \cite{0804.1773,0911.5085,1009.6087} for partition function
of (super-) gravity and higher spin theory in
$AdS_3$ also includes the
effect of graviton loops. However in 3 dimensions there
are no dynamical degrees of freedom in the graviton and
hence only the boundary modes associated with
asymptotic symmetries contribute to the
partition function.
}

We would like to emphasize that the limit of charges we
consider is different from the Cardy limit which, in the
present context, would amount to taking one of the charges
representing momentum along a compact direction
to infinity keeping
the other charges fixed. This was studied in detail in
\cite{1009.3226}. The coefficient
of the $\log \Delta$ term in this limit can also be read out
from the general expression for the microscopic entropy, and 
is given by $-2$ for $\NN=8$ supersymmetric string
theory, and $-(m+2)/4$ for $\NN=4$ supersymmetric 
string theories, $m$ being the total number of matter
multiplets in the theory. Thus for type IIB string theory on
$K3\times T^2$ the coefficient will be $-(22+2)/4=-6$.

Since our analysis will be somewhat technical we shall
now give a brief description of our analysis and the results.
The one loop contribution to $Z_{AdS_2}$ arises from two
sources. First, 
the integration over each eigenmode of the kinetic operator
carrying non-zero
eigenvalue gives a contribution to $Z_{AdS_2}$ through the
determinant of the kinetic operator. 
The logarithm of this determinant can be expressed as integral
over the proper time parameter $s$, with the integrand given
by the trace of the heat kernel\cite{gilkey,0306138} after removing
the contribution due to the zero modes. 
This typically will be proportional
to the infinite volume of $AdS_2\times S^2$ and hence is 
apparently infrared
divergent. But we use the trick of 
\cite{0805.0095,0809.3304}
to express the $AdS_2$ volume as $c_1 \beta +c_2$ where
$c_1$ and $c_2$ are finite constants and $\beta$ is the
(divergent) length of the boundary. The coefficient of
$\beta$ can be
absorbed into a redefinition of the ground state
energy  and the $\beta$ independent term gives the
correction to the black hole
entropy $S_{BH}$. This leaves us with an
infrared finite
contribution to the entropy. 
There is also ultraviolet divergence which comes
from the lower limit of integration of the parameter $s$. This is
regulated by setting the lower limit of $s$
integration to be the string scale $l_s^2$.
The resulting integration over $s$
goes as $ds/s$ in the range $l_s^2 << s << a^2$ where $a$ is the
radius of curvature of $AdS_2$ and $S^2$. This
gives a term of order $\ln (a^2/l_s^2)$ which can be
reinterpreted as $\ln\sqrt\Delta$ using \refb{edelta}.

The other source of logarithmic corrections is the integration
over the zero modes. These zero modes represent eigenmodes
of the kinetic operator with zero eigenvalues and arise due to
the asymptotic symmetries of the euclidean near horizon 
geometry. To find the result of integration over these zero
modes we first make a change of integration variable from
the zero modes to parameters labelling the supergroup of
asymptotic symmetries. The supergroup is parametrized in a
way that its volume is manifestly independent of $a$, -- the
radius of curvature of $AdS_2$ and $S^2$. Thus the net 
$a$ dependence of the zero mode integration arises from the
Jacobian associated with the
change of variables from the field modes to the supergroup
parameters. The $a$ dependence of the Jacobian
can be calculated explicitly and leads to additional corrections
to the entropy proportional to $\ln a$.\footnote{The left-over 
integration over
the supergroup includes integration over both, fermionic and
bosonic modes. With the help of 
localization\cite{0608021,0905.2686,
1012.0265,dabappear} it 
can be shown that the infinite contribution from integration
over the bosonic variables cancel the zero contribution due to
integration over the fermionic modes, leaving behind a finite
result\cite{0905.2686}.}

\begin{table} { 
\begin{center}\def\st{\vrule height 3ex width 0ex}
\begin{tabular}{|l|l|l|l|l|l|l|l|l|l|l|} \hline 
The theory  & non-zero mode contribution & zero
mode contribution & total contribution 
\st\\[1ex] \hline \hline
$\NN=4$ & ${1\over 4}(6+m)\ln\Delta$    & 
$-{1\over 4}(6+m)\ln\Delta$   
& 0 
\st\\[1ex] \hline
$\NN=8$ &   ${5}\ln\Delta$ & $-{7}\ln\Delta$ 
& $-{2}\ln\Delta$ 
\st\\[1ex] 
\hline
\hbox{STU model} &   ${}2\ln\Delta$ & $-\ln\Delta$ 
& $\ln\Delta$ 
\st\\[1ex] \hline
 \hline 
\end{tabular}
\caption{The logarithmic correction to the black hole entropy in
$\NN=4$ and $\NN=8$ supersymmetric string theories
in four
dimensions
and the STU model. We have displayed separately the
contributions from the non-zero modes and the 
zero modes.
$m$ denotes the number of matter multiplet fields in the
$\NN=4$ supergravity theory.
The difference between the zero mode contributions in the
different theories arise solely due to the different number of
gauge fields they have, -- the contributions from the
graviton and gravitino zero modes cancel in all the
theories.
} \label{t1}
\end{center} }
\end{table}

For both $\NN=4$ and $\NN=8$ supergravity theories
we find that the final results of the macroscopic analysis,
after adding up the contribution from the zero mode and
the non-zero mode integration,
are in perfect agreement with the microscopic results
\refb{en4} and \refb{en8}. Even for the $\NN=4$ 
supersymmetric theory where the net contribution 
vanishes, the individual contributions from
the zero modes and the non-zero modes are non-trivial.
This has been illustrated in table \ref{t1} where we have
displayed separately the contributions from the zero modes
and the non-zero modes.

Our analysis can also be extended to compute the
logarithmic correction to the entropy of half BPS black
holes in $\NN=2$ supersymmetric 
STU model\cite{9508064,9901117}. 
The low energy effective action of this theory
is a truncation of the $\NN=4$ supergravity theory, and
furthermore the black hole of the $\NN=4$ supergravity
theory for which we carry out the analysis can be
embedded in this theory. Thus 
eigenmodes and the eigenvalues of the kinetic operator
in the near horizon geometry of the black hole solution 
in the STU model are a subset of the corresponding
eigenmodes and eigenvalues in the $\NN=4$ supersymmetric
string theory.  As a result the
coefficient of the $\ln\Delta$ term in the STU model 
can be found by examining the contribution to the logarithmic
correction to  $\NN=4$ black hole entropy from
this subset. From this we arrive at the
following prediction for the asymptotic growth of black
hole entropy
in the STU model:
\be \label{estu}
\pi\sqrt{\Delta} +  \ln\Delta + \OO(1) \, .
\ee
This logarithmic correction is in apparent violation
of the proposal of \cite{0711.1971}
for the index of half BPS states in the STU model,
and more generally, with the one loop correction to
OSV integral\cite{0405146} 
proposed in \cite{0808.2627}. However
\refb{estu} is consistent with the measure proposed in
\cite{0702146} if we assume that this formula 
is valid for weak topological string
coupling.

The rest of the paper is organized as follows. In \S\ref{seigen}
we review some results on the eigenfunctions and eigenvalues
of the Laplacian operator in $AdS_2\times S^2$ acting on fields
carrying different spins.  \S\ref{s2} we review the general
procedure for computing the
logarithmic correction to the black hole entropy. 
\S\ref{squad} we describe the action, to quadratic order, of the
fluctuations of the gravity multiplet fields of $\NN=4$
supergravity around the near
horizon geometry of the black hole. Most of the results in these
sections were already discussed in \cite{1005.3044}.
\S\ref{s3}, \S\ref{sferm}, \S\ref{szero}, \S\ref{n8} and \S\ref{sstu}
contain the  new results. In \S\ref{s3} we compute the
contribution to the heat kernel (and hence to the
logarithmic correction to the entropy) 
due to the bosonic non-zero modes of the gravity multiplet of the
$\NN=4$ supergravity theory.
\S\ref{sferm} contains the contribution from the fermionic
non-zero modes of the gravity multiplet of the
$\NN=4$ supergravity theory.
In \S\ref{szero} we augment the results by computing the
contribution due to the integration over the zero modes, and
show that the net contribution to the coefficient of
$\ln\Delta$ 
vanishes. In \S\ref{n8} we include the contribution due to
the extra fields which are present in the 
$\NN=8$ supergravity theory
and show that they give the result $-2\ln\Delta$. 
These results
are in agreement with
the microscopic results \refb{en4} and \refb{en8}.
In \S\ref{sstu} we compute logarithmic corrections to the
black hole entropy in the STU model leading to
\refb{estu}.

We conclude this introduction by commenting on the
method we use to compute the heat kernel, and
an
alternative. We compute
the heat kernel by explicitly constructing the eigenstates
and eigenvalues of the kinetic operator in the near
horizon geometry, but we could
also compute the relevant terms by simply computing
the one loop contribution to the trace anomaly 
in the near horizon 
geometry\cite{gilkey,0306138}.
Indeed in the presence of background gravitational
field the contribution to the heat kernel in supersymmetric
theories was computed in 
\cite{christ-duff1,christ-duff2,duffnieu,duffroc}. In order to
apply it to the present problem we either need to repeat
the analysis in the presence of background gauge fields,
or use supersymmetry to determine the possible
structure of the one loop counterterms and hence
the trace anomaly.  Neither of this is 
completely straightforward. Furthermore the trace anomaly
method includes the contribution from the zero modes as
well which, as we have described, need to be analyzed
separately. Thus even if we use the trace anomaly methods
for computing the heat kernel, we need to find separately
the zero modes of the kinetic operator, remove their
contribution from the heat kernel
and then separately evaluate their contribution to the
entropy. To whatever extent we have tested, the two methods
lead to the same result.

\sectiono{Eigenfunctions of Laplacians on $AdS_2$
and $S^2$} \label{seigen}

In this section we shall review the results on 
eigenfunctions and
eigenvalues of the Laplacian operator 
$\square\equiv g^{\mu\nu}D_\mu D_\nu$ 
on $AdS_2$ and $S^2$ for
different tensor and spinor fields.
These have been studied 
extensively in \cite{campo,camhig1,campo2,camhig2},
and also discussed in the context of near horizon geometry
of black holes in \cite{1005.3044}.
We consider the background $AdS_2\times S^2$ space 
with a metric of the form:
\be \label{emets}
ds^2 = a^2 (d\eta^2 +\sinh^2\eta \, d\theta^2) + a^2 (d\psi^2 +
\sin^2\psi d\phi^2)\, .
\ee
We shall denote by
$x^m$ the coordinates $(\eta,\theta)$
on $AdS_2$ and by $x^\alpha$ the
coordinates $(\psi,\phi)$ on $S^2$
and introduce
the invariant antisymmetric tensors $\vareps_{\alpha\beta}$
on $S^2$ and $\vareps_{mn}$ on $AdS_2$
respectively, computed with the background metric
\refb{emets}:
\be \label{edefve}
\ve_{\psi\phi} = a^{2}\, \sin\psi, \qquad \ve_{\eta\theta}
= a^{2} \, \sinh\eta\, .
\ee
All indices will be raised and lowered with the background metric $g_{\mu\nu}$ defined in
\refb{emets}.

We shall first review the construction of the eigenstates and
eigenvalues of the Laplacian acting on individual fields in
$AdS_2$ and $S^2$ separately. First consider
the Laplacian acting on the
scalar fields.
On $S^2$ the normalized
eigenfunctions of $-\square$ are just the usual spherical harmonics
$Y_{lm}(\psi,\phi)/a$ with eigenvalues $l(l+1)/a^2$. 
On the other hand on $AdS_2$ the $\delta$-function
normalized eigenfunctions
of $-\square$ are given by\cite{camhig1}\footnote{Although often
we shall give the basis states in terms of complex functions, we
can always work with a real basis by choosing the real and
imaginary parts of the function.}
\ben \label{e5p}
f_{\lambda,\ell}(\eta,\theta)
&=& {1\over \sqrt{2\pi\, a^2}}\, {1\over 2^{|\ell|} (|\ell|)!}\, \left|
{\Gamma\left(i\lambda +{1\over 2} + |\ell|\right)\over
\Gamma(i\lambda)}\right|\, 
e^{i\ell\theta} \sinh^{|\ell|}\eta\nn
&& F\left(i\lambda +{1\over 2}+|\ell|, -i\lambda 
+{1\over 2}+|\ell|; |\ell|+{1}; -\sinh^2{\eta\over 2}\right), \nn
&& \qquad \qquad \qquad \qquad
\ell\in \ZZZ, \qquad 0<\lambda<\infty\, ,
\een
with eigenvalue $\left({1\over 4}+\lambda^2\right)/a^2$. 
Here $F$ denotes hypergeometric function.

The normalized basis of vector fields
on $S^2$ may be taken as
\be \label{ebasis}
{1\over \sqrt{\kappa_1^{(k)}}} \, \p_\alpha U_k, \qquad
{1\over \sqrt{\kappa_1^{(k)}}} \, \vareps_{\alpha\beta} 
\p^\beta U_k, \, ,
\ee
where $\{U_k\}$ denote normalized eigenfunctions of the scalar
Laplacian with eigenvalue $\kappa_1^{(k)}$. The basis states
given in \refb{ebasis} have eigenvalue of $-\square$ equal to
$\kappa_1^{(k)} - a^{-2}$. Note that for $\kappa_1^{(k)}=0$,
\i.e.\ for $l=0$, $U_k$ is a constant and $\p_m U_k$ vanishes.
Hence these modes do not exist for $l=0$.

Similarly 
a normalized basis of vector fields
on $AdS_2$ may be taken as
\be \label{ebasistwo}
{1\over \sqrt{\kappa_2^{(k)}}} \, \p_m W_k, \qquad
{1\over \sqrt{\kappa_2^{(k)}}} \, \vareps_{mn} \p^n W_k, \, ,
\ee
where $W_k$ are the $\delta$-function
normalized eigenfunctions of the scalar
Laplacian with eigenvalue $\kappa_2^{(k)}$. The basis states
given in \refb{ebasistwo} has eigenvalues of $-\square$ equal to
$\kappa_2^{(k)} + a^{-2}$. There are also
additional square integrable modes of eigenvalue
$a^{-2}$, given 
by\cite{camhig1}
\be \label{e24p}
A = d\Phi, \qquad \Phi = {1\over \sqrt{2\pi |\ell|}}\,
\left[ {\sinh\eta \over 1+\cosh\eta}\right]^{|\ell|} e^{i\ell\theta},
\quad \ell = \pm 1, \pm 2, \pm 3, \cdots\, .
\ee
These are not included in \refb{ebasistwo} since the $\Phi$ given in
\refb{e24p} is not normalizable. 
$d\Phi$ given in \refb{e24p} is
self-dual or anti-self-dual depending on the sign of $\ell$.
Thus we do not get independent eigenfunctions
from $*d\Phi$. However we can also work with a real basis
in which we take $d Re(\Phi)$ and $d Im(\Phi)\propto *d Re(\Phi)$
as the independent basis states for $\ell>0$.

A similar choice of basis can be made for a symmetric rank two
tensor representing the graviton fluctuation. For example on $S^2$
we can choose a basis of these modes to be
\be \label{ebasisgrav}
{1\over \sqrt 2}\, g_{\alpha\beta} U_k, \qquad 
{1\over  \sqrt{2(\kappa_1^{(k)} - 2 a^{-2})}} \, \left[
D_\alpha\xi_\beta + D_\beta \xi_\alpha - D^\gamma 
\xi_\gamma \, g_{\alpha\beta}\right]\, ,
\ee
where $\xi_\alpha$ denotes one of the two vectors given in
\refb{ebasis}. Note that for $\kappa_1^{(k)}=2a^{-2}$, \i.e.\ for
$l=1$, the second set of states given in \refb{ebasisgrav}
vanishes since the corresponding $\xi_\alpha$'s label the
conformal Killing vectors of the sphere.

On $AdS_2$ the basis states for a symmetric rank two
tensor may be chosen as
\be \label{ebasgravtwo}
{1\over \sqrt 2}\, g_{mn} W_k, \qquad 
{1\over  \sqrt{2(\kappa_2^{(k)} + 2 a^{-2})}} \, \left[
D_m\wh\xi_n + D_n \wh\xi_m - D^p 
\wh\xi_p \, g_{mn}\right]\, ,
\ee
where $\wh\xi_m$ denotes one of the two vectors given in
\refb{ebasistwo},  or the vector given in
\refb{e24p}. Besides these there is another set of
square integrable modes of eigenvalue $2a^{-2}$ of
$-\square$, given by\cite{camhig1}
\ben \label{ehmn}
h_{mn} dx^m dx^n &=&
{a\over  \sqrt{\pi}} \, \left[ {  |\ell| 
(\ell^2-1)\over
2} \right]^{1/2} \, {(\sinh\eta)^{|\ell|-2} \over (1 +\cosh\eta)^{|\ell|}}
\, e^{i\ell\theta}\, (d\eta^2 + 2 \, i\, \sinh\eta\, d\eta d\theta
- \sinh^2\eta \, d\theta^2) \nn
&& \quad \ell\in \ZZZ, \quad |\ell|\ge 2\, .
\een
Locally these can be regarded as deformations generated by
a diffeomorphism on $AdS_2$, 
but these diffeomorphisms 
themselves are not square integrable. 

We can construct
the basis states of various fields on $AdS_2\times S^2$ by taking
the product of the basis states on $S^2$ and $AdS_2$.
For example for a scalar field the basis states will be given by
the product of $Y_{lm}(\psi,\phi)$ with the states
given in \refb{e5p}, and satisfy
\be \label{eigen1}
\square\, f_{\lambda,k}(\eta,\theta) \, Y_{lm}(\psi, \phi)
= -{1\over a^2}\, 
\left\{l(l+1) + \lambda^2 +{1\over 4} \right\}
\, f_{\lambda,k}(\eta,\theta) \, Y_{lm}(\psi, \phi)\, .
\ee
For a vector field
on $AdS_2\times S^2$ the basis
states will contain two sets. One set
will be given by the product of $Y_{lm}(\psi,\phi)$ and
\refb{ebasistwo} or \refb{e24p}. The other set will
contain the product of
the functions \refb{e5p} on $AdS_2$ and the
vector fields \refb{ebasis} on $S^2$.

Finally we turn to the basis states for the fermion fields.
Consider a Dirac spinor\footnote{Even if the spinors satisfy Majorana/Weyl
condition, we shall compute their heat kernel by first computing the result
for a Dirac spinor and then taking appropriate square roots.}
on $AdS_2\times S^2$. It decomposes
into a product of a Dirac spinor on $AdS_2$ and a Dirac spinor
on $S^2$. 
We use the following conventions for the vierbeins and the gamma
matrices
\be \label{evier1}
e^0= a\, \sinh\eta \, d\theta, \quad e^1 = a\, d\eta, \quad
e^2 = a\, \sin\psi\, d\phi, \quad e^3 = a\, d\psi\, ,
\ee
\be \label{egam1}
\gamma^0 = -\sigma_3\otimes \tau_2, \quad \gamma^1 = \sigma_3
\otimes \tau_1, \quad \gamma^2 = -\sigma_2\otimes I_2, \quad 
\gamma^3 = \sigma_1\otimes I_2\, ,
\ee
where $\sigma_i$ and $\tau_i$ are two dimensional Pauli matrices
acting on different spaces and $I_2$ is $2\times 2$ identity
matrix. In this convention 
the Dirac operator on
$AdS_2\times S^2$ can be written as
\be \label{ek1}
\not \hskip-4pt D_{AdS_2\times S^2}
= \not \hskip-4pt D_{S^2} + \sigma_3 \, 
\not \hskip-4pt D_{AdS_2}\, ,
\ee
where
\be \label{ed1}
\not \hskip -4pt D_{S^2} = a^{-1}\left[ -
\sigma^2\, {1\over \sin\psi} \p_\phi
+ \sigma^1 \, \p_\psi +{1\over 2}\, \sigma^1\, \cot\psi\right]\, ,
\ee
and
\be \label{ed1a}
\not \hskip -4pt D_{AdS_2} =
a^{-1}\left[ -\tau^2\, {1\over \sinh\eta} \p_\theta
+ \tau^1 \, \p_\eta +{1\over 2}\, \tau^1\, \coth\eta\right]\, .
\ee

First let us analyze the eigenstates of $\not\hskip -4pt D_{S^2}$.
They
are given by\cite{9505009}
\ben \label{ed2}
\chi_{l,m}^{\pm} &=& {1\over \sqrt{4\pi a^2}}\,
{\sqrt{(l-m)!(l+m+1)!}\over l!}\,
e^{ i\left(m+{1\over 2}\right)\phi} 
\pmatrix{ i\, \sin^{m+1}{\psi\over 2}\cos^m {\psi\over 2}
P^{\left(m+1, m\right)}_{l-m}(\cos\psi)
\cr
\pm \sin^{m}{\psi\over 2}\cos^{m+1} {\psi\over 2}
P^{\left(m, m+1\right)}_{l-m}(\cos\psi)
}, \nn
\eta_{l,m}^{\pm} &=& {1\over \sqrt{4\pi a^2}}\,
{\sqrt{(l-m)!(l+m+1)!}\over l!}\,
e^{ -i\left(m+{1\over 2}\right)\phi} 
\pmatrix{ \sin^{m}{\psi\over 2}\cos^{m+1} {\psi\over 2}
P^{\left(m, m+1\right)}_{l-m}(\cos\psi)
\cr
\pm i \, \sin^{m+1}{\psi\over 2}\cos^m {\psi\over 2}
P^{\left(m+1, m\right)}_{l-m}(\cos\psi)}, 
\nn
&& 
\qquad l,m\in \ZZZ, \quad l\ge 0, \quad 0\le m\le l\, ,
\een
satisfying
\be \label{ed3}
\not \hskip -4pt D_{S^2} \chi_{l,m}^\pm =\pm i\, a^{-1}\,
\left(l +1\right) 
\chi_{l,m}^\pm\, , \qquad
\not \hskip -4pt D_{S^2} \eta_{l,m}^\pm =\pm i\, a^{-1}\,
\left(l +1\right) 
\eta_{l,m}^\pm\, .
\ee
Here $P^{\alpha,\beta}_n(x)$ are the Jacobi Polynomials:
\be \label{ed4}
P_n^{(\alpha,\beta)}(x) = { (-1)^n\over 2^n \, n!} (1-x)^{-\alpha}
(1+x)^{-\beta} {d^n\over dx^n} \left[ (1-x)^{\alpha+n}
(1+x)^{\beta+n}\right]\, .
\ee
$\chi^\pm_{l,m}$ and $\eta^\pm_{l,m}$ provide an
orthonormal set of basis functions, {\it e.g.}
\be \label{edefchinorm}
a^2 \int_{S^2} \left(\chi^{\pm}_{l,m}\right)^\dagger \, 
\chi^{\pm}_{l',m'} \, \sin\psi \,
d\psi\, d\phi = \delta_{ll'}\delta_{mm'}
\ee
etc.

The eigenstates of $\not\hskip -4pt D_{AdS_2}$
are given by the analytic continuation of the
eigenstates given in \refb{ed2}\cite{9505009}, making the
replacement $\psi\to i\eta$, $l\to -i\lambda -1$,
\ben \label{ed2a}
\chi_{k}^{\pm}(\lambda) &=& {1\over \sqrt{4\pi a^2}}\,
\left|{\Gamma\left( {1} + k + i\lambda\right)
\over \Gamma(k+1) \Gamma\left({1\over 2}+i\lambda\right)}\right|\,
e^{ i\left(k+{1\over 2}\right)\theta}  \nn
&& \qquad 
\pmatrix{ i \, {\lambda\over k+1}\, 
\cosh^{k}{\eta\over 2}\sinh^{k+1} {\eta\over 2}
F\left(k+1+i\lambda, k+1-i\lambda; k+2;-\sinh^2{\eta\over 2}\right)
\cr
\pm \cosh^{k+1}{\eta\over 2}\sinh^k {\eta\over 2}
F\left(k+1+i\lambda, k+1-i\lambda; k+1;-\sinh^2{\eta\over 2}\right)}, \nn \cr \cr
\eta_{k}^{\pm}(\lambda) &=& {1\over \sqrt{4\pi a^2}}\,
\left|{\Gamma\left( {1} + k + i\lambda\right)
\over \Gamma(k+1) \Gamma\left({1\over 2}+i\lambda\right)}\right|\,
e^{ -i\left(k+{1\over 2}\right)\theta}\nn
&&  \qquad 
\pmatrix{ \cosh^{k+1}{\eta\over 2}\sinh^k {\eta\over 2}
F\left(k+1+i\lambda, k+1-i\lambda; k+1;-\sinh^2{\eta\over 2}\right)
\cr
\pm i \, {\lambda\over k+1}\, 
\cosh^{k}{\eta\over 2}\sinh^{k+1} {\eta\over 2}
F\left(k+1+i\lambda, k+1-i\lambda; k+2;-\sinh^2{\eta\over 2}\right)
}, 
\nn \cr && 
\qquad  k\in \ZZZ, \quad 0\le k<\infty, \quad 0<\lambda<\infty\, ,
\een
satisfying
\be \label{eadsev}
\not \hskip -4pt D_{AdS_2} \chi_{k}^{\pm}(\lambda)=\pm i\, a^{-1}\,
\lambda\,  \chi_{k}^{\pm}(\lambda)
\, , \qquad
\not \hskip -4pt D_{AdS_2} \eta_{k}^{\pm}(\lambda)
=\pm i\, a^{-1}\,
\lambda\,  \eta_{k}^{\pm}(\lambda)
\, .
\ee
$\chi_k^\pm(\lambda)$ and $\eta_k^\pm(\lambda)$
provide an orthonormal set of basis functions on $AdS_2$,
{\it e.g.} 
\be \label{eadsnorm}
a^2 \int \sinh\eta\, d\eta\, d\theta\, 
(\chi_k^\pm(\lambda))^\dagger\,
\chi_{k'}^\pm(\lambda') = \delta_{kk'} \delta(\lambda-
\lambda')\, ,
\ee
etc.

The basis of spinors on $AdS_2\times S^2$ can be constructed
by taking the direct product of the spinors given in
\refb{ed2} and \refb{ed2a}. Suppose that
$\psi_1$ denotes an eigenstate of $\not \hskip-4pt D_{S^2}$
with eigenvalue $i\zet_1$ and
$\psi_2$ denotes an eigenstate of $\not \hskip-4pt D_{AdS_2}$
with eigenvalue $i\zet_2$:
\be \label{ek2}
\not \hskip-4pt D_{S^2} \psi_1 = i\zet_1 \, \psi_1,
\qquad \not \hskip-4pt D_{AdS_2} \psi_2 = i\zet_2 \, \psi_2.
\ee
We have $\zet_1=\pm a^{-1}(l+1)$ and $\zet_2=\pm a^{-1}
\lambda$.
Since $\sigma_3$ anti-commutes with 
$\not \hskip-4pt D_{S^2}$ and commutes with 
$\not \hskip-4pt D_{AdS_2}$, we
have, using \refb{ek1},
\ben \label{ek3}
\not \hskip-4pt D_{AdS_2\times S^2}\, \psi_1\otimes \psi_2
&=&  i\zet_1 \psi_1\otimes \psi_2 +
i\zet_2 \sigma_3 \, \psi_1\otimes \psi_2 \, , \nn
\not \hskip-4pt D_{AdS_2\times S^2} \, \sigma_3\,
\psi_1\otimes \psi_2
&=& i\zet_2  \, \psi_1\otimes \psi_2 
 -i\zet_1 \sigma_3 \, \psi_1\otimes \psi_2
\, .\nn
\een
Diagonalizing the $2\times 2$ matrix we see that
$\not \hskip-4pt D_{AdS_2\times S^2}$ has eigenvalues
$\pm i\sqrt{\zet_1^2 + \zet_2^2}$. Thus the square
of the eigenvalue of $\not \hskip-4pt D_{AdS_2\times S^2}$ is
given by the sum of squares of the eigenvalues of
$\not \hskip-4pt D_{AdS_2}$ and $\not \hskip-4pt D_{S^2}$.

By introducing the `charge conjugation operator' $\wt C=\sigma_2
\otimes \tau_1$ and defining $\bar \psi =\psi^T \wt C$, we can express
the orthonormality relations \refb{edefchinorm}, \refb{eadsnorm} 
as
\be \label{enewnorm}
\int d^4 x \, \sqrt{\det g}\, \, \overline{\left(\chi^+_{l,m} \otimes 
\chi^+_k(\lambda)\right)} \, 
\left(\eta^+_{l',m'}\otimes \eta^-_{k'}(\lambda')\right)
= i \,
\delta_{l,l'}\delta_{m,m'}\delta_{k,k'}\delta(\lambda-\lambda')\, ,
\ee
etc. This is important since eventually we shall be dealing with fields
satisfying appropriate reality conditions for which $\bar\psi$ will be
defined as $\psi^T \wt C$ as far as the $SO(4)$ Clifford algebra
associated with $AdS_2\times S^2$ is concerned (see \refb{emajo}).

In our analysis we shall also need to find a basis in which we can
expand the Rarita-Schwinger field $\Psi_\mu$. 
Let us denote by $\chi$ the spinor $\psi_1\otimes \psi_2$
where $\psi_1$ and $\psi_2$ are eigenstates of
$\not \hskip-4pt D_{S^2}$ and $\not \hskip-4pt D_{AdS_2}$ with
eigenvalues $i\zet_1$ and $i\zet_2$ respectively.
Then a (non-orthonormal set of) 
basis states for expanding $\Psi_\mu$ on $AdS_2\times S^2$
can
be chosen as follows:
\ben \label{ers5}
&&\Psi_\alpha =\gamma_\alpha \chi, \quad \Psi_m=0\, , \nn
&& \Psi_\alpha =0, \quad \Psi_m=\gamma_m \chi,  \nn
&&\Psi_\alpha =D_\alpha  \chi, \quad \Psi_m=0,  \nn
&& \Psi_\alpha =0, \quad \Psi_m= D_m 
\chi \, . 
\een
By including all possible eigenstates $\chi$ of 
$\not \hskip-4pt D_{S^2}$ and $\not \hskip-4pt D_{AdS_2}$
we shall generate the complete set of basis states for
expanding the Rarita-Schwinger field barring the subtleties
mentioned below.

There are two additional points which will be important for our
analysis. First of all 
we have the relations 
\be \label{eindep}
D_\alpha\chi^\pm_{0,0} = 
\pm {i\over 2} \, a^{-1}\, 
\gamma_\alpha\chi^\pm_{0,0}, \qquad
D_\alpha\eta^\pm_{0,0} = 
\pm {i\over 2} \, a^{-1}\, \gamma_\alpha\eta^\pm_{0,0}\, .
\ee
Thus if we take $\chi=\psi_1\otimes \psi_2$ 
where $\psi_1$ corresponds to any of the states
$\chi^\pm_{0,0}$ or $\eta^\pm_{0,0}$, and $\psi_2$ is any eigenstate of
$\not\hskip -4pt D_{AdS_2}$, then 
the basis vectors appearing in \refb{ers5} are not all independent, --
the modes in the third row of \refb{ers5} are related to those in
the first row.
The second point is that the modes given in \refb{ers5}
do not exhaust all the modes of the Rarita Schwinger
operator; there are some additional discrete modes of the form
\be \label{eadd1}
\xi_m^{(k)\pm}\equiv
\psi_1\otimes \left(D_m\pm {1\over 2a} \sigma_3
\gamma_m\right)\chi^\pm_k(i),
\qquad \wh\xi_m^{(k)\pm}\equiv
\psi_1\otimes \left(D_m\pm {1\over 2a}  \sigma_3
\gamma_m\right)
\eta^\pm_k(i), \quad k=1,\cdots \infty \, ,
\ee
where $\chi^\pm_k(\lambda)$ and $\eta^\pm_k(\lambda)$ have been defined
in \refb{ed2a}.
Since $\chi^\pm_k(i)$ and $\eta^\pm_k(i)$ are
not square integrable,
these states are not included in the set  given in \refb{ers5}.
However the modes described in \refb{eadd1} are square
integrable and hence they must be included among the eigenstates
of the Rarita-Schwinger operator.
These modes can be shown to satisfy the chirality projection
condition
\ben \label{eimp1}
\tau_3 \left(D_m\pm {1\over 2a}  \sigma_3
\gamma_m\right) \chi^\pm_k(i)
&=& - \left(D_m\pm {1\over 2a}  \sigma_3
\gamma_m\right) \chi^\pm_k(i), \nn 
\tau_3 \left(D_m\pm {1\over 2a}  \sigma_3
\gamma_m\right)\eta^\pm_k(i)
&=& \left(D_m\pm {1\over 2a}  \sigma_3
\gamma_m\right) \eta^\pm_k(i)\, .
\een

\sectiono{Logarithmic correction
to the black hole entropy} \label{s2}

In this section we shall review the general 
procedure for computing the
logarithmic correction to the extremal black hole entropy.
Suppose
we have an extremal black hole with near horizon geometry
$AdS_2\times S^2$, with equal radius of
curvature 
$a$ of $AdS_2$ and $S^2$. Then
the Euclidean near horizon metric takes the form
given in \refb{emets}. As in \cite{1005.3044}, 
we shall make use of the
flat directions of the classical 
entropy function to choose the near
horizon parameters such that $a$ is the only parametrically
large number, all other parameters {\it e.g.} the string
coupling or the size of the compact space remains fixed
as we take the large charge limit.
Let $Z_{AdS_2}$ denote the partition function of string
theory in the near horizon geometry, evaluated by
carrying out functional integral over all the string fields
weighted by the exponential of the Euclidean action
$\SSS$, 
with boundary conditions such that asymptotically the field
configuration approaches the near horizon geometry
of the black hole.\footnote{Since in $AdS_2$ the asymptotic
boundary conditions fix the electric fields, or equivalently
the charges carried by the black hole, and let the
constant modes of the gauge fields to fluctuate, we need
to include in the path integral a boundary term
$\exp(-i\ointop \sum_k q_k A^{(k)}_\mu dx^\mu)$ where
$A^{(k)}_\mu$ are the gauge fields and $q_k$ are the
corresponding electric charges carried by the black 
hole\cite{0809.3304}.
This term plays a crucial role in establishing that the
classical contribution to the black hole entropy computed
via \refb{eads2cft1} gives us the Wald entropy, but will
not play any 
role in the computation of logarithmic corrections.
}
Then $AdS_2/CFT_1$ correspondence
tells us that the full quantum corrected entropy $S_{BH}$
is related to $Z_{AdS_2}$ via\cite{0809.3304,0903.1477}:
\be \label{eads2cft1}
e^{S_{BH} - E_0 \beta} =  Z_{AdS_2}\, ,
\ee
where $E_0$ is the energy of the ground state
of the black hole carrying a given set of charges,
and $\beta$ denotes the length of the boundary of $AdS_2$
in a regularization scheme that renders the
volume of $AdS_2$ finite by putting an infrared
cut-off $\eta\le \eta_0$.
Let $\Delta\LL_{eff}$ denote the one loop correction to the four
dimensional effective lagrangian density evaluated in the background
geometry \refb{emets}. Then the one loop correction to 
$Z_{AdS_2}$ is given by
\be \label{e2}
\exp\left[
 \int\, \sqrt{\det g} \, d\eta \, d\theta  \, d\psi\,
 d\phi \, \Delta\LL_{eff}\right]
= \exp\left[8\pi^2 \, a^4 \, (\cosh\eta_0-1) \, 
\Delta\LL_{eff}\right]\, .
\ee
The term proportional to
$\cosh\eta_0$ in the exponent
has the interpretation of $-\beta \Delta E_0
+\OO\left(\beta^{-1}\right)$
where $\beta=2\pi a \sinh\eta_0$ is the
length of the boundary of $AdS_2$
parametrized by $\theta$ and $\Delta E_0=-4\pi a^3
\Delta\LL_{eff}$
is the shift in the
ground state energy.
The rest of the contribution in the exponent
can be interpreted as the
one loop correction to the black hole 
entropy\cite{0809.3304,0903.1477}. Thus we have
\be \label{e3}
\Delta S_{BH} = -8\pi^2 a^4\, \Delta\, \LL_{eff}\, .
\ee
While the term in the exponent proportional to $\beta$
and hence $\Delta E_0$ can get further corrections
from boundary terms in the action, the finite part
$\Delta S_{BH}$ is defined unambiguously.
This reduces the problem of computing one loop correction to
the black hole entropy to that of computing one
loop correction to $\LL_{eff}$.
We shall now describe the general procedure for calculating
$\Delta\LL_{eff}$.

The near horizon geometry of the black hole  
has background flux of various
electromagnetic fields through $S^2$ and $AdS_2$. 
In this section we shall ignore the effect of this background
flux, leaving the full problem for later
sections.
Then the dynamics of various fields is controlled essentially
by their coupling to the background metric \refb{emets}.
First consider the case of a massless scalar field.
If we denote the eigenvalues of the scalar laplacian by $\{-\kappa_n\}$
and the corresponding normalized
eigenfunctions by $f_n(x)$ then the heat
kernel $K^s(x,x';s)$ of the scalar Laplacian is defined as
(see \cite{gilkey,0306138} and references therein)
\be \label{eh1}
K^s(x,x';s) = \sum_n \, e^{-\kappa_n\, s} \, f_n(x)\,
f_n(x')\, .
\ee
The superscript $s$ on $K$ reflects that the laplacian acts on
the scalar fields.
In \refb{eh1} we have assumed that we are working in a basis
in which the eigenfunctions are real; if this is not the case then we
need to replace $f_n(x')$ by $f_n^*(x')$.
The contribution of this scalar field to the one loop effective
action can now be expressed as
\be \label{e4}
\Delta\SSS =-{1\over 2}\, \sum_n \ln\kappa_n =
{1\over 2} \int_\eps^\infty {ds\over s} \sum_n e^{-\kappa_n s}
=
{1\over 2}\, \int_{\eps}^\infty\, {ds\over s} \, 
\int d^4 x \, \sqrt{\det g}\,  K^s(x,x; s)\, ,
\ee
where $g_{\mu\nu}$ is the $AdS_2\times S^2$ metric 
and $\eps$
is an ultraviolet cut-off.
Identifying this as\break\noindent 
$\int d^4 x \sqrt{\det g}\, \Delta\LL_{eff}$ we get
\be \label{e3ab}
\Delta\LL_{eff} = {1\over 2} \, \int_{\eps}^\infty\, {ds\over s} \, 
K^s(0;s)\, ,
\ee
where $K^s(0;s)\equiv K^s(x, x; s)$. Note that
using the fact that $AdS_2$ and $S^2$ are homogeneous spaces
we have dropped the dependence on $x$ from $K^s(x,x;s)$.

For higher spin fields the field will carry an extra index (say $a$).
Then we can define the heat kernel $K_{ab}(x,x';s)$
by generalizing \refb{eh1} and the contribution to $\Delta\LL_{eff}$
from these fields will be given by
\be \label{e3abc}
{1\over 2} \, \int_{\eps}^\infty\, {ds\over s} \, 
K_{aa}(x,x;s)\, .
\ee
For notational simplicity
we shall refer to $K_{aa}$ as the heat kernel
and denote it by $K$,
but it should be kept in mind that for higher spin fields
this refers to the trace of the heat kernel.
For fermions there will be an extra minus sign since the fermionic
integral produces a positive power of the determinant.
We shall
choose the convention in which this extra factor is absorbed into the
definition of $K$. Also the fermionic kinetic operator is linear in
derivatives; we shall  
find it convenient to define the heat kernel
using the square of the fermionic kinetic operator, and then include an
extra factor of half in the definition of $K$ to account for the final
square root that we need to take. 

Let us now return to the computation of $K^s(0;s)$.
It follows from
\refb{eh1} and the fact that
$\square_{AdS_2\times S^2}=\square_{AdS_2}+\square_{S^2}$
that the heat kernel of a massless scalar field
on $AdS_2\times S^2$ 
is given by the product
of the heat kernels on $AdS_2$ and $S^2$, and in the
$x'\to x$ limit takes the form\cite{campo}
\be \label{e4a}
 K^s(0;s) = K^s_{AdS_2}(0;s) K^s_{S^2}(0;s)\, .
\ee
$K^s_{S^2}$ and $K^s_{AdS_2}$ in turn can be calculated using
\refb{eh1} since we know the eigenfunctions and the eigenvalues of the
Laplace operator on these respective spaces. 
The eigenfunction on $AdS_2$ are described in \refb{e5p}.
Since $f_{\lambda,\ell}$  vanish at
$\eta=0$ for $\ell\ne 0$, only the $\ell=0$ 
eigenfunctions will contribute to
$K^s_{AdS_2}(0;s)$. At $\eta=0$ the $\ell=0$ 
eigenfunction
has the value
$\sqrt{\lambda\tanh(\pi\lambda)}/\sqrt{2\pi a^2}$.
Thus \refb{eh1} gives
\be \label{e5q}
K^s_{AdS_2}(0;s) =  {1\over 2\pi\, a^2}
\int_0^\infty \, d\lambda \, \lambda
\tanh(\pi\lambda) \, 
\exp\left[- s\left(\lambda^2 +{1\over 4}\right)/a^2\right]\, 
\, .
\ee
On $S^2$ the eigenfunctions are $Y_{lm}(\psi,\phi)/a$
and the corresponding eigenvalues are $-l(l+1)/a^2$.
Since $Y_{lm}$
vanishes at $\psi=0$ for $m\ne 0$, and $Y_{l0}=\sqrt{2l+1}/\sqrt{4\pi}$ 
at $\psi=0$ we have
\be \label{e6p}
K^s_{S^2}(0;s) = {1\over 4\pi a^2} 
\sum_{l} e^{-sl(l+1)/a^2} (2l+1)\, .
\ee
We can bring this to a form similar to
\refb{e5q} by expressing it as
\be \label{ess1}
{1\over 4\pi i\, a^2} \, e^{s/4a^2}\, 
\ointop \, d\wt\lambda \, \wt\lambda \, 
\tan(\pi\wt\lambda)\, e^{- s\wt\lambda^2/a^2}\, ,
\ee
where $\ointop$ denotes integration along a contour that travels
from  $\infty$ to 0 staying below the real axis and returns to
$\infty$ staying above the real axis. By deforming the integration
contour to a pair of straight lines through the origin --
one at an angle $\kappa$ below the positive real
axis and the other at an angle $\kappa$ above the positive
real axis -- we get
\be \label{ess2}
K^s_{S^2}(0;s) ={1\over 2\pi a^2} e^{s/4a^2}
\, Im \, \int_0^{e^{i\kappa}\times \infty}
\, \wt\lambda \, d\wt\lambda\, \tan(\pi\wt\lambda) \, e^{-s\wt\lambda^2/a^2}
\, , \qquad 0<\kappa<< 1\, .
\ee
Combining \refb{e6p} and \refb{e5q} we get the heat kernel of a
scalar field on $AdS_2\times S^2$:
\ben \label{ecomb1}
K^s(0;s) 
&=& {1\over 8\pi^2 a^4}\,
\sum_{l=0}^\infty (2l+1) \int_0^\infty \, d\lambda \, \lambda
\tanh(\pi\lambda) \, \exp\left[ -\bar s \lambda^2 - \bar 
s \left(l+{1\over 2}
\right)^2
\right]\nn
&=& {1\over 4\pi^2 a^4}\,
\int_0^\infty \, d\lambda \, \lambda
\tanh(\pi\lambda) \, 
\, Im \, \int_0^{e^{i\kappa}\times \infty}
\, \wt\lambda \, d\wt\lambda\, \tan(\pi\wt\lambda)\,
\exp\left[ -\bar s \lambda^2 - \bar 
s \wt\lambda^2
\right]\, , \nn
\een
where
\be \label{e7}
\bar s = s/a^2\, .
\ee

We can in principle evaluate the full one loop correction to $S_{BH}$ 
due to massless fields using \refb{e3}, 
\refb{e3ab} and \refb{ecomb1}, 
but our goal is to extract the piece proportional
to $\ln a$ for large $a$. 
Such contributions come from the region of integration
$1  << s << a^2$ or equivalently $a^{-2} << \bar s << 1$. Thus
we need to study the behaviour of  \refb{e5q},
\refb{e6p} for
small $\bar s$. We shall now describe a general procedure for
carrying out this small $\bar s$ expansion, not just for the integrals
appearing in \refb{ecomb1} but for a more general class of integrals
where we insert some powers of $\lambda$ and $\tilde\lambda$ into
the integrand.
For this  we first write
\be \label{expand}
\tanh(\pi\lambda) = 1 + 2 \sum_{k=1}^\infty (-1)^k\, e^{-2\pi k\lambda},
\qquad 
\tan(\pi\wt\lambda) = i \left[ 1 + 2 \sum_{k=1}^\infty (-1)^k\, 
e^{2\pi i k\wt\lambda}\right]\, .
\ee
In the term proportional to 1 in the expression for $\tanh(\pi\lambda)$
($\tan\pi\lambda$) we change the integration variable
in \refb{ecomb1}
from $\lambda$ ($\wt\lambda$) to $u\equiv \bar s \lambda^2$
($v=\bar s \wt\lambda^2$). These integrals can be performed
exactly in terms of $\Gamma$ functions. 
On the other hand in the term
proportional to $e^{-2\pi k \lambda}$ ($e^{2\pi i k \wt\lambda}$)
in the expansion of $\tanh(\pi\lambda)$ ($\tan(\pi\wt\lambda)$)
we change the variable of integration to $u=2\pi k \lambda$
($v=2\pi i k\wt\lambda$) and then expand 
the $e^{-\bar s\lambda^2}$ ($e^{-\bar s\wt\lambda^2}$)
term in a power 
series in $u$ (or $v$).
After performing the integrals and a resummation over $k$
we get
\ben\label{esa1}
&& \int_0^\infty d\lambda \, \lambda\, \tanh(\pi\lambda)\,
e^{-\bar s \lambda^2} \, \lambda^{2n}\cr
&=& {1\over 2} \bar s^{-1 - n} 
\Gamma(1+n) + 2 \sum_{m=0}^\infty \bar s^m 
{(2m+2n+1)!\over m!} \, (2\pi)^{-2(m+n+1)} \, (-1)^m\cr
&& \qquad \qquad \qquad \qquad \qquad \qquad
(2^{-2m -2n-1}-1)\, \zeta(2(m+n+1))\, .
\een
\ben\label{esa2}
&& Im \int_0^{e^{i\kappa}\times \infty} 
d\wt\lambda \, \wt\lambda\, \tan(\pi\wt\lambda)\,
e^{-\bar s \wt\lambda^2} \, \wt\lambda^{2n}\cr
&=& {1\over 2} \bar s^{-1 - n} 
\Gamma(1+n) + 2 \sum_{m=0}^\infty \bar s^m 
{(2m+2n+1)!\over m!} \, (2\pi)^{-2(m+n+1)} (-1)^{n+1}\cr
&& \qquad \qquad \qquad \qquad \qquad \qquad
(2^{-2m -2n-1}-1)\, \zeta(2(m+n+1))\, .
\een
This leads to the
following expression for $K^s_{AdS_2}(0;s)$ and $K^s_{S^2}(0;s)$:
\ben \label{e8}
K^s_{AdS_2}(0;s) &=& {1\over 4\pi a^2\, \bar s}\,
e^{-\bar s / 4} \, \left[ 1 + 
\sum_{n=0}^\infty {(-1)^n\over n!} (2n+1)! {\bar s^{n+1}
\over \pi^{2n+2}} {1\over 2^{2n}} \left(2^{-2n-1}-1\right)
\zeta(2n+2)\right]\nn
&=& {1\over 4\pi a^2\, \bar s}\,
e^{-\bar s / 4} \, \left(1 -{1\over 12} \bar s+ 
{7\over 480} \bar s^2 + \OO(\bar s^3) \right)\, ,
\een
\ben \label{e9}
K^s_{S^2}(0;s) &=& {1\over 4\pi a^2\, \bar s}\,
e^{\bar s / 4} \, \left[ 1 -
\sum_{n=0}^\infty {1\over n!} (2n+1)! {\bar s^{n+1}
\over \pi^{2n+2}} {1\over 2^{2n}} \left(2^{-2n-1}-1\right)
\zeta(2n+2)\right]\nn
&=& {1\over 4\pi a^2\, \bar s}\,
e^{\bar s / 4} \, \left(1 +{1\over 12} \bar s+ 
{7\over 480} \bar s^2 + \OO(\bar s^3) \right)\, .
\een
Substituting \refb{e8} and \refb{e9} into \refb{e4a} we get
\be \label{e10}
K^s(0;s) =  
{1\over 16\pi^2 a^4\, \bar s^2} 
\left( 1 +{1\over 45} \bar s^2 +\OO(s^4)\right)\, .
\ee
Eq.\refb{e3ab} now gives
\be \label{e12}
\Delta\LL_{eff}= {1\over 32\pi^2 a^4} 
\, \int_{\eps/a^2}^\infty \, {d\bar s\over \bar s^3} \, 
\left( 1 +{1\over 45} \bar s^2 +\OO(\bar s^4)\right)\, .
\ee
This integral has a quadratically divergent piece proportional to
$1/\eps^2$. This can be thought of as a renormalization of the
cosmological constant and will cancel against 
contribution from other fields in a supersymmetric theory in which
the cosmological constant is not renormalized. 
Even otherwise in string theory there is a physical cut-off set
by the string scale.\footnote{Typically in a string theory there are
multiple scales {\it e.g.} string scale, Planck scale, scale set by
the mass of the D-branes etc. We shall consider near horizon
background where the string coupling constant as well as all the
other parameters describing the shape, size and the various background
fields along the six compact directions are of order unity. In this
case all these length scales will be of the same order. \label{f1}}
Our main interest is in
the logarithmically divergent piece which comes from the 
order $\bar s^2$ term inside the parentheses. This is given by
\be \label{e13}
{1\over 1440\pi^2 a^4} \, \ln (a^2/\eps)\, ,
\ee
and, according to \refb{e3} gives a contribution to the entropy
\be \label{e14}
\Delta S_{BH} = -{1\over 180} \ln (a^2 /\eps)\, .
\ee

Computation for the higher spin fields follows in a similar manner.
We use the basis described in \S\ref{seigen} to construct
the heat kernel. For evaluating $K(0;s)$ we need to compute
$u(x)^2$ at $x=0$ where $u$ is a generic basis element for the
higher spin fields. This can of course be done using the explicit
form of the basis functions given in \S\ref{seigen} but here we
shall suggest a useful
shortcut. Consider for example the state of the
form $(\kappa_2^{(k)})^{-1} \p_m W_k$ given in \refb{ebasistwo}.
Since $K_{\alpha\alpha}(x,x;s)$ is independent
of $x$ due to the homogeneity of $AdS_2$ and $S^2$, we can
replace the contribution from every term 
to $K_{\alpha\alpha}(x,x;s)$ by the volume average
of the term. Now since $W_k$ and
$(\kappa_2^{(k)})^{-1} \p_m W_k$ are both
$\delta$-function 
normalized states, the volume average of the square of
$(\kappa_2^{(k)})^{-1} \p_m W_k$ over $AdS_2$ is the same as
that of the square of $W_k$; hence we can replace the
square of $(\kappa_2^{(k)})^{-1} \p_m W_k$ by
$W_k(x)^2$ while computing $K_{\alpha\alpha}(x,x;s)$.
The contribution to the heat kernel from this set of
modes will have the same form as \refb{e5q}, except
that the $\exp[-s (\lambda^2 + 1/4)/a^2]$ term will be
replaced by $\exp[-s\gamma(\lambda)]$ where 
$\gamma(\lambda)$ is a function of $\lambda$ that
gives the eigenvalue of the kinetic operator acting on
this state.
Similar remark holds for all other basis states which are obtained
by acting suitable differential operators on the eigenfunctions
of the scalar Laplacian.  This will be illustrated in detail in
\S\ref{s3}.

As we shall see in the later sections, 
in the presence of non-trivial background gauge fields
the individual basis states introduced {\it e.g.} in
\refb{eigen1} and similar basis states for higher spin fields
no longer remain eigenstates of the kinetic
operators. Instead
the kinetic operator is   represented as
a matrix on such basis states for fields of
different spin. The matrix however is still block diagonal, with
each block spanned by basis states built by the action of various
differential operators on the $Y_{lm}(\psi,\phi) f_{\lambda,k}(\eta,
\theta)$ for fixed $(l,\lambda,m,k)$.
In
this case we have to replace the 
$e^{-\bar s \lambda^2 -\bar s l(l+1)}$
factor in the integrand by
$\sum_i \exp[-s\gamma_i(l,\lambda)]$ where the sum over
$i$ runs over all the eigenvectors of this matrix  
and $\gamma_i(l,\lambda)$ represent the 
corresponding eigenvalues.

For the discrete modes given in
\refb{e24p} and \refb{ehmn} we need to evaluate the contribution
explicitly. This can be done by noting that at $\eta=0$ only the
$\ell=\pm 1$ modes in \refb{e24p} are non-vanishing and 
only the $\ell=\pm 2$ modes in \refb{ehmn} are non-vanishing.
This allows us to explicitly evaluate the contribution from the
discrete modes to $K_{AdS_2}(0;s)$ for the vector and the
symmetric tensor fields:
\ben \label{ezcont}
\hbox{vector} &:& {1\over 2\pi a^2} \nn
\hbox{symmetric tensor} &:& {3\over 2\pi a^2}\, .
\een
Again in the presence of non-trivial background field
there can be mixing between the discrete modes of
various fields, carrying the same $l$ label, 
under the action of the kinetic
operator. In this case we have to 
find the eigenvalues $\gamma_i(l)$ of the corresponding
matrix,
and include factors of 
$\exp[-s\gamma_i(l)]$ in the summand
in computing the contribution to the heat kernel from the discrete
modes. 

This procedure for computing the heat kernel for
higher spin fields from that of scalars
does not work for 
fermions since the eigenfunctions  of the fermionic kinetic
operator are not given by simple differential operators
acting on the eigenfunctions of the scalar kinetic operator.
However since the eigenfunctions of the fermionic kinetic
operator are given in \refb{ed2}, \refb{ed2a}, we can use 
this to explicitly compute the heat kernel of a fermion on
$AdS_2\times S^2$. This was done in 
\cite{1005.3044} and the result
for a Dirac fermion is
\be \label{ediracf}
-{1\over 2\pi^2 a^4} 
\int_0^\infty d\lambda e^{-\bar s\lambda^2}
\, \lambda \, \coth(\pi\lambda)
 \sum_{l=0}^\infty
(2l+2)
\, e^{-s\left(l+1\right)^2/a^2}\, .
\ee
Since the basis for the expansion of a spin 3/2 field is given
by various operators acting on the eigenmodes of the spin 1/2
Dirac operator, we can use the previous trick to compute
the heat kernel for spin 3/2 field in terms of the heat kernel
of the spin 1/2 field. This will be illustrated in \S\ref{sferm}.

One final issue that enters the computation is the
following. Typically for higher spin fields 
the heat kernel on $AdS_2\times S^2$
also receives contribution from
zero modes, -- discrete modes representing 
eigenfunctions of the kinetic
operator with zero eigenvalue.\footnote{Note
that here we are refering to the zero eigenvalues of
the full kinetic operator on $AdS_2\times S^2$, taking
into acount the effect of background gauge fields,
and not eigenvalues of the kinetic operators on $AdS_2$
and $S^2$ separately.}
These give $s$
independent contribution to the heat kernel.
Integration over these zero modes cannot be
represented as a determinant of the kinetic operator and
must be computed separately. For this reason we need
to identify in the final expression for the heat kernel
on $AdS_2\times S^2$ the $s$-independent contribution
from the discrete modes and subtract it from the full
heat kernel. We then have to evaluate separately
the contribution due to integraton over these zero modes.
 
It follows from \refb{e3}, \refb{e3ab} and \refb{edelta}
that if the total contribution to
$K(0;s)$ after removing the contribution due to
the zero modes is given by $c/\pi^2 a^4$ for some
constant $c$, then the net logarithmic correction to the
black hole entropy from the non-zero modes will be given by
\be \label{elogfor}
-4c \ln a^2 = -2c\ln \Delta\, .
\ee

\sectiono{Quadratic action of gravity multiplet fields
in $\NN=4$ supergravity} 
\label{squad}

We consider type II string theory compactified on $K3\times
T^2$ or equivalently heterotic string theory on $T^6$. In the
language of heterotic string theory the black hole
solution we consider contains momentum and winding
charge along one of the circles of $T^6$ denoted by the
coordinate $x^4$, and Kaluza-Klein monopole and H-monopole
charges associated with another circle of $T^6$ denoted by
$x^5$. The other compact directions will be
denoted by $x^6,\cdots x^9$.
The quadratic action involving fluctuations of the
various massless fields of $\NN=4$ supergravity around the
near horizon geometry of the black hole was
analyzed in \cite{1005.3044}. The result of this paper shows
that the dependence on the charges can be scaled out by
a simple field redefinition and the final action depends on
the charges only through an overall length parameter $a$
describing the radius of curvature of $AdS_2$ and $S^2$.
The relation between $a$ and the charges has been given in
\refb{edelta}.
It was further found that the net logarithmic correction to
the black hole entropy due to fields in the matter
multiplet vanish. Thus we shall focus on fields in the
gravity multiplet.

We shall use the convention in which the indices $\mu,\nu$
run over all the four coordinates of $AdS_2\times S^2$, the
indices $\alpha,\beta$ run over the coordinates of $S^2$ and
the indices $m,n$ run over the coordinates of $AdS_2$. In
this convention the gravity multiplet fluctuations around the
near horizon geometry are labelled by a set of six vector
fields $\AAA_\mu^{(a)}$ ($1\le a\le 6$), 
a spin two field $h_{\mu\nu}$,
two scalars $\chi_1$ 
and $\chi_2$ describing fluctuations of the axion-dilaton
field, and four gravitino and four dilatino fields.
It is natural to combine the four dilatino fields into a 16
component right-handed Majorana-Weyl spinor $\Lambda$
of the ten
dimensional Lorentz group, and the four gravitino fields into
a set of fields $\{\psi_\mu\}$ where for each $\mu$ ($0\le\mu
\le 3$) $\psi_\mu$ is a 
16 component left-handed 
Majorana-Weyl spinor of the
ten dimensional Lorentz group.\footnote{Our conventions 
here are somewhat different
from that of \cite{1005.3044}, where $\Lambda$ refered
to the dilatino field in ten dimensions, and the four 
dimensional
dilatino, obtain after dimensional reduction, was denoted
by $\lambda$. The latter field is being called $\Lambda$ here,
and we shall not make any reference to the ten dimensional
fields before dimensional reduction.}
In the harmonic gauge the quadratic part of the action
involving these fluctuating fields is given 
by\cite{1005.3044}
\be \label{esbf}
S = S_b + S_f= \int d^4 x \, \sqrt{\det g} (\LL_b +
\LL_f)\, ,
\ee
\ben \label{sboson}
\LL_b &=&  
-{1\over 4} h_{\mu\nu} 
\left(\wt \Delta
h\right)^{\mu\nu}+{1\over 2}\, \chi_1\, \square\, \chi_1
+ {1\over 2}\, \chi_2\, \square\, \chi_2+ {1\over 2} \sum_{a=1}^{6}
\AAA_\mu^{(a)} (g^{\mu\nu}\square  - R^{\mu\nu}) 
\AAA_\nu^{(a)}\nn
&& + {1\over 2} \, a^{-2}\, \left(h^{mn} h_{mn}
- h^{\alpha\beta} h_{\alpha\beta}
+ 2\, \chi_2\, (  h^m_{~m}
-  h^\alpha_{~\alpha})\right) + {\sqrt 2\over  a}
\left[  i \vareps^{mn}\, f^{(1)}_{\alpha m} h^\alpha_{~n}
+  \vareps^{\alpha\beta}\, f^{(2)}_{\alpha m} h_\beta^{~m}
\right]\nn
&& + {1\over 2\sqrt 2 \, a}\, \left[ i \vareps^{mn} f^{(1)}_{mn} 
\left( -2\chi_2 + h^\gamma_{~\gamma} - h^p_{~p}\right)
- \vareps^{\alpha\beta} f^{(2)}_{\alpha\beta}\, 
\left(  -2\chi_2 + h^p_{~p} - h^\gamma_{~\gamma}\right)
\right] \nn
&& +{1\over a\sqrt 2} \, \chi_1 \, \left( i\ve^{mn} f^{(2)}_{mn}
+ \ve^{\alpha\beta} f^{(1)}_{\alpha\beta}\right)
\, ,
\een
and
\ben \label{efermi}
\LL_f &=& -{1\over 2}\Bigg[ \bar \psi_\mu \Gamma^{\mu\nu\rho} D_\nu
\psi_\rho + \bar \Lamb \Gamma^\mu D_\mu \Lamb \nn
&& 
+  {1\over 4\sqrt 2}\, \bar\psi_\mu 
\left[-\Gamma^{\mu\nu\rho\sigma}
+ 2 g^{\mu\sigma} g^{\nu\rho} + 2 \Gamma^{\mu\rho\nu}
\Gamma^\sigma + \Gamma^{\mu\nu}\Gamma^{\rho\sigma}
\right] \left(\bar F^1_{\rho\sigma} \Gamma^4 
+\bar F^2_{\rho\sigma} 
\Gamma^5\right)  \psi_\nu \nn &&
+ {1\over 4} \left[ \bar\psi_\mu \Gamma^{\rho\sigma}
\Gamma^\mu \left(\bar F^1_{\rho\sigma} \Gamma^4 
+\bar F^2_{\rho\sigma} 
\Gamma^5\right) \Lamb
-\bar\Lamb \left(\bar F^1_{\rho\sigma} \Gamma^4 
+\bar F^2_{\rho\sigma} 
\Gamma^5\right) \Gamma^\mu
\Gamma^{\rho\sigma} \psi_\mu
\right] 
\nn
&& - {1\over 2} 
\bar \psi_\mu \Gamma^\mu \Gamma^\nu 
D_\nu\Gamma^\rho
 \psi_\rho \Bigg]\, .
\een
Here
\be \label{edeffmn}
f^{(a)}_{\mu\nu} \equiv \p_\mu \AAA^{(a)}_\nu
-\p_\nu \AAA^{(a)}_\mu\, ,
\ee
\ben \label{e35}
\left(\wt\Delta h\right)_{\mu\nu} 
&=& -\square h_{\mu\nu} - R_{\mu\tau} h^\tau_{~\nu}
- R_{\nu\tau} h_\mu^{~\tau} - 2 R_{\mu\rho\nu\tau} h^{\rho\tau}
+{1\over 2} \, g_{\mu\nu} \, 
g^{\rho\sigma} \, \square\, h_{\rho\sigma}
\nn && + R\, h_{\mu\nu} 
+ \left(g_{\mu\nu} R^{\rho\sigma} + R_{\mu\nu}
g^{\rho\sigma}\right) h_{\rho\sigma} -{1\over 2}\, R\, 
g_{\mu\nu} \, g^{\rho\sigma}\, h_{\rho\sigma}
\, ,
\een
and $\vareps^{\alpha\beta}$ and $\vareps^{mn}$ have been
defined in \refb{edefve}. All indices are raised and lowered by
the background metric $g_{\mu\nu}$ given in
\refb{emets}. $R_{\mu\nu\rho\sigma}$ is the
Riemann tensor on $AdS_2\times S^2$ constructed from the
background metric \refb{emets} and 
$\bar F^1_{\rho\sigma}$ and $\bar F^2_{\rho\sigma}$ are
background gauge field strengths whose non-vanishing
components are
\be \label{efnon}
\bar F^1_{mn} = -{i\over \sqrt 2 a} \ve_{mn}, \qquad
\bar F^2_{\alpha\beta} = {1\over \sqrt 2 a} \ve_{\alpha\beta}
\, .
\ee
$\Gamma^M$ are ten dimensional gamma matrices
chosen as follows:
\be \label{etengamma}
\Gamma^\mu=\gamma^\mu\otimes I_8, \qquad
\Gamma^m=\sigma_3\otimes \tau_3\otimes
\wh\Gamma^m, \qquad 0\le\mu\le 3, \quad
4\le m\le 9,
\ee
where $\gamma^\mu$'s have been defined in 
\refb{egam1} and $\wh \Gamma^m$ are $8\times 8$ $SO(6)$
gamma matrices.
$\bar\psi_\mu$, $\bar\Lambda$ are defined as 
\be \label{emajo}
\bar \psi_\mu \equiv\psi^{T}_\mu\, C, \quad 
\bar\Lamb \equiv
\Lamb^T\, C\, ,
\ee
where $T$ denotes transpose
and $C$ is the $SO(10)$ charge 
conjugation matrix satisfying
\be \label{echco}
(C\Gamma^A)^T = C\Gamma^A, \qquad C^T  = -C\, .
\ee
Our choice for $C$ will be:
\be \label{egamma}
C=\sigma_2\otimes \tau_1\otimes \wh C,
\ee
where  $\wh C$ is the
$SO(6)$ charge conjugation matrix satisfying
\be \label{esogamma}
 (\wh C\wh\Gamma^{p})^T 
=-\wh C\wh\Gamma^{p}, \qquad \wh C^T =\wh C\, .
\ee
We can use the vielbeins to convert
the tangent space indices to coordinate indices and vice versa.
We shall use the same symbol $\Gamma$ for labelling the
gamma matrices carrying coordinate indices. 

The Lagrangian densities given in \refb{sboson} and 
\refb{efermi} includes gauge fixing terms of the form:
\be \label{egauge}
-{1\over 2} 
g^{\rho\sigma}\,
\left(D^\mu h_{\mu\rho} -{1\over 2} D_\rho \, h^\mu_{~\mu}\right)
\left( D^\nu \, h_{\nu\sigma} -{1\over 2} 
D_\sigma h^\nu_{~\nu}\right)
 - {1\over 2} D^\mu \AAA^{(a)}_\mu D^\nu \AAA^{(a)}_\nu
 + {1\over 4} 
\bar \psi_\mu \Gamma^\mu \Gamma^\nu 
D_\nu\Gamma^\rho
 \psi_\rho\, .
\ee
Gauge fixing also leads to a set of ghost fields.
Let us denote by $b_\mu$ and $c_\mu$ the ghosts associated
with diffeomorphism invariance, by $b^{(a)}$ and $c^{(a)}$ the ghosts
associated with the U(1) gauge invariances,
and by $\wt b$, $\wt c$ the ten dimensional left-handed
Majorana-Weyl
bosonic ghosts associated with local supersymmetry. 
Quantization of the gravitino also requires the introduction
of a third ten dimensional right-handed 
Majorana-Weyl bosonic 
ghost field which we shall denote by 
$\tilde e$.\footnote{This comes from the
special nature of the gauge fixing term given 
in \refb{egauge}; to get this
term we first insert into the path integral
the gauge fixing term 
$\delta(\Gamma^\mu \psi_\mu - \xi(x))$
for some
arbitrary space-time dependent spinor $\xi(x)$; 
and then average over all $\xi(x)$ with a weight factor of
$\exp(-\int \sqrt{\det g}
\, \bar \xi \not \hskip -4pt D \xi)$. The integration over
$\xi$ introduces an extra factor of $\det \not \hskip -4pt D$ which
needs to be canceled by an additional spin half bosonic ghost
with the standard kinetic operator
proportional to $\not \hskip -4pt D$.}
Then the total
ghost action is given by\cite{1005.3044}
\be \label{emghost}
\LL_{ghost} = 
\left[b^\mu \left(g_{\mu\nu}\square + R_{\mu\nu}\right) 
 c^\nu + b^{(a)}\square c^{(a)}
- 2 \, b^{(a)} \bar F^a_{\mu\nu} \, D^\mu c^\nu\right]
+ \bar{\tilde b} \, \Gamma^\mu D_\mu \tilde c
+ \bar{\tilde e} \, \Gamma^\mu D_\mu \tilde e\, .
\ee
Our goal will be to compute the one loop contribution to
$\LL_{eff}$ due to these
fields and use this to compute the correction to the black hole
entropy.

\sectiono{Contribution from the integer spin fields} \label{s3}

In this section we shall compute the contribution to the heat kernel
due to the gravity multiplet fields of integer spin -- both physical
fields and the ghosts. 
We begin with the physical bosonic fields which include the
fluctuations $h_{\mu\nu}$, $\AAA^{(a)}_\mu$ for $1\le a\le 6$ and the
scalar fields $\chi_1$ and $\chi_2$. 
{}From the structure of $\LL_{b}$ given in \refb{sboson} we see that
the fields $\AAA^{(a)}_\mu$ for $3\le a\le 6$ are not affected by
the presence of the background flux. Hence their contribution to
the heat kernel is given by that of four regular vector fields in
$AdS_2\times S^2$. This was computed 
in \cite{1005.3044}, but we shall
review the analysis since the method it uses 
will be of use for other fields as well.
As explained in \S\ref{s2}, the general strategy is to 
express
various fields as derivatives of scalar fields and then express the
scalar fields as linear combinations of complete set eigenstates
of the $-\square_{S^2}$ and $-\square_{AdS_2}$ operator. For example
we can write\footnote{Note that we are pretending 
that the eigenvalues are
discrete whereas in reality the eigenvalues of $-\square_{AdS_2}$
are continuous and hence the $u$'s are delta function normalized.
But this does not affect the diagonalization of the kinetic operator.}
\ben \label{ewr1}
\AAA^{(a)}_\alpha &=& \sum_{k} \left[ {1\over \sqrt {\kappa_1^{(k)}}}
\left( P^{(k)}_a \p_\alpha \, u_k + Q^{(k)}_a \ve_{\alpha\beta}
\p^\beta \, u_k\right)\right]\, , \nn
\AAA^{(a)}_m &=&  \sum_{k} \left[ {1\over \sqrt {\kappa_2^{(k)}}}
\left( R^{(k)}_a \p_m      \, u_k + S^{(k)}_a \ve_{mn}
\p^n \, u_k\right)\right], \quad \hbox{for $3\le a\le 6$}\, ,
\een
where $\{u_k\}$ are a complete set of scalar
functions with eigenvalue $\kappa_1^{(k)}=l(l+1)/a^2$ 
of $-\square_{S^2}$ and 
$\kappa_2^{(k)}= \lambda^2 + {1\over 4}$ 
of $-\square_{AdS_2}$ and $P^{(k)}_a$'s, 
$Q^{(k)}_a$'s, $R^{(k)}_a$'s, and $S^{(k)}_a$'s, are
constants. 
Upon substiting \refb{ewr1}  into 
\refb{sboson}, and integrating over
$AdS_2\times S^2$, we shall get an expression
quadratic in the coefficients  $P,Q,R,S$. Orthonormality of the
$u_k$'s guarantee that the quadratic term is block diagonal, with
different blocks labelled by different $k$,
\i.e.\ different $(l,\lambda)$. Thus for each $(l,\lambda)$
we shall have
a finite dimensional matrix to diagonalize.
If we denote the eigenvalues of this
matrix by $\gamma_i(l,\lambda)$, then the
net contribution to the heat kernel will be given by
\be \label{egenfor}
{1\over 8\pi^2 a^4} 
\int_0^\infty d\lambda \, \lambda\, \tanh(\pi\lambda)\, 
\sum_{l=0}^\infty \, (2l+1)\, \sum_i \, 
e^{-\bar s \gamma_i(l,\lambda)} + K_{\rm discrete}\, 
\, ,
\ee
where $K_{\rm discrete}$ denotes the contribution from the
discrete modes given by the product of
$Y_{lm}(\psi,\phi)$ with  \refb{e24p}. 
This can be computed in a
similar way using the fact that the discrete modes of
each vector gives a contribution of $1/2\pi a^2$ to
the $AdS_2$ heat kernel. The corresponding contribution
will involve only a sum over $l$ but
no integration over $\lambda$.
In order to avoid proliferation of indices we shall from now on work
in a fixed $k$ sector and drop the superscript $k$ and the
subscript $a$ from all
subsequent formul\ae. Then the part of the action involving the
coefficients $P,Q,R,S$ is given by
\be \label{eactionpart}
-{1\over 2} (\kappa_1+\kappa_2) (P^2 + Q^2 + R^2 + S^2)\, .
\ee
For $\kappa_1=0$, \i.e.\ for $l=0$ the modes $P$ and $Q$ are
absent since the corresponding $u$ is constant on $S^2$ and
hence $\p_\alpha u$ vanishes. Thus we have four eigenvalues
of the form $\kappa_1+\kappa_2
=[\lambda^2 +(l+{1\over 2})^2]/a^2$ for $l\ge 1$ and two eigenvalues
of the form $\lambda^2+{1\over 4}$. Finally for $\kappa_2=0$,
\i.e.\ $\lambda = i/2$ we have some additional discrete modes.
The net contribution from all these modes to the trace
of the heat kernel is given by:
\be \label{efourgauge}
{4\over 8\pi^2 a^4} \bigg[ e^{-\bar s/4}
\int_0^\infty d\lambda \, \lambda\, \tanh(\pi\lambda)\, 
e^{-\bar s \lambda^2}\, 
\sum_{l=0}^\infty \, (2l+1)\, e^{-\bar s l (l+1)}\, 
(4 - 2\delta_{l,0})
+ \sum_{l=0}^\infty \, (2l+1)\, e^{-\bar s l (l+1)}
\bigg]\, .
\ee
The last term without an integration over $\lambda$ 
represents the 
contribution from the discrete
modes. 

We now turn to the rest of the physical bosonic fields which
include the gauge fields $\AAA_\mu^{(a)}$ for $a=1,2$, the
graviton $h_{\mu\nu}$ and the scalars $\chi_1$ and $\chi_2$.
The analysis proceeds in a similar manner by expanding various
fields in a basis obtained from derivatives of $u_k$.
As before we work in a fixed $k$ sector
and drop the index $k$ since there is no mixing between
sectors with different $k$.
We take the following expansion for different
fields:
\ben \label{ewr2}
&& \AAA^{(1)}_\alpha = {1\over \sqrt {\kappa_1}}
\left( C _1 \p_\alpha \, u + C _2 \ve_{\alpha\beta}
\p^\beta \, u\right)\, , \qquad
\AAA^{(1)}_m =  {1\over \sqrt {\kappa_2 }}
\left( C _3 \p_m      \, u + C _4 \ve_{mn}
\p^n \, u\right)\, , \nn
&& \AAA^{(2)}_\alpha = {1\over \sqrt {\kappa_1}}
\left( C_5 \p_\alpha \, u + C _6 \ve_{\alpha\beta}
\p^\beta \, u\right)\, , \qquad
\AAA^{(2)}_m =  {1\over \sqrt {\kappa_2 }}
\left( C _7 \p_m      \, u + C _8 \ve_{mn}
\p^n \, u\right)\, , \nn
&& h_{m\alpha} = {1\over \sqrt {\kappa_1\kappa_2}}
\left(B_1 \, \p_\alpha \p_m \, u + B_2 \, \ve_{mn}  \p_\alpha \p^n u
+ B_3 \, \ve_{\alpha\beta} \, \p^\beta \p_m u +
B_4 \, \ve_{\alpha\beta} \, \ve_{mn} \, \p^\beta \p^n u\right)\, , \nn
&& h_{\alpha\beta} = {1\over \sqrt 2} \, (i\, B_5 
+ B_6)\, g_{\alpha\beta} \, u + {1\over \sqrt{
\kappa_1 - 2 a^{-2}}}
\left(D_\alpha\xi_\beta + D_\beta \xi_\alpha 
- g_{\alpha\beta} \, D^\gamma\xi_\gamma\right),
 \nn
&& h_{mn} = {1\over \sqrt 2} \, (i\, B_5 
- B_6)\, g_{mn} \, u + {1\over 
\sqrt{\kappa_2 + 2 a^{-2}}}
\left(D_m\wh\xi_n + D_n \wh\xi_m - g_{mn} \, D^p\wh\xi_p\right),
\nn
&& \qquad  \quad \xi_\alpha = {1\over \sqrt{\kappa_1}}\left(
B_7 \p_\alpha \, u 
+ B_8 \, \ve_{\alpha\beta}
\p^\beta \, u\right)\, ,\qquad \wh\xi_m = 
{1\over \sqrt{\kappa_2}}\, \left(B_9 \p_m      \, u + B_{0} 
\, \ve_{mn}
\p^n \, u\right)\, , \nn
&& \chi_1 = C_9 u, \qquad \chi_2 = C_0 u\, ,
\een
where $B_0,\cdots B_9$ and $C_0,\cdots C_9$ are arbitrary coefficients.
The normalizations of the coefficients have been chosen such that the
deformations parametrized by individual coefficients are correctly
normalized and the deformations parametrized by different coefficients
have vanishing inner product.
For reasons to be explained later we shall first consider deformations
associated with $u_k$'s for which $\kappa^{(1)}_k > 2 a^{-2}$ 
(\i.e. $l>1$). 
The need for restricting to modes with $\kappa_1>2 a^{-2}$
is clear
from the denominator factor of $\kappa_1-2a^{-2}$ 
in the expansion \refb{ewr2} of $h_{\alpha\beta}$.
We also exclude all the discrete modes 
on $AdS_2$ corresponding to $\kappa^{(2)}_k=0$
from the initial analysis; they will be incorporated
later.
Note the $i$ multiplying $B_5$, -- we have taken into account that the
conformal factor of the metric, parametrized by $B_5$, has wrong sign 
kinetic
term, and hence must be rotated to lie along the imaginary axis to make
the path integral well defined.
Substituting \refb{ewr2} into \refb{sboson} and
integrating over $AdS_2\times S^2$ using the 
orthonormality of the
basis states we 
get the contribution to the action from the
$\kappa_1> 2 a^{-2}$, $\kappa_2> 0$ modes to be
\ben \label{eget1}
&& 
 - {1\over 2} (\kappa_1 +\kappa_2) \, \left[\sum_{i=0}^9 C_i^2
+ \sum_{i=1}^6 B_i^2 \right] -{1\over 2} (\kappa_1+\kappa_2
- 4\, a^{-2}) (B_7^2 + B_8^2) \nn &&
 -{1\over 2} (\kappa_1+\kappa_2 +4a^{-2})
(B_9^2 + B_0^2) \nonumber \\ 
&& + a^{-2}\, (B_9^2 + B_0^2)
- a^{-2}\, (B_7^2 + B_8^2) - 2\, i \,a^{-2} B_5 B_6
- 2\sqrt 2 \, a^{-2} B_6 C_0
\nn
&& + i \sqrt 2 a^{-1} \left[ -\sqrt{\kappa_1} C_3 B_2 + \sqrt{\kappa_1}
C_4 B_1 + \sqrt{\kappa_2} C_1 B_2 + \sqrt{\kappa_2} C_2 B_4
\right] \nn
&& + \sqrt 2 a^{-1} \left[ -\sqrt{\kappa_1} C_7 B_3 - \sqrt{\kappa_1}
C_8 B_4 + \sqrt{\kappa_2} C_5 B_3 - \sqrt{\kappa_2}
C_6 B_1
\right] \nn
&& + i\sqrt 2 a^{-1} \sqrt{\kappa_2} \, (-C_0 + \sqrt 2 B_6) C_4
+ \sqrt 2 a^{-1} \sqrt{\kappa_1} (C_0 + \sqrt 2 B_6) C_6 \nn
&& 
+ \sqrt 2 a^{-1} (\sqrt{\kappa_1} C_2 + i \sqrt{\kappa_2} C_8)
C_9 \, . 
\een
This needs to be further integrated over $\lambda$ and summed over $l$
to get the full action but we shall work in a sector with fixed $l$ and
$\lambda$ as before.

We can now diagonalize the kinetic operator by analyzing various blocks. First
of all note that $B_7$, $B_8$, $B_9$ and $B_0$ do not have any cross
terms. Hence the eigenvalues in these sectors can be read out immediately
from \refb{eget1}. We get
\ben \label{eget3}
B_7, B_8&:& \qquad \kappa_1 + \kappa_2 - 2 \, a^{-2} \, , \nn
B_9, B_0&:& \qquad \kappa_1 + \kappa_2 + 2 \, a^{-2} \, .
\een
Next we note that the parameters $B_2$, $C_3$ and $C_1$ mix among 
themselves but do not mix with any parameter outside this set. In this
three dimensional subspace the kinetic operator takes the form
\be \label{ekin1}
\pmatrix{ \kappa_1 + \kappa_2 &  ia^{-1}\sqrt{2\kappa_1}
& -ia^{-1}\sqrt{2\kappa_2} \cr ia^{-1}\sqrt{2\kappa_1} & 
\kappa_1 + \kappa_2 & 0\cr -ia^{-1}\sqrt{2\kappa_2} & 0
& \kappa_1 + \kappa_2}\, .
\ee
Diagonalizing this matrix we find the eigenvalues in this sector to be
\be \label{ekin2}
\kappa_1 +\kappa_2, \quad 
\kappa_1 + \kappa_2 \pm ia^{-1}\sqrt{
2(\kappa_1 + \kappa_2)}\, .
\ee
Similarly we find from \refb{eget1} that the parameters
$B_3$, $C_7$, $C_5$ mix among themselves but do not mix with any 
parameters outside this set. In this subspace the kinetic operator
takes the form:
\be \label{ekin3}
\pmatrix{ \kappa_1 + \kappa_2 & a^{-1}\sqrt{2\kappa_1}
& -a^{-1}\sqrt{2\kappa_2} \cr a^{-1}\sqrt{2\kappa_1} & 
\kappa_1 + \kappa_2 & 0\cr -a^{-1}\sqrt{2\kappa_2} & 0
& \kappa_1 + \kappa_2}\, .
\ee
The eigenvalues of this matrix are given by
\be \label{ekin4}
\kappa_1 +\kappa_2, \quad
\kappa_1 + \kappa_2 \pm a^{-1}\sqrt{
2(\kappa_1 + \kappa_2)}\, .
\ee
The parameters $B_4$, $C_2$, $C_8$, $C_9$ mix among themselves but do
not mix with any other parameter. In this four dimensional subspace
the kinetic operator is given by
\be \label{ekin5}
\pmatrix{\kappa_1 + \kappa_2 & -ia^{-1}\sqrt{2\kappa_2}
& a^{-1}\sqrt{2\kappa_1} & 0 \cr -ia^{-1}\sqrt{2\kappa_2}
& \kappa_1 + \kappa_2 & 0 & -a^{-1}\sqrt{2\kappa_1} \cr
a^{-1}\sqrt{2\kappa_1} & 0 & \kappa_1 + \kappa_2 & 
-ia^{-1}\sqrt{2\kappa_2} \cr 0 & -a^{-1}\sqrt{2\kappa_1}
& -ia^{-1}\sqrt{2\kappa_2} & \kappa_1 + \kappa_2
}\, .
\ee
The eigenvalues are
\ben \label{ekin6}
\kappa_1 + \kappa_2 + a^{-1}\sqrt{2(\kappa_1-\kappa_2)}
, \quad 
\kappa_1 + \kappa_2 + a^{-1}\sqrt{2(\kappa_1-\kappa_2)}, \nn
\kappa_1 + \kappa_2 - a^{-1}\sqrt{2(\kappa_1-\kappa_2)}, 
\quad \kappa_1 + \kappa_2 - a^{-1}\sqrt{2(\kappa_1-\kappa_2)}\, .
\een
Finally the remaining parameters $B_1$, $C_4$, $C_6$, $C_0$,
$B_6$, $B_5$ all mix among themselves and produce a kinetic
operator
\be \label{efinkin}
\pmatrix{\kappa_1 + \kappa_2 & -ia^{-1}\sqrt{2\kappa_1} &
a^{-1}\sqrt{2\kappa_2} & 0 & 0 & 0\cr 
-ia^{-1}\sqrt{2\kappa_1} & 
\kappa_1 + \kappa_2 & 0 & i a^{-1}\sqrt{2\kappa_2} &
- 2 ia^{-1}\sqrt{\kappa_2} & 0\cr
a^{-1}\sqrt{2\kappa_2} & 0 & \kappa_1 + \kappa_2 & -
a^{-1}\sqrt{2\kappa_1} 
& - 2a^{-1}\sqrt{\kappa_1}
& 0\cr 
0 & i a^{-1}\sqrt{2\kappa_2} & - a^{-1}\sqrt{2\kappa_1} 
& \kappa_1 + \kappa_2 & 2\sqrt 2\, a^{-2} & 0\cr
0 & - 2 ia^{-1}\sqrt{\kappa_2} & - 2
a^{-1}\sqrt{\kappa_1} &2\sqrt 2\, a^{-2} & 
\kappa_1 + \kappa_2 & 2i a^{-2}\cr
0 & 0 & 0 & 0 & 2ia^{-2} & \kappa_1 + \kappa_2
}\, .
\ee
We shall denote
the eigenvalues 
of this matrix by 
\be \label{edeffi}
\kappa_1+\kappa_2+a^{-2}
f_i(l,\lambda), \qquad 1\le i\le 6\, , \quad 
\kappa_1\equiv l(l+1)/a^2, \quad \kappa_2 \equiv {1\over 4a^2}
+{\lambda^2\over a^2}\, .
\ee
For $\kappa_1=2a^{-2}$ (\i.e.\ $l=1$) 
the modes parametrized by $B_7$ and $B_8$ are absent since the
vectors $\p_\alpha u$ and $\ve_{\alpha\beta}\p^\beta u$ are the
conformal Killing vectors of $S^2$ and hence do not generate any
deformation of the metric.
The rest of the modes are not affected. Thus we get the same set
of eigenvalues except the ones given in the first line of \refb{eget3}.
The net contribution to the heat kernel from the
$l\ge 1$ modes is then given by
\ben \label{enet1}
&& {1\over 8\pi^2 a^4}\, e^{-\bar s/4}\,
\int_0^\infty d\lambda \, \lambda\, \tanh(\pi\lambda)\,
\, e^{-\bar s \lambda^2}
\Bigg[
\sum_{l=1}^\infty  \, (2l+1)
\, e^{-\bar s l(l+1)} \, \bigg\{ 2 + 
2 e^{2\bar s} + 2 e^{-2\bar s}  - 2 \, e^{2\bar s}
\delta_{l,1}
 \nn && \qquad
+ e^{i\bar s \sqrt{2\lambda^2 + 2l(l+1) + {1\over 2}}}
+ e^{-i\bar s \sqrt{2\lambda^2 +2 l(l+1) + {1\over 2}}}
+ e^{\bar s \sqrt{2\lambda^2 +2 l(l+1) + {1\over 2}}}
+ e^{-\bar s \sqrt{2\lambda^2 +2 l(l+1) + {1\over 2}}} 
\nn && \qquad
+2\, e^{\bar s \sqrt {2 l(l+1)- 2\lambda^2 - {1\over 2}}}
+2\, e^{-\bar s \sqrt {2 l(l+1)- 2\lambda^2 - {1\over 2}}}
+ \sum_{i=1}^6 e^{-\bar s f_i(l,\lambda)}
\bigg\}\Bigg]\, .
\een

For $\kappa_1=0$ (\i.e.\ $l=0$)
the function $u$ is a constant on $S^2$,
and as a result all the modes which involve a derivative with
respect to a coordinate of $S^2$ are absent. This will require us
to set to zero the  modes corresponding to $C_1$, $C_2$, $C_5$,
$C_6$, $B_1$, $B_2$, $B_3$, $B_4$, $B_7$ and $B_8$.
The net contribution to the action from the rest of the modes
is given by
\ben \label{eremain}
&&
- {1\over 2} \kappa_2\, \sum_{i=0,3,4,7,8,9} C_i^2- {1\over 2} \kappa_2\,  
\sum_{i=5}^6 B_i^2  
-{1\over 2} (\kappa_2 +2a^{-2})
(B_9^2 + B_0^2) - 2\sqrt 2 \, a^{-2} \, B_6 C_0  \nn
&&
  + i\sqrt 2 a^{-1} \sqrt{\kappa_2} \, (-C_0 + \sqrt 2 B_6) C_4
  + a^{-1}  i \sqrt{2\kappa_2} C_8
C_9- 2\, i \,a^{-2} B_5 B_6 \, . 
\een
Since $C_3$, $C_7$, $B_9$ and $B_0$ do not mix with other fields,
they produce the following eigenvalues of the kinetic
operator:
\be \label{eev1}
\kappa_2, \quad \kappa_2, \quad \kappa_2+2\, a^{-2}, \quad
\kappa_2+2\, a^{-2}\, .
\ee
$C_8$ and $C_9$ mix with each other but not with others,
producing eigenvalues:
\be \label{eev2}
\kappa_2 \pm i \, a^{-1} \, \sqrt{2\kappa_2}\, .
\ee
Finally $C_4$, $C_0$, $B_6$, $B_5$ mix with each other
producing the matrix:
\be \label{eev3}
\pmatrix{  \kappa_2 & i a^{-1}\sqrt{2\kappa_2} &
- 2 ia^{-1}\sqrt{\kappa_2} & 0\cr
i a^{-1}\sqrt{2\kappa_2}  
&  \kappa_2 & 2\sqrt 2 \, a^{-2} & 0\cr
- 2 ia^{-1}\sqrt{\kappa_2}  & 2\sqrt 2 \, a^{-2}  & 
\kappa_2 & 2ia^{-2}\cr
0 & 0 & 2ia^{-2} & \kappa_2
}\, .
\ee
We shall denote the eigenvalues of this matrix by
\be \label{edefgi}
\kappa_2+
a^{-2}\,
g_i(\lambda), \quad 1\le  i\le 4, \quad \kappa_2 \equiv {1\over 4a^2}
+{\lambda^2\over a^2}\, .
\ee
Thus the net contribution to the heat kernel from the $l=0$ modes
is given by
\be \label{enet0}
{1\over 8\pi^2 a^4}\, e^{-\bar s/4}\,
\int_0^\infty d\lambda \, \lambda\, \tanh(\pi\lambda)\,
\, e^{-\bar s \lambda^2}\Bigg[2 + 2\, e^{-2\bar s} + 
e^{i\bar s
\sqrt{2\lambda^2 +{1\over 2}}}
+ e^{-i\bar s\sqrt{2\lambda^2 +{1\over 2}}}
+\sum_{i=1}^4 e^{-\bar s \, g_i(\lambda)}
\Bigg]\, .
\ee

We can combine the contributions \refb{enet1} and \refb{enet0}
as follows. We first extend the sum in \refb{enet1} all the way to
$l=0$ and subtract explicitly the extra contribution due to the
$l=0$ terms. This includes in particular the terms involving 
$f_i(0,\lambda)$. Now it is easy to see that for $l=0$, \i.e.\
$\kappa_1=0$, the $6\times 6$ matrix given in \refb{efinkin}
takes a block diagonal form, with $B_1$ and $C_6$ forming
a $2\times 2$ block with eigenvalues $\kappa_2 \pm a^{-1}
\sqrt{2\kappa_2}$, and $C_4, C_0, B_6, B_5$ forming a
$4\times 4$ block that is identical to the matrix given in
\refb{eev3}. Thus the corresponding $f_i(0, \lambda)$'s
coincide with $\kappa_2 \pm a^{-1}
\sqrt{2\kappa_2}$ and the four $g_i(\lambda)$'s. 
Using this result
we can express the sum of \refb{enet1} and \refb{enet0}
as
\ben \label{enettot}
&& {1\over 8\pi^2 a^4}\, e^{-\bar s/4}\,
\int_0^\infty d\lambda \, \lambda\, \tanh(\pi\lambda)\,
\, e^{-\bar s \lambda^2}
\Bigg[
\sum_{l=0}^\infty  \, (2l+1)
\, e^{-\bar s l(l+1)} \, \bigg\{ 2 + 
2 e^{2\bar s} + 2 e^{-2\bar s}   \nn && \qquad
+ e^{i\bar s \sqrt{2\lambda^2 + 2l(l+1) + {1\over 2}}}
+ e^{-i\bar s \sqrt{2\lambda^2 +2 l(l+1) + {1\over 2}}}
+ e^{\bar s \sqrt{2\lambda^2 +2 l(l+1) + {1\over 2}}}
+ e^{-\bar s \sqrt{2\lambda^2 +2 l(l+1) + {1\over 2}}} 
\nn && \qquad
+2\, e^{\bar s \sqrt {2 l(l+1)- 2\lambda^2 - {1\over 2}}}
+2\, e^{-\bar s \sqrt {2 l(l+1)- 2\lambda^2 - {1\over 2}}}
+ \sum_{i=1}^6 e^{-\bar s f_i(l,\lambda)}
\bigg\}\Bigg] \nn
&& -  {1\over 8\pi^2 a^4}\, e^{-\bar s/4}\,
\int_0^\infty d\lambda \, \lambda\, \tanh(\pi\lambda)\,
\, e^{-\bar s \lambda^2} \Bigg[ 6 + 2 e^{2\bar s} 
+ 2 e^{i\bar s \sqrt{2\lambda^2 + {1\over 2}}}
+ 2 e^{-i\bar s \sqrt{2\lambda^2 + {1\over 2}}}\nn &&
\qquad 
+ 2 e^{\bar s \sqrt{2\lambda^2 + {1\over 2}}}
+ 2 e^{-\bar s \sqrt{2\lambda^2 + {1\over 2}}}
\Bigg]\, .
\een

Finally we need to consider the discrete modes associated with
square integrable wave-functions of various fields on $AdS_2$.
These involve the discrete modes of the vector fields on
$AdS_2$ described in \refb{e24p}, and also the discrete modes
of the symmetric rank two tensor on $AdS_2$ described
in \refb{ehmn}. We can take the product of these modes with any
mode of $S^2$ to describe deformations of vector and symmetric
rank 2 tensors on $AdS_2\times S^2$. Let us denote by 
$\{v^{(k)}_m, \vareps_{mn}v^{(k)n}\}$ 
a real basis of vector fields obtained from the product of the real and
imaginary parts of
\refb{e24p} and a spherical harmonic on $S^2$ with eigenvalue 
$\kappa^{(k)}_1$
of $-\square_{S^2}$ and 
by $w^{(k)}_{mn}$ a real basis for 
symmetric rank two tensors obtained from the product of real and
imaginary parts of
\refb{ehmn} and a spherical harmonic on $S^2$  
 with
eigenvalue $\kappa_1^{(k)}$ of $-\square_{S^2}$.
Note that the eigenvalues of $\square_{AdS_2}$
are already fixed for these modes; so we do not need to specify them.
We shall choose $w^{(k)}_{mn}$ and $v^{(k)}_m$ to be real.
As before we shall drop the superscript $(k)$ and consider the
following deformations for each $k$:
\ben \label{edefo1}
&& \AAA^{(1)}_m = E_1 v_m + \wt E_1 \ve_{mn} v^n, 
\quad \AAA^{(2)}_m 
= E_2 v_m + \wt E_2 \ve_{mn} v^n, \nn
&& h_{m\alpha} = {1\over \sqrt{\kappa_1}}\,
\left(E_3 \p_\alpha v_m + \wt E_3 \ve_{mn} \p_\alpha v^n
+
E_4 \ve_{\alpha\beta} \p^\beta v_m
+ \wt E_4 \ve_{\alpha\beta} \ve_{mn} \p^\beta v^n \right)
\nn
&& h_{mn} ={a\over \sqrt 2}\,
 \left(D_m \wh\xi_n + D_n \wh\xi_m 
- g_{mn} D^p \wh\xi_p\right) \, , \quad 
\wh\xi_m = E_5 v_m + \wt E_5 \ve_{mn} v^n
\, ,
\een
and
\be \label{edefo2.5}
h_{mn} = E_6 w_{mn}\, .
\ee
Note that for $\kappa_1=0$ the modes $E_3,\wt E_3,
E_4, \wt E_4$ are absent; this will be taken care of
in the computation. 
These parameters describe a set of orthonormal deformations as long
as the $v_m$ and $w_{mn}$ are correctly normalized. Also
orthonormality of the various modes on $AdS_2$
guarantee that the modes given in \refb{edefo1}, \refb{edefo2.5}
do not
mix with each other and 
the modes analyzed earlier. Substituting the modes given in
\refb{edefo1} into
\refb{sboson} we arrive
at the following contribution to the action from these modes
\be \label{edefo2}
-{1\over 2}
\kappa_1 \sum_{i=1}^4 (E_i^2 + \wt E_i^2)-
{1\over 2} 
\left(\kappa_1 + 2 a^{-2}\right)\, (E_5^2 + \wt E_5^2)
-  a^{-1}\, \sqrt {2\kappa_1} \, (i E_1 \wt E_3 - i \wt E_1 E_3
+ E_2 E_4 + \wt E_2 \wt E_4)\, .
\ee
The ten eigenvalues of the kinetic operator are
\ben \label{eve1}
&& \kappa_1+2\, a^{-2}, \quad 
\kappa_1+2\, a^{-2},  \quad
\kappa_1 \pm a^{-1}\sqrt{2\kappa_1},
\quad
\kappa_1 \pm a^{-1}\sqrt{2\kappa_1},
\nn
&& \kappa_1 \pm i \, a^{-1}\sqrt{2\kappa_1},
\quad \kappa_1 \pm i \, a^{-1}\sqrt{2\kappa_1}\, .
\een
Note that for $\kappa_1 = 2/a^2$, \i.e.\ for $l=1$ we have
a pair of zero eigenvalues. Physically these arise due
to the fact that the dimensional reduction of the metric
on $S^2$ produces a massless SU(2) gauge field on
$AdS_2$, and these, like the U(1) gauge fields, have
zero modes on $AdS_2$. 
For $\kappa_1=0$ the modes $E_3$, $\wt E_3$, $E_4$ 
and $\wt E_4$ are absent and
we get six eigenvalues
\be \label{eve2}
0, \quad 0, \quad 0, \quad 0, \quad 2a^{-2}, \quad 2a^{-2}\, .
\ee
Finally the modes described in \refb{edefo2.5} does not
mix with anything and  
describes a mode with eigenvalue $\kappa_1$
of the kinetic operator, leading to a contribution
\be \label{elastnew}
-{1\over 2}\, \kappa_1 \, E_6^2\, 
\ee
to the action.
Combining these results and recalling the coefficient of the
contribution from the discrete modes of $AdS_2$ given
in \refb{ezcont} ($1/2\pi a^2$ for the discrete mode of the 
vector\footnote{In carrying out this computation we have to
take into account the fact that we have chosen a real basis
in which $v_m$ and $\ve_{mn} v^n$ are independent
vectors. This gives an extra factor of ${1\over 2}$, leading
to a contribution of $1/4\pi a^2$ per mode from $AdS_2$.
For example the zero modes of a 
free gauge field which does not couple to the
background flux will be described by the modes $E_1$
and $\wt E_1$ with action $\kappa_1(E_1^2 +
\wt E_1^2)$, and together they will give a contribution
of $1/2\pi a^2$ from the $AdS_2$ part.}
and $3/2\pi a^2$ for the discrete mode of the symmetric rank two
tensor) 
we get the net contribution from the discrete modes
to be:
\ben \label{edis1}
&& \hskip -30pt
{1\over 8\pi^2 a^4} \, \Bigg[ \sum_{l=1}^\infty e^{-\bar s l(l+1)}
\, (2l+1)\,
\bigg\{ e^{-2\bar s} +  e^{-\bar s \sqrt{2l(l+1)} }
+ e^{\bar s \sqrt{2l(l+1)}}
+  e^{-i\bar s \sqrt{2l(l+1)} }
+ e^{i\bar s \sqrt{2l(l+1)}}+ 3 \bigg\} \nn 
&& \qquad \qquad + 2 + e^{-2\bar s} +3
\Bigg]
\nn && \hskip -30pt =
{1\over 8\pi^2 a^4} \, \Bigg[ \sum_{l=0}^\infty e^{-\bar s l(l+1)}
\, (2l+1)\,
\bigg\{3+ e^{-2\bar s} +  e^{-\bar s \sqrt{2l(l+1)} }
+ e^{\bar s \sqrt{2l(l+1)}}
+  e^{-i\bar s \sqrt{2l(l+1)} }
+ e^{i\bar s \sqrt{2l(l+1)}} \bigg\} \nn 
&& \qquad \qquad -2
\Bigg]
\, .
\een
The $2 + e^{2\bar s}$ in the second line represents the contribution to the heat
kernel from the product of the
discrete modes of the vector field
with $l=0$ mode on $S^2$ (\i.e.\
with eigenvalues given in \refb{eve2}) and 
the 3 represents the contribution from the modes of
the metric given by the product of
$l=0$ 
modes in $S^2$ and 
$w_{mn}$ in $AdS_2$.

We must also include in our list of bosonic fields in the gravity multiplet
the ghosts which arise during the gauge fixing of the six U(1)
gauge groups and the diffeomorphism group. 
The Lagrangian density
for the ghost fields has been given in \refb{emghost}.
In particular the kinetic term has the form:
\be \label{ekgh1}
-\pmatrix{b^\mu & b^{(a)}} \pmatrix{-g_{\mu\nu}\square 
- R_{\mu\nu} & 0\cr
2 \, \bar F^a_{\rho\nu} \, D^\rho &  -\square} \pmatrix{c^\nu \cr c^{(a)}}\, .
\ee
Since this has a lower triangular form, the off diagonal term
does not affect the eigenvalues. 
Thus the scalar ghosts have the standard
kinetic operator $-\square$ and the twelve 
scalar ghosts arising from U(1) gauge invariance 
gives a contribution $-12K^s(0;s)$:
\ben\label{eghost0}
&& -12\, {1\over 8\pi^2 a^4}\, e^{-\bar s/4}\,
\int_0^\infty d\lambda \, \lambda\, \tanh(\pi\lambda)\,
\, e^{-\bar s \lambda^2}
\Bigg[
\sum_{l=0}^\infty  \, (2l+1)
\, e^{-\bar s l(l+1)} 
\Bigg]\, .\nn
\een
The contribution from the
vector ghosts $b_\mu$, $c_\mu$ can be analyzed by decomposing
them into various modes as {\it e.g.} in \refb{ewr2} for $\kappa_1>0$:
\ben \label{eghdecomp}
b_\alpha &=& A {1\over \sqrt{\kappa_1}}\,
\p_\alpha u + B {1\over \sqrt{\kappa_1}}\,
\ve_{\alpha\beta}\p^\beta u, \nn
b_m &=& C {1\over \sqrt{\kappa_2}}\,
\p_m u + D {1\over \sqrt{\kappa_2}}\,
\ve_{mn} \p^n u, \nn
c_\alpha &=& E {1\over \sqrt{\kappa_1}}\,
\p_\alpha u + F {1\over \sqrt{\kappa_1}}\,
\ve_{\alpha\beta}\p^\beta u, \nn
c_m &=& G {1\over \sqrt{\kappa_2}}\,
\p_m u + H {1\over \sqrt{\kappa_2}}\, \ve_{mn} 
\p^n u\, .
\een
Substituting this into \refb{ekgh1} we get the following action:
\be \label{eghac}
(\kappa_1 + \kappa_2 - 2 a^{-2}) (AE + BF)
+ (\kappa_1 + \kappa_2 + 2 a^{-2})(CG + DH)\, .
\ee
This has four eigenvalues of magnitude
$(\kappa_1 + \kappa_2 - 2 a^{-2})$ and four eigenvalues of magnitude
$(\kappa_1 + \kappa_2 + 2 a^{-2})$.
For $\kappa_1=0$ \i.e.\ $l=0$ 
the modes corresponding to $A,B,E,F$ are missing
and we get the action to be
\be \label{eghac2}
(\kappa_2 + 2 a^{-2})(CG + DH)\, .
\ee
This has four eigenvalues of magnitude $(\kappa_2 + 2 a^{-2})$. The
net
contribution to the trace of the heat kernel from these modes is
\ben \label{eghost1}
&& -{1\over 8\pi^2 a^4}\, e^{-\bar s/4}\,
\int_0^\infty d\lambda \, \lambda\, \tanh(\pi\lambda)\,
\, e^{-\bar s \lambda^2}
\Bigg[
\sum_{l=1}^\infty  \, (2l+1)
\, e^{-\bar s l(l+1)} \bigg\{
4 e^{-2\bar s} + 4 e^{2\bar s}
\bigg\} + 4 e^{-2\bar s} 
\Bigg]\nn
&=& -{1\over 8\pi^2 a^4}\, e^{-\bar s/4}\,
\int_0^\infty d\lambda \, \lambda\, \tanh(\pi\lambda)\,
\, e^{-\bar s \lambda^2}
\Bigg[
\sum_{l=0}^\infty  \, (2l+1)
\, e^{-\bar s l(l+1)} \bigg\{
4 e^{-2\bar s} + 4 e^{2\bar s}
\bigg\} - 4 e^{2\bar s} 
\Bigg]\, .
\nn
\een
To this we must include the contribution from the additional
modes obtained by taking the product of the discrete modes
for vector fields on $AdS_2$
given in \refb{e24p} and the eigenstates of the scalar Laplacian on
$S^2$. These modes have eigenvalues $\kappa_1 + 2 a^{-2}$
and hence gives a contribution to the heat kernel of the form:
\be\label{edisghost}
-{1\over 4\pi^2 a^4} 
\sum_{l=0}^\infty  \, (2l+1)
\, e^{-\bar s l(l+1)} e^{-2\bar s}\, .
\ee

Adding \refb{efourgauge}, \refb{enettot}, 
\refb{edis1}, \refb{eghost0}, \refb{eghost1} and \refb{edisghost}
 we get the following expression for the
total heat kernel from the bosonic sector and the scalar and vector ghost
fields of the
gravity multiplet:
\ben \label{eboson}
K^B_{gravity}(0;s) &=& {1\over 8\pi^2 a^4}\, e^{-\bar s/4}\,
\int_0^\infty d\lambda \, \lambda\, \tanh(\pi\lambda)\,
\, e^{-\bar s \lambda^2}
\Bigg[
\sum_{l=0}^\infty  \, (2l+1)\, e^{-\bar s l(l+1)} \, \nn
&& \times \, \bigg\{ 6 -
2 e^{2\bar s} - 2 e^{-2\bar s}   \nn &&
+ e^{i\bar s \sqrt{2\lambda^2 + 2l(l+1) + {1\over 2}}}
+ e^{-i\bar s \sqrt{2\lambda^2 +2 l(l+1) + {1\over 2}}}
+ e^{\bar s \sqrt{2\lambda^2 +2 l(l+1) + {1\over 2}}}
+ e^{-\bar s \sqrt{2\lambda^2 +2 l(l+1) + {1\over 2}}} \nn &&
+2\, e^{\bar s \sqrt {2 l(l+1)- 2\lambda^2 - {1\over 2}}}
+2\, e^{-\bar s \sqrt {2 l(l+1)- 2\lambda^2 - {1\over 2}}}
+ \sum_{i=1}^6 e^{-\bar s f_i(l,\lambda)}\bigg\}\nn &&
- \bigg\{ 14  -2 e^{2\bar s} + 2 e^{i\bar s
\sqrt{2\lambda^2 +{1\over 2}}}
+ 2 e^{-i\bar s\sqrt{2\lambda^2+{1\over 2}}} + 
2 e^{\bar s
\sqrt{2\lambda^2 +{1\over 2}}}
+ 2 e^{-\bar s\sqrt{2\lambda^2+{1\over 2}}}
\bigg\}\Bigg]\nn &&
+ {1\over 8\pi^2 a^4}\, \sum_{l=0}^\infty\, (2l+1)\, e^{-\bar s l (l+1)}
\, \bigg\{ 7  -  e^{-2\bar s} 
+  e^{\bar s \sqrt{2l(l+1)}}
+  e^{-\bar s \sqrt{2l(l+1)}} \nn && \qquad \qquad
+ e^{i\bar s \sqrt{2l(l+1)}}
+  e^{-i\bar s \sqrt{2l(l+1)}} 
\bigg\} - {1\over 4 \pi^2 a^4}\, .   \een
Following the trick leading to \refb{ess1} we can express this 
as\footnote{Note that although the individual terms in the
sum have branch points on the real $\wt\lambda$ axis, the
sum of all the terms inside each curly bracket is free from
such branch point singularities.}
\ben \label{ebosona}
K^B_{gravity}(0;s) &=& {1\over 8\pi^2 a^4}\, e^{-\bar s/4}\,
\int_0^\infty d\lambda \, \lambda\, \tanh(\pi\lambda)\,
\, e^{-\bar s \lambda^2}
\Bigg[ e^{\bar s/4} Im \int_0^{e^{i\kappa}\times\infty} 
d\wt\lambda \, \wt\lambda \tan(\pi\wt\lambda)\, 
\, e^{-\bar s \wt\lambda^2} \, \nn
&& \times 2\times \bigg\{ 6 -
2 e^{2\bar s} - 2 e^{-2\bar s}   \nn &&
+ e^{i\bar s \sqrt{2\lambda^2 + 2\wt\lambda^2}}
+ e^{-i\bar s \sqrt{2\lambda^2 + 2\wt\lambda^2}}
+ e^{\bar s \sqrt{2\lambda^2 + 2\wt\lambda^2}}
+ e^{-\bar s \sqrt{2\lambda^2 + 2\wt\lambda^2}} \nn &&
+2\, e^{\bar s \sqrt {2\wt\lambda^2- 2\lambda^2 - {1}}}
+2\, e^{-\bar s \sqrt {2\wt\lambda^2- 2\lambda^2 - {1}}}
+ \sum_{i=1}^6 e^{-\bar s f_i(\wt\lambda -{1\over 2},\lambda)}
\bigg\}\nn &&
- \bigg\{ 14  -2 e^{2\bar s} + 2 e^{i\bar s
\sqrt{2\lambda^2 +{1\over 2}}}
+ 2 e^{-i\bar s\sqrt{2\lambda^2+{1\over 2}}} + 
2 e^{\bar s
\sqrt{2\lambda^2 +{1\over 2}}}
+ 2 e^{-\bar s\sqrt{2\lambda^2+{1\over 2}}}
\bigg\}\Bigg]\nn &&
+ {1\over 8\pi^2 a^4}\, e^{\bar s/4} Im \int_0^{e^{i\kappa}\times\infty} 
d\wt\lambda \, \wt\lambda \tan(\pi\wt\lambda)\, 
\, e^{-\bar s \wt\lambda^2} \, \nonumber \\
&& \bigg\{ 14  -  2e^{-2\bar s} 
+  2e^{\bar s \sqrt{2\wt\lambda^2 -{1\over 2}}}
+  2e^{-\bar s \sqrt{2\wt\lambda^2 -{1\over 2}}} + 2e^{i\bar s \sqrt{2\wt\lambda^2 -{1\over 2}}}
+  2e^{-i\bar s \sqrt{2\wt\lambda^2 -{1\over 2}}} 
\bigg\} \nn && 
- {1\over 4 \pi^2 a^4}\, .   \een

Our goal is to extract the behavior of this expression in the region
$a^{-2}<<\bar s<<1$ since the logarithmic correction to the entropy
from the non-zero modes come from this domain. This is done using the
same trick as in \S\ref{s2}. 
First we expand all terms in \refb{ebosona} other than the
$e^{-\bar s\lambda^2}$ and $e^{-\bar s\wt\lambda^2}$
factors in a power
series expansion in $\bar s$. 
The only additional subtlety in this analysis comes from the fact
that the eigenvalues $f_i(l,\lambda)$ are not given explicitly. 
However we can use the expansion
\be \label{efexpan}
\sum_{i=1}^6 e^{-\bar s f_i} = \sum_{n=0}^\infty \, {1\over n!}\, 
(-1)^n \bar s^n \sum_{i=1}^6 (f_i)^n
\ee
to reduce the problem to the computation of $\sum_{i=1}^6 (f_i)^n$.
Since $f_i/a^2$'s are the eigenvalues of the matrix 
$M$ obtained by
removing the diagonal $\kappa_1+\kappa_2$ 
terms from the
matrix given in \refb{efinkin}, 
$\sum_{i=1}^6 (f_i)^n$ is given by 
${{a^{2n}}Tr(M^n)}$ which can be easily
computed. In our analysis we need the result 
for $n\le 4$. The results are
\ben \label{empowers}
&& \sum_i f_i 
=0, \quad \sum_i (f_i)^2 = 4(2\wt\lambda^2 -2\lambda^2 +1),
\quad \sum_i (f_i)^3 =
48(\wt\lambda^2 + \lambda^2), \nonumber \\
&& \sum_i (f_i)^4 = -28 - 112 \lambda^2 
+ 80 \lambda^4 + 112 \wt\lambda^2 + 96 \lambda^2 
\wt\lambda ^2 + 80 \wt\lambda^4
\een
This allows us to express 
the right hand side of \refb{ebosona}
in terms of products of factors of the form
\be \label{efactorform}
\int_0^\infty d\lambda \, \lambda\, \tanh(\pi\lambda)\,
e^{-\bar s \lambda^2} \, \lambda^{2n}, \quad
\hbox{and} \quad 
Im \int_0^{e^{i\kappa}\times \infty} 
d\wt\lambda \, \wt\lambda\, \tan(\pi\wt\lambda)\,
e^{-\bar s \wt\lambda^2} \, \wt\lambda^{2n}
\ee
Using eqs.\refb{esa1}, \refb{esa2} we can express 
the right hand side of \refb{ebosona} in a power series
expansion in $\bar s$.
We need to compute up to order $s^0$ term in this expansion
for computation of logarithmic correction to the entropy.
Collecting all the terms of order $s^0$ we get
\be \label{efgrav}
K^B_{gravity}(0;s) = -{13\over 90\pi^2 a^4}
+ {2\over 3\pi^2 a^4} + {2\over 3\pi^2 a^4}
-{1\over 4\pi^2 a^4} +\cdots =
{169 \over 180\pi^2 a^4} + \cdots\, ,
\ee
where $\cdots$ denote terms proportional to $\bar s^{-2}$
and $\bar s^{-1}$ as well as positive powers of $\bar s$.
In the central expression in \refb{efgrav} the four terms
represent respectively the contributions from the terms inside
the three curly brackets in \refb{ebosona} and the last term in
\refb{ebosona}.

Finally we need to remove from this the contribution due to the
zero modes.
To identify the zero modes we
can look for the $s$
independent terms 
in the contribution to the heat kernel from
various discrete modes. These consist of the following:
\begin{enumerate}
\item The $l=0$ modes in the last term in \refb{efourgauge},
giving a contribution of $1/2\pi^2 a^4$ to $K(0;s)$.'These 
represent the zero modes of the four gauge fields
$\AAA^{(a)}_m$ for $3\le a\le 6$.
\item The third term inside $\{~\}$ in the first
line of \refb{edis1} for $l=1$,
giving a contribution of $3/8\pi^2 a^4$. These represent the
zero modes of the SU(2) gauge fields arising out of
dimensional reduction on $S^2$.
\item The 2 and 3 in the second line of \refb{edis1}. The
first one gives a contribution of $2/8\pi^2 a^4$ and
represent the zero modes of the two gauge fields
$\AAA^{(a)}_m$ for $1\le a\le 2$. The second one
gives a contribution of $3/8\pi^2 a^4$ and
represent the zero modes of the metric associated with
the asymptotic symmetries of $AdS_2$.
\end{enumerate}
Thus the net contribution to $K^B_{gravity}(0;s)$ 
from all the zero modes
is given by
\be \label{epx1}
{1\over 2\pi^2 a^4} +{3\over 8\pi^2 a^4} 
+ {1\over 4\pi^2 a^4}  +{3\over 8\pi^2 a^4} 
 = {3\over 2\pi^2 a^4}\, .
\ee
Subtracting
\refb{epx1} from \refb{efgrav} we get the net contribution to
the $s$ independent part of
$K^B_{gravity}(0;s)$ from the non-zero modes:
\be \label{efgravnz}
-{101\over 180 \pi^2 a^4}\, .
\ee

\sectiono{Contribution from the half
integer spin fields} \label{sferm}

Next we must analyze the fermionic contribution to the
heat kernel. For this we express
the gravity multiplet 
part of fermionic action given in
\refb{efermi} as
\be \label{ezxa}
\LL_f =  -{1\over 2} \left(\bar 
\Lamb K^{(1)}
+ \bar\psi^\alpha K^{(2)}_\alpha + \bar \psi^m K^{(3)}_m\right)\, ,
\ee
where
\ben \label{ezya}
K^{(1)} &=& (\not \hskip -4pt D_{S^2} +\sigma_3 \, \not \hskip -4pt D_{AdS_2})
\Lamb
+ {1\over 2\sqrt 2\, 
a} (\sigma_3 \wh \Gamma^4 - i\tau_3 \wh \Gamma^5)
(\Gamma^\beta \psi_\beta - \Gamma^n\psi_n)\, , \nn
K^{(2)}_\alpha &=& -  {1\over 2\sqrt 2\, a} \, \Gamma_\alpha \,
(\sigma_3 \wh \Gamma^4 - i\tau_3 \wh \Gamma^5)\, \Lamb 
- {1\over 2} \Gamma^n (\not \hskip -4pt D_{S^2} +\sigma_3 \, \not \hskip -4pt D_{AdS_2}) \Gamma_\alpha \psi_n \nn
&& - \left({1\over 2}\Gamma^\beta\left(\not \hskip -4pt D_{S^2} 
+\sigma_3 \, \not \hskip -4pt D_{AdS_2}\right) \Gamma_\alpha
+{i \over 2a } \sigma_3
\ve_\alpha^{~\beta} \left(\sigma_3\wh\Gamma^4
-i\tau_3\wh\Gamma^5\right)\right)
\psi_\beta \nn
K^{(3)}_m &=& {1\over 2\sqrt 2\, a} \, \Gamma_m\,
(\sigma_3 \wh \Gamma^4 - i\tau_3 \wh \Gamma^5)\, \Lamb
- {1\over 2} \Gamma^\beta (\not \hskip -4pt D_{S^2} 
+\sigma_3 \, \not \hskip -4pt D_{AdS_2}) \Gamma_m\psi_\beta
\nn &&
+\left( -
{1\over 2}\Gamma^n\left(\not \hskip -4pt D_{S^2} 
+\sigma_3 \, \not \hskip -4pt D_{AdS_2}\right) \Gamma_m
+{i \over 2a}\tau_3 \ve_m^{~n} \left(
\sigma_3\wh\Gamma^4 - i\tau_3\wh\Gamma^5
\right) \right) \psi_n \, .\nn
\een
Let us denote by $\DD$ the differential opeartor such that 
\refb{ezya} may be expressed as
\be \label{ezyb}
\pmatrix{K^{(1)}\cr K^{(2)}_\alpha \cr K^{(3)}_m}
= \DD \, \pmatrix{\Lamb \cr \psi_\alpha\cr \psi_m}\, .
\ee
Our goal will be to calculate the eigenvalues of $\DD$ (or more
precisely $\DD^2$) since these will appear in the expression of the
heat kernel. For this we follow the same
 strategy as in the bosonic case, \i.e.\
instead of working in the infinite dimensional space of
fermionic deformations,
we identify finite dimensional subspaces such that modes inside
one subspace do not mix with the modes outside this subspace
under the action of $\DD$.
Let us pick one
particular basis state $\chi$ for the spinor,
given by the direct product
of ($\chi^+_{lm}$ or $\eta^+_{lm}$) with
($\chi^+(\lambda)$ or $\eta^+(\lambda)$) defined in \refb{ed2},
\refb{ed2a}, and
an arbitrary spinor in the representation
of the Clifford 
algebra generated by $\wh\Gamma^4,\cdots \wh\Gamma^9$
carrying $\wh\Gamma^{45}$ eigenvalue $i$.
Then $\chi$ satisfies
\be \label{ek2arep}
\wh\Gamma^{45}\chi = i\chi, \quad 
\not \hskip-4pt D_{S^2} \chi= i\zet_1 \, \chi,
\quad \not \hskip-4pt D_{AdS_2}\chi= i\zet_2 \, \chi,
\quad \zet_1 > 0, \quad \zet_2\ge 0\, .
\ee
{}From this we can derive the identities:
\be \label{ezx2}
\ve_{\alpha\beta} D^\beta\chi = -i \sigma_3 D_\alpha\chi
-\zet_1 \sigma_3 \Gamma_\alpha\chi\, ,
\quad 
\ve_{mn} D^n\chi = -i\tau_3 D_m\chi - \zet_2 
\tau_3 \sigma_3 \Gamma_m\chi
\, .
\ee
The set of states $\chi$ constructed this way 
do not form a complete set of basis states since we have
left out the states with $\zeta_1<0$ and/or $\zeta_2<0$ and those
with $\wh\Gamma^{45}$ eigenvalue $-i$. We
shall overcome the first two problems 
by including in the basis the states
$\sigma_3\chi$, 
$\tau_3\wh\Gamma^4\wh\Gamma^5\chi=
i\tau_3\chi$ and
$\sigma_3 \tau_3\wh\Gamma^4\wh\Gamma^5\chi=
i \sigma_3 \tau_3\chi$.
Since $\chi$ and 
$\sigma_3\chi$ have 
opposite $\not\hskip -4pt D_{S^2}$ eigenvalues
and 
$\chi$ and 
$\tau_3\chi$ 
have opposite $\not\hskip -4pt D_{AdS_2}$ eigenvalues,
this amounts to including in the basis states with
$\zeta_1<0$ and/or $\zeta_2<0$. To overcome the
last problem
we add four more states in the basis obtained by acting
$\wh\Gamma^4$ on the states already included. 
We now 
consider the subspace consisting of the following fermionic
deformations:
\ben \label{ezx1}
\Lamb &=& a_1 \chi + a_2 \sigma_3\wh \Gamma^4\chi + a_3 \tau_3
\wh\Gamma^5\chi 
+ a_4 \sigma_3 \tau_3\wh \Gamma^4\wh\Gamma^5\chi \nn &&
+ \sigma_3 \left[a_1' \chi + a_2' \sigma_3\wh \Gamma^4\chi 
+ a_3' \tau_3
\wh\Gamma^5\chi 
+ a_4' \sigma_3 \tau_3\wh \Gamma^4\wh\Gamma^5\chi
\right]\, , \nn
\psi_\alpha &=& b_1 \Gamma_\alpha\chi + b_2 \sigma_3 \wh \Gamma^4
\Gamma_\alpha\chi
+ b_3 \tau_3\wh\Gamma^5\Gamma_\alpha\chi  + b_4
\sigma_3 \tau_3 \wh \Gamma^4 \wh\Gamma^5
\Gamma_\alpha\chi \nn && +
b_5 D_\alpha\chi + b_6 \sigma_3 \wh \Gamma^4 D_\alpha\chi
+ b_7 \tau_3 \wh\Gamma^5 D_\alpha\chi  + b_8
\sigma_3 \tau_3  \wh \Gamma^4 \wh\Gamma^5 D_\alpha\chi 
\nn &&
+\sigma_3 \bigg[b_1' \Gamma_\alpha\chi + b_2' \sigma_3 \wh \Gamma^4
\Gamma_\alpha\chi
+ b_3' \tau_3\wh\Gamma^5\Gamma_\alpha\chi  + b_4'
\sigma_3 \tau_3 \wh \Gamma^4 \wh\Gamma^5
\Gamma_\alpha\chi \nn && +
b_5' D_\alpha\chi + b_6' \sigma_3 \wh \Gamma^4 D_\alpha\chi
+ b_7' \tau_3 \wh\Gamma^5 D_\alpha\chi  + b_8'
\sigma_3 \tau_3  \wh \Gamma^4 \wh\Gamma^5 D_\alpha\chi
\bigg] \nn
\psi_m &=& c_1 \Gamma_m \chi + c_2 \sigma_3 \wh \Gamma^4
\Gamma_m \chi
+ c_3 \tau_3\wh\Gamma^5  \Gamma_m \chi  + c_4
\sigma_3 \tau_3 \wh \Gamma^4 \wh\Gamma^5 \Gamma_m \chi \nn 
&& +
c_5 \sigma_3\, D_m \chi + c_6 \wh \Gamma^4 D_m \chi
+ c_7 \sigma_3 \tau_3 \wh\Gamma^5 D_m \chi  + c_8
\tau_3 \wh \Gamma^4 \wh\Gamma^5 D_m \chi 
\nn && 
+\sigma_3\bigg[
c_1' \Gamma_m \chi + c_2' \sigma_3 \wh \Gamma^4
\Gamma_m \chi
+ c_3' \tau_3\wh\Gamma^5  \Gamma_m \chi  + c_4'
\sigma_3 \tau_3 \wh \Gamma^4 \wh\Gamma^5 \Gamma_m \chi \nn 
&& +
c_5' \sigma_3 D_m \chi + c_6' \wh \Gamma^4 D_m \chi
+ c_7' \tau_3 \sigma_3 \wh\Gamma^5 D_m \chi  + c_8'
\tau_3 \wh \Gamma^4 \wh\Gamma^5 D_m \chi\bigg]
\een
where $a_i$'s, $b_i$'s, $c_i$'s, $a'_i$'s, $b'_i$'s and $c_i'$'s
are arbitrary grassman variables and $\chi$ is a fixed spinor
satisfying \refb{ek2arep}. We shall see that the action of
$\DD$ keeps us inside this subspace.

Before we proceed some comments are in order. First 
note that the basis states used in \refb{ezx1} are not orthonormal.
As we shall discuss shortly, this will not affect our analysis.
Second, due to the
relation \refb{eindep} the  basis states 
used in the expansion
\refb{ezx1} are not all independent for $\zet_1=1/a$.
For this reason we shall for now consider the case 
$\zet_1>1/a$. The $\zet_1=1/a$ case will be analyzed
separately. Finally there are additional set of states associated
with the discrete modes described in \refb{eadd1}, -- these
will also be discussed separately.

Using \refb{ezya}, \refb{ezx1} and
\refb{ezx2}
we can 
express $K^{(1)}$, $K^{(2)}$ and $K^{(3)}$ in the form
\ben \label{ezxb}
K^{(1)} &=& A_1 \chi + A_2 \sigma_3\wh \Gamma^4\chi + A_3 \tau_3
\wh\Gamma^5\chi 
+ A_4 \sigma_3 \tau_3\wh \Gamma^4\wh\Gamma^5\chi \nn &&
+ \sigma_3 \left[A_1' \chi + A_2' \sigma_3\wh \Gamma^4\chi 
+ A_3' \tau_3
\wh\Gamma^5\chi 
+ A_4' \sigma_3 \tau_3\wh \Gamma^4\wh\Gamma^5\chi
\right]\, , \nn
K^{(2)}_\alpha &=& 
B_1 \Gamma_\alpha\chi + B_2 \sigma_3 \wh \Gamma^4
\Gamma_\alpha\chi
+ B_3 \tau_3\wh\Gamma^5\Gamma_\alpha\chi  + B_4
\sigma_3 \tau_3 \wh \Gamma^4 \wh\Gamma^5
\Gamma_\alpha\chi \nn && +
B_5 D_\alpha\chi + B_6 \sigma_3 \wh \Gamma^4 D_\alpha\chi
+ B_7 \tau_3 \wh\Gamma^5 D_\alpha\chi  + B_8
\sigma_3 \tau_3  \wh \Gamma^4 \wh\Gamma^5 D_\alpha\chi 
\nn &&
+\sigma_3 \bigg[B_1' \Gamma_\alpha\chi + B_2' \sigma_3 \wh \Gamma^4
\Gamma_\alpha\chi
+ B_3' \tau_3\wh\Gamma^5\Gamma_\alpha\chi  + B_4'
\sigma_3 \tau_3 \wh \Gamma^4 \wh\Gamma^5
\Gamma_\alpha\chi \nn && +
B_5' D_\alpha\chi + B_6' \sigma_3 \wh \Gamma^4 D_\alpha\chi
+ B_7' \tau_3 \wh\Gamma^5 D_\alpha\chi  + B_8'
\sigma_3 \tau_3  \wh \Gamma^4 \wh\Gamma^5 D_\alpha\chi
\bigg] \nn
K^{(3)}_m &=& C_1 \Gamma_m \chi + C_2 \sigma_3 \wh \Gamma^4
\Gamma_m \chi
+ C_3 \tau_3\wh\Gamma^5  \Gamma_m \chi  + C_4
\sigma_3 \tau_3 \wh \Gamma^4 \wh\Gamma^5 \Gamma_m \chi \nn 
&& +
C_5 \sigma_3\, D_m \chi + C_6 \wh \Gamma^4 D_m \chi
+ C_7 \sigma_3 \tau_3 \wh\Gamma^5 D_m \chi  + C_8
\tau_3 \wh \Gamma^4 \wh\Gamma^5 D_m \chi 
\nn && 
+\sigma_3\bigg[
C_1' \Gamma_m \chi + C_2' \sigma_3 \wh \Gamma^4
\Gamma_m \chi
+ C_3' \tau_3\wh\Gamma^5  \Gamma_m \chi  + C_4'
\sigma_3 \tau_3 \wh \Gamma^4 \wh\Gamma^5 \Gamma_m \chi \nn 
&& +
C_5' \sigma_3 D_m \chi + C_6' \wh \Gamma^4 D_m \chi
+ C_7' \tau_3 \sigma_3 \wh\Gamma^5 D_m \chi  + C_8'
\tau_3 \wh \Gamma^4 \wh\Gamma^5 D_m \chi\bigg]\nn
\een
where\footnote{These relations were derived without
using the fact that $\chi$ has $\wt\Gamma^{45}$
eigenvalue $i$ or that $\zeta_1$ and $\zeta_2$ are
positive.}
\ben \label{ezxd}
A_1 &=& i\zet_1 a_1  -{b_2\over \sqrt 2 a}  -
{i b_3\over \sqrt 2 a} - {i\zet_1 b_6\over 2\sqrt 2 a}  +
{\zet_1 b_7\over 2\sqrt 2 a}  -  {c_2\over \sqrt 2 a} 
-  {i c_3\over \sqrt 2 a}- {i\zet_2 c_6\over 2\sqrt 2 a} 
+ {\zet_2  c_7\over 2\sqrt 2 a}+i\zet_2 a_1'
\nn 
A_2 &=& -i\zet_1 a_2 +{b_1\over \sqrt 2 a}  -
{i b_4\over \sqrt 2 a} + {i\zet_1 b_5\over 2\sqrt 2 a}  +
{\zet_1 b_8\over 2\sqrt 2 a}  -  {c_1\over \sqrt 2 a} 
+ {i c_4\over \sqrt 2 a}- {i\zet_2 c_5\over 2\sqrt 2 a} 
- {\zet_2  c_8\over 2\sqrt 2 a}+i\zet_2 a_2' 
\nn
A_3 &=& i\zet_1 a_3  -{i b_1\over \sqrt 2 a}  -
{b_4\over \sqrt 2 a} + {\zet_1 b_5\over 2\sqrt 2 a}  -
{i\zet_1 b_8\over 2\sqrt 2 a}  + {ic_1\over \sqrt 2 a} 
+ {c_4\over \sqrt 2 a}- {\zet_2 c_5\over 2\sqrt 2 a} 
+ {i\zet_2  c_8\over 2\sqrt 2 a}-i\zet_2 a_3'
\nn 
A_4 &=& -i\zet_1 a_4  -{i b_2\over \sqrt 2 a}  +
{b_3\over \sqrt 2 a} + {\zet_1 b_6\over 2\sqrt 2 a}  +
{i\zet_1 b_7\over 2\sqrt 2 a}  -  {ic_2\over \sqrt 2 a} 
+ { c_3\over \sqrt 2 a}+ {\zet_2 c_6\over 2\sqrt 2 a} 
+ {i\zet_2  c_7\over 2\sqrt 2 a}-i\zet_2 a_4'
\nn 
A_1' &=& i\zet_2 a_1  -i\zet_1 a_1'+{b_2'\over \sqrt 2 a}  +
{i b_3'\over \sqrt 2 a} + {i\zet_1 b_6'\over 2\sqrt 2 a}  -
{\zet_1 b_7'\over 2\sqrt 2 a}  -  {c_2'\over \sqrt 2 a} 
-  {i c_3'\over \sqrt 2 a}- {i\zet_2 c_6'\over 2\sqrt 2 a} 
+ {\zet_2  c_7'\over 2\sqrt 2 a}
\nn 
A_2' &=&i\zet_2 a_2  + i\zet_1 a_2'-{b_1'\over \sqrt 2 a}  +
{i b_4'\over \sqrt 2 a} - {i\zet_1 b_5'\over 2\sqrt 2 a}  -
{\zet_1 b_8'\over 2\sqrt 2 a}  -  {c_1'\over \sqrt 2 a} 
+ {i c_4'\over \sqrt 2 a}- {i\zet_2 c_5'\over 2\sqrt 2 a} 
- {\zet_2  c_8'\over 2\sqrt 2 a}
\nn
A_3' &=& -i\zet_2 a_3  -i\zet_1 a_3'+{i b_1'\over \sqrt 2 a}  +
{b_4'\over \sqrt 2 a} - {\zet_1 b_5'\over 2\sqrt 2 a}  +
{i\zet_1 b_8'\over 2\sqrt 2 a}  + {ic_1'\over \sqrt 2 a} 
+ {c_4'\over \sqrt 2 a}- {\zet_2 c_5'\over 2\sqrt 2 a} 
+ {i\zet_2  c_8'\over 2\sqrt 2 a}
\nn 
A_4' &=& -i\zet_2 a_4  + i\zet_1 a_4'+{i b_2'\over \sqrt 2 a}  -
{b_3'\over \sqrt 2 a} - {\zet_1 b_6'\over 2\sqrt 2 a}  -
{i\zet_1 b_7'\over 2\sqrt 2 a}  -  {ic_2'\over \sqrt 2 a} 
+ { c_3'\over \sqrt 2 a}+ {\zet_2 c_6'\over 2\sqrt 2 a} 
+ {i\zet_2  c_7'\over 2\sqrt 2 a}
\nn 
B_1 &=& -{a_2\over 2\sqrt 2 a} + {i a_3\over 2\sqrt 2 a}
\nn &&
- i \zet_1 b_1 +{1\over 2a} b_2 -{i\over 2a}b_3
+ \left(\wt\zet_1^2-{1\over 2} \zet_1^2 + K
\right) b_5 + {i \zet_1\over 2a} b_6 +{\zet_1
\over 2a} b_7 +  {1\over 2} \zet_1\zet_2
 b_5'
 \nn &&
 + i \zet_1 c_1 
-{1\over 2} \zet_1\zet_2  c_5 
+ \left(\wt\zet_2^2-{1\over 2} \zet_2^2
\right) c_5' 
\nn 
B_2 &=& {a_1\over 2\sqrt 2 a} + {i a_4\over 2\sqrt 2 a} 
\nn &&
+{1\over 2a} b_1+ i \zet_1 b_2  +{i\over 2a}b_4
+ {i \zet_1\over 2a} b_5 
- \left(\wt\zet_1^2-{1\over 2} \zet_1^2 + K
\right) b_6 -{\zet_1
\over 2a} b_8 + {1\over 2} \zet_1\zet_2 b_6'
\nn &&
+ i \zet_1 c_2 -  {1\over 2} \zet_1\zet_2  c_6
+\left( 
{1\over 2} \zet_2^2 - \wt\zet_2^2
 \right)c_6' 
\nn 
B_3 &=& {i a_1\over 2\sqrt 2 a} - {a_4\over 2\sqrt 2 a}
\nn &&
-{i\over 2a}b_1- i \zet_1 b_3 +{1\over 2a} b_4 
+{\zet_1
\over 2a} b_5
+ \left(\wt\zet_1^2-{1\over 2} \zet_1^2 + K
\right) b_7 + {i \zet_1\over 2a} b_8  - {1\over 2} \zet_1\zet_2 
b_7'
\nn && - i \zet_1 c_3 + {1\over 2} \zet_1\zet_2 
c_7
+ \left(
\wt\zet_2^2
- {1\over 2} \zet_2^2\right)
c_7'
\nn 
B_4 &=& {i a_2\over 2\sqrt 2 a} + {a_3\over 2\sqrt 2 a}
\nn &&
+{i\over 2a}b_2+{1\over 2a} b_3 + i \zet_1 b_4 
-{\zet_1
\over 2a} b_6+ {i \zet_1\over 2a} b_7
- \left(\wt\zet_1^2-{1\over 2} \zet_1^2 + K
\right) b_8   - {1\over 2} \zet_1\zet_2 b_8'
\nn && 
- i \zet_1 c_4 
+ {1\over 2} \zet_1\zet_2  c_8
+\left( {1\over 2} \zet_2^2 
- \wt\zet_2^2 
 \right) c_8'
\nn 
B_5 &=& 
-{1\over 2a} b_6 +{i\over 2a}b_7 + i\zet_2 b_5'
-2 c_1 - i \zet_2 c_5 
\nn 
B_6 &=& -{1\over 2a} b_5 -{i\over 2a}b_8 + i\zet_2 b_6' 
- 2 c_2 - i\zet_2 c_6 
\nn 
B_7 &=&  {i\over 2a}b_5-{1\over 2a} b_8  - i\zet_2 b_7'
+ 2 c_3 + i\zet_2 c_7
\nn 
B_8 &=& -{i\over 2a}b_6-{1\over 2a} b_7  - i\zet_2 b_8'
+2 c_4 + i\zet_2 c_8
\nn 
B'_1 &=& {a_2'\over 2\sqrt 2 a} - {i a_3'\over 2\sqrt2 a}
\nn &&
+{1\over 2} \zet_1\zet_2 b_5 
+ i\zet_1 b_1'+{1\over 2a} b_2' -{i\over 2a} b_3'
- \left(\wt\zet_1^2-{1\over 2} \zet_1^2 + K
\right) b_5'
+{i\zet_1\over 2a} b_6' +{\zet_1\over 2a} b_7'
\nn && +\left( {1\over 2} \zet_2^2 - \wt\zet_2^2 
 \right) c_5
+ i \zet_1 c_1' 
-  {1\over 2} \zet_1\zet_2 c_5'
\nn 
B'_2 &=& -{a_1'\over 2\sqrt 2 a} - {i a_4'\over 2\sqrt 2 a} 
\nn &&
+{1\over 2} \zet_1\zet_2 b_6
+{1\over 2a} b_1' - i\zet_1 b_2'+{i\over 2a} b_4'
+{i\zet_1\over 2a} b_5' 
+ \left(\wt\zet_1^2-{1\over 2} \zet_1^2 + K
\right) b_6'
-{\zet_1\over 2a} b_8' 
\nn &&
+  \left(\wt\zet_2^2 
-{1\over 2} \zet_2^2 \right) c_6
+ i \zet_1 c_2' 
- {1\over 2} \zet_1\zet_2 c_6'
\nn 
B'_3 &=& -{i a_1'\over 2\sqrt 2 a} + {a_4'\over 2\sqrt 2 a}
\nn &&
-{1\over 2} \zet_1\zet_2 b_7 
-{i\over 2a} b_1' + i\zet_1 b_3'+{1\over 2a} b_4' 
+{\zet_1\over 2a} b_5'
- \left(\wt\zet_1^2-{1\over 2} \zet_1^2 + K
\right) b_7'
+{i\zet_1\over 2a} b_8' 
\nn &&
+ \left( {1\over 2} \zet_2^2 
-  \wt\zet_2^2 
  \right)
c_7
- i \zet_1 c_3' 
+ {1\over 2} \zet_1\zet_2 c_7'
\nn 
B_4' &=& -{i a_2'\over 2\sqrt 2 a} - {a_3'\over 2\sqrt 2 a}
\nn &&
-{1\over 2} \zet_1\zet_2 b_8 
+{i\over 2a} b_2'+{1\over 2a} b_3' - i\zet_1 b_4'
-{\zet_1\over 2a} b_6'+{i\zet_1\over 2a} b_7' 
+ \left(\wt\zet_1^2-{1\over 2} \zet_1^2 + K
\right) b_8'
\nn &&
+\left(\wt\zet_2^2 
-{1\over 2} \zet_2^2 \right)c_8
- i \zet_1 c_4' 
+  {1\over 2} \zet_1\zet_2 c_8'
\nn 
B'_5 &=& 
 i\zet_2 b_5 - {1\over 2a} b_6' +{i\over 2a} b_7'
- 2 c_1' - i\zet_2 c_5'
\nn 
B'_6 &=& 
 i\zet_2 b_6 - {1\over 2a} b_5' -{i\over 2a} b_8'
- 2 c_2' - i\zet_2 c_6'
\nn 
B'_7 &=& 
- i\zet_2 b_7 - {1\over 2a} b_8' +{i\over 2a} b_5'
+ 2 c_3' + i\zet_2 c_7' \nn 
B'_8 &=&- i\zet_2 b_8 - {1\over 2a} b_7' -{i\over 2a} b_6'
+  2 c_4' + i\zet_2 c_8'
\nn 
C_1 &=& {a_2\over 2\sqrt 2 a} - {i a_3\over 2\sqrt 2 a} 
+ \left(\wt\zet_1^2 -{1\over 2}\zet_1^2\right) b_5
- i \zet_2 b_1' +{1\over 2} \zet_1 \zet_2 b_5'
\nn &&
-{1\over 2a} c_2 +{i\over 2a} c_3-{1\over 2}\zet_1
\zet_2 c_5 - {i\zet_2\over 2a} c_6
 -{\zet_2
\over 2a} c_7
-i\zet_2 c_1'  +\left( \wt\zet_2^2 -{1\over 2}\zet_2^2
+L\right) c_5' 
\nn 
C_2 &=& {a_1\over 2\sqrt 2 a} + {i a_4\over 2\sqrt 2 a} 
+ \left(\wt\zet_1^2 -{1\over 2}\zet_1^2\right) b_6
+ i \zet_2 b_2' -{1\over 2} \zet_1 \zet_2 b_6'
\nn &&
-{1\over 2a} c_1 -{i\over 2a} c_4
- {i\zet_2\over 2a} c_5 
+{1\over 2}\zet_1
\zet_2 c_6 
+{\zet_2
\over 2a} c_8
-i\zet_2 c_2'  +\left( \wt\zet_2^2 -{1\over 2}\zet_2^2
+L\right) c_6' 
\nn 
C_3 &=& {ia_1\over 2\sqrt 2 a} - {a_4\over 2\sqrt 2 a} 
- \left(\wt\zet_1^2 -{1\over 2}\zet_1^2\right) b_7
- i \zet_2 b_3' +{1\over 2} \zet_1 \zet_2 b_7'
\nn &&
+{i\over 2a} c_1-{1\over 2a} c_4 
 -{\zet_2
\over 2a} c_5
-{1\over 2}\zet_1
\zet_2 c_7 - {i\zet_2\over 2a} c_8
+i\zet_2 c_3'  -\left( \wt\zet_2^2 -{1\over 2}\zet_2^2
+L\right) c_7' 
\nn 
C_4 &=& -{ia_2\over 2\sqrt 2 a} - {a_3\over 2\sqrt 2 a} 
- \left(\wt\zet_1^2 -{1\over 2}\zet_1^2\right) b_8
+ i \zet_2 b_4' -{1\over 2} \zet_1 \zet_2 b_8'
\nn &&
-{i\over 2a} c_2-{1\over 2a} c_3 
+{\zet_2
\over 2a} c_6
- {i\zet_2\over 2a} c_7 
+{1\over 2}\zet_1
\zet_2 c_8
+i\zet_2 c_4'  -\left( \wt\zet_2^2 -{1\over 2}\zet_2^2
+L\right) c_8' 
\nn 
C_5 &=& 2 b_1' + i\zet_1 b_5'
- i\zet_1 c_5 +{1\over 2a} c_6 -{i\over 2a} c_7
\nn 
C_6 &=& -2 b_2' - i\zet_1 b_6'
+{1\over 2a} c_5 + i\zet_1 c_6  +{i\over 2a} c_8
\nn 
C_7 &=& 2 b_3' + i\zet_1 b_7'
-{i\over 2a} c_5 - i\zet_1 c_7 +{1\over 2a} c_8 
\nn 
C_8 &=& -2 b_4' - i\zet_1 b_8'
+{i\over 2a} c_6 +{1\over 2a} c_7 + i\zet_1 c_8
\nn 
C_1' &=& {a_2'\over 2\sqrt 2 a} - {i a_3'\over 2\sqrt 2 a} 
+ i\zet_2 b_1 -{1\over 2}\zet_1 \zet_2 b_5
+ \left(\wt\zet_1^2 - {1\over 2} \, \zet_1^2\right) b_5'
\nn &&
-i\zet_2 c_1 +\left( \wt\zet_2^2 -{1\over 2}
\zet_2^2 +L\right) c_5
-{1\over 2a} c_2' 
+{i\over 2a} c_3' +{1\over 2} \zet_1\zet_2 c_5'
-{i\zet_2\over 2a} c_6'-{\zet_2\over 2a} c_7' \nn 
C_2' &=& {a_1'\over 2\sqrt 2 a} + {i a_4'\over 2\sqrt 2 a} 
-
 i\zet_2 b_2 +{1\over 2}\zet_1 \zet_2 b_6
+ \left(\wt\zet_1^2 - {1\over 2} \, \zet_1^2\right) b_6'
\nn &&
-i\zet_2 c_2 +\left( \wt\zet_2^2 -{1\over 2}
\zet_2^2 +L\right) c_6
-{1\over 2a} c_1' 
-{i\over 2a} c_4' -{i\zet_2\over 2a} c_5'
-{1\over 2} \zet_1\zet_2 c_6'+{\zet_2\over 2a} c_8' 
\nn 
C_3' &=& {ia_1'\over 2\sqrt 2 a} - {a_4'\over 2\sqrt 2 a} 
+ i\zet_2 b_3 -{1\over 2}\zet_1 \zet_2 b_7
- \left(\wt\zet_1^2 - {1\over 2} \, \zet_1^2\right) b_7'
\nn &&
+i\zet_2 c_3 -\left( \wt\zet_2^2 -{1\over 2}
\zet_2^2 +L\right) c_7
+{i\over 2a} c_1' -{1\over 2a} c_4' 
-{\zet_2\over 2a} c_5' +{1\over 2} \zet_1\zet_2 c_7'
-{i\zet_2\over 2a} c_8'\nn 
C_4' &=& -{ia_2'\over 2\sqrt 2 a} - {a_3'\over 2\sqrt 2 a} 
- i\zet_2 b_4 + {1\over 2}\zet_1 \zet_2 b_8
- \left(\wt\zet_1^2 - {1\over 2} \, \zet_1^2\right) b_8'
\nn &&
+i\zet_2 c_4 -\left( \wt\zet_2^2 -{1\over 2}
\zet_2^2 +L\right) c_8
-{i\over 2a} c_2' -{1\over 2a} c_3' 
+{\zet_2\over 2a} c_6' 
-{i\zet_2\over 2a} c_7'
-{1\over 2} \zet_1\zet_2 c_8'
\nn 
C'_5 &=& -2b_1 - i\zet_1 b_5
+i\zet_1 c_5' +{1\over 2a} c_6' -{i\over 2a} c_7'
\nn 
C'_6 &=& 2 b_2 + i\zet_1 b_6
+{1\over 2a} c_5' -i\zet_1 c_6' +{i\over 2a} c_8'
\nn 
C'_7 &=& - 2 b_3 - i\zet_1 b_7
-{i\over 2a} c_5'+i\zet_1 c_7' +{1\over 2a} c_8' 
\nn 
C'_8 &=& 2 b_4 + i\zet_1 b_8
+{i\over 2a} c_6'+{1\over 2a} c_7' -i\zet_1 c_8' 
\een
where
\be\label{edeftildekappa}
\wt\zet_1^2 = \zet_1^2 - {1\over 2a^2}\, ,
\qquad \wt\zet_2^2 = \zet_2^2 + {1\over 2a^2},
\qquad K = {1\over 2a^2}, \qquad L = -{1\over 2a^2}
\, ,
\ee
and we have used
\be \label{edefketc}
-D_\alpha D^\alpha \chi = \wt\zet_1^2\chi,
\quad 
-D_mD^m \chi = \wt\zet_2^2 \chi, \quad
\Gamma^\beta[D_\beta, D_\alpha] \chi = K\,
\Gamma_\alpha\chi, \quad
\Gamma^m [D_m, D_n] \chi = L\, 
\Gamma_n \chi\, .
\ee

We can express \refb{ezxd} as
\be \label{ezxc}
\pmatrix{\vec A\cr \vec B\cr \vec C\cr \vec A'\cr \vec
B'\cr \vec C'} = \MM \, \pmatrix{\vec a\cr
\vec b\cr \vec c\cr \vec a'\cr \vec b'\cr \vec c'}\, ,
\ee
where $\MM$ is a $40\times 40$ matrix. 
The eigenvalues of
$\MM^2$ will determine the heat kernel in 
the fermionic sector of
the gravity multiplet.

Let us now discuss the possible complication that could arise
due to the fact that we have chosen to expand the various fields
in a non-orthonormal set of basis functions. If we did use an
orthonormal basis then the resulting matrix $\MM$ will be related
to the one appearing in \refb{ezxc} by a similarity
transformation. This however will not affect the eigenvalues
of $\MM^2$.
Since our final result will be expressed in terms of the eigenvalues
of $\MM^2$, the non-orthonormality of our basis vectors will not
affect the result.

To proceed we introduce a matrix $\MM_1$ through
\be \label{esx1}
\MM^2 = - (\zet_1^2 + \zet_2^2) I_{40} + 
a^{-2} \MM_1 \, ,
\ee
where $I_{40}$ denotes the $40\times 40$ identity matrix.
It is easy to see that in the limit of large 
$\zet_1$, $\zet_2$ the dominant contribution to the
eigenvalues come from the first term. Let us denote
by $\beta_k$ for $1\le k\le 40$ the 40 eigenvalues of the
matrix $\MM_1$, and
introduce variables $\lambda$ and $l$ through:
\be \label{esx1.5}
\zet_1 = (l+1)/a, \qquad \zet_2 = \lambda/a
\ee
Then the contribution to the heat kernel
from the fermionic modes for $\zet_1 >1/a$, 
$\zet_2\ge 0$ will be given by
\be \label{esx2}
K^f_{(1)}(0;s) = -{1\over 8\pi^2 a^4}\,
\sum_{l=1}^\infty (2l+2) \int_0^\infty d\lambda \lambda
\coth(\pi\lambda) e^{-\bar s (l+1)^2 -\bar s\lambda^2}
\sum_{k=1}^{40} e^{\bar s \beta_k}\, .
\ee
The overall minus sign reflects the fact that we are dealing
with fermions. The normalization factor is fixed by noting that
since the four gravitinoes represented by the sixteen
component field $\psi_\mu$ for $0\le \mu\le 3$
give effectively $4\times 4=16$ Majorana
fermions, and the dilatino, represented by the sixteen
component field $\Lamb$, gives 4 Majorana fermions
in four dimensions, 
we have in total 20 Majorana or 10 Dirac fermions in
four dimensions. Thus the 
heat kernel should agree with that
of 10 free Dirac fermions in the limit of small $\bar s$
when the effect of background flux can be ignored, \i.e.\
$\bar s\beta_k$ can be set equal to 0. Comparing
\refb{esx2} with \refb{ediracf} we see that we indeed have the equivalent
of ten Dirac fermions.

The contribution from the $\zet_1=1/a$,
\i.e.\ $l=0$ term has to be
evaluated separately. For this the basis states used in the
\refb{ezx1} are not independent, since we have
$D_\alpha\chi = {i\over 2a} 
\Gamma_\alpha\chi$. Using this
we can choose the coefficients $b_5,\cdots b_8$ and
$b'_5,\cdots b'_8$ to zero in \refb{ezx1}. Furthermore in 
eqs.\refb{ezxb} we can
make the replacement $D_\alpha\chi\to {i\over 2a}
\Gamma_\alpha\chi$, which amounts to replacing
in 
\refb{ezxd}
the expressions for $B_k$ by that of
$B_k + {i\over 2a} B_{k+4}$ and of
$B'_k$ by that of
$B'_k + {i\over 2a} B'_{k+4}$ for $1\le k\le 4$ and then
drop the expressions for $B_{k+4}$ and $B'_{k+4}$
for $1\le k\le 4$. This gives a $32\times 32$ matrix 
$\wt \MM$ relating $(A_1,\cdots A_4, A_1',\cdots A_4',
B_1,\cdots B_4, B_1',\cdots B_4',C_1,\cdots C_8,C_1',
\cdots C_8'$) to
$(a_1,\cdots a_4, a_1',\cdots a_4',
b_1,\cdots b_4, b_1',\cdots b_4',c_1,\cdots c_8,c_1',
\cdots c_8'$). Let us now define 
a matrix $\wt\MM_1$ through
\be \label{esx1.9}
\wt\MM^2 = -(a^{-2}+\zet_2^2) I_{32} + a^{-2} \wt\MM_1 \, ,
\ee
where $I_{32}$ denotes the $32\times 32$ identity matrix.
If $\wt\beta_k$'s are the eigenvalues of $\wt\MM_1$
then the contribution from the $l=0$ modes to the
heat kernel may be expressed as
\be \label{esy2}
K^f_{(2)}(0;s) = -{1\over 4\pi^2 a^4}\,
\int_0^\infty d\lambda \lambda
\coth(\pi\lambda) e^{-\bar s  -\bar s\lambda^2}
\sum_{k=1}^{32} e^{\bar s \wt\beta_k}\, .
\ee
We can combine \refb{esx2} and \refb{esy2} to write
\be \label{esy3}
K^f_{(1)}(0;s) +  K^f_{(2)}(0;s) = \wt
K^f_{(1)}(0;s) +  \wt K^f_{(2)}(0;s)\, ,
\ee
where
\ben \label{esy4a}
\wt K^f_{(1)}(0;s) &=&  -{1\over 8\pi^2 a^4}\,
\sum_{l=0}^\infty (2l+2) \int_0^\infty d\lambda \lambda
\coth(\pi\lambda) e^{-\bar s (l+1)^2 -\bar s\lambda^2}
\sum_{k=1}^{40} e^{\bar s \beta_k}\nn
&=& -{1\over 4\pi^2 a^4} \, Im \,
\int_0^{e^{i\kappa}\times \infty} 
d\wt\lambda \, \wt\lambda \, 
\cot(\pi\wt\lambda)\, \int_0^\infty d\lambda \lambda
\coth(\pi\lambda) e^{-\bar s\wt\lambda^2-\bar s\lambda^2}
\sum_{k=1}^{40} e^{\bar s \beta_k|_{l+1 \to \wt\lambda}}\nn
\een
\be \label{esy4b}
\wt K^f_{(2)}(0;s) = -{1\over 4\pi^2 a^4}\,
\int_0^\infty d\lambda \lambda
\coth(\pi\lambda) e^{-\bar s  -\bar s\lambda^2}
\left[\sum_{k=1}^{32} e^{\bar s \wt\beta_k}
- \sum_{k=1}^{40} e^{\bar s \beta_k|_{l=0}}\right]\, .
\ee
In the second step in \refb{esy4a} we have used a trick
similar to that described in \refb{ess1}, \refb{ess2} 
to convert the
sum over $l$ to integral of $\wt\lambda$.

We also need to compute the contribution 
due to the discrete
modes described in \refb{eadd1}.  For this 
we set the fields $\Lamb$ and $\psi_\alpha$ to
0, and expand $\psi_m$ as in \refb{ezx1} with $c_{k+4}=2c_ka$,
$c'_{k+4} = 2 c'_ka$ for $1\le k \le 4$, with $\zet_2=i/a$,
$\zet_1\ge 1/a$, \i.e.\ $l\ge 0$.\footnote{This
basis is still overcomplete since, as discussed in
\refb{eimp1}, the action of $\tau_3$ on the basis states
is fixed once we choose $\chi$ to be $\chi^+_k(i)$ or
$\eta^+_k(i)$. Thus we could work with either the $C_i$'s
or the $C'_i$'s. But we shall proceed by including both sets
and include a factor of 1/2 in the expression for the heat
kernel.} 
It can be seen that with this choice
$A_i$, $A'_i$, $B_i$, $B'_i$ computed from 
\refb{ezxd} vanish and we have $C_{k+4}=2C_ka$,
$C'_{k+4} = 2 C'_ka$ for $1\le k \le 4$.
Thus we can express
these relations as
\be \label{eexx1}
\pmatrix{C_1\cr C_2\cr C_3\cr C_4\cr C_1'\cr C_2'\cr
C_3'\cr C_4'} = \wh\MM \, 
\pmatrix{c_1\cr c_2\cr c_3\cr c_4\cr c_1'\cr c_2'\cr
c_3'\cr c_4'} \, ,
\ee
for some $8\times 8$ matrix $\wh\MM$. If $\wh\beta_k$
denote the eigenvalues of 
\be \label{edefm1hat}
\wh\MM_1 \equiv a^2\{\wh \MM^2 + (\zet_1^2
-a^{-2}) I_8 \}\, ,
\ee
 then the 
contribution to $K(0;s)$ from these modes is given by
\ben \label{eexx2}
K^f_{(3)}(0;s) &=& -{1\over 8\pi^2 a^4}\, 
\sum_{l=0}^\infty (2l+2) e^{\bar s-\bar s(l+1)^2} \sum_{k=1}^8
e^{\bar s \wh\beta_k}\nn
&=& -{1\over 4\pi^2 a^4} \, Im \,
\int_0^{e^{i\kappa}\times \infty} 
d\wt\lambda \, \wt\lambda \, 
\cot(\pi\wt\lambda)\, e^{\bar s-\bar s \wt\lambda^2} 
\sum_{k=1}^8
e^{\bar s \wh\beta_k|_{l+1\to\wt\lambda}}\, .
\een

Finally the three sets of bosonic ghosts 
$\wt b$, $\wt c$ and $\wt e$ associated with gauge
fixing of local supersymmetry, each of which gives rise
to four Majorana fermions in four dimensions, 
contributes
\ben \label{efghx1}
K^f_{ghost} &=& {3\over \pi^2 a^4}\,
\sum_{l=0}^\infty (2l+2) \int_0^\infty d\lambda \lambda
\coth(\pi\lambda) e^{-\bar s (l+1)^2 -\bar s\lambda^2}
\nn
&=& {6\over \pi^2 a^4} \, Im \,
\int_0^{e^{i\kappa}\times \infty} 
d\wt\lambda \, \wt\lambda \, 
\cot(\pi\wt\lambda)\, \int_0^\infty d\lambda \lambda
\coth(\pi\lambda) e^{-\bar s\wt\lambda^2-\bar s\lambda^2}\, ,
\een
to
$K(0;s)$.

To evaluate the right hand sides of  \refb{esy4a},
\refb{esy4b}, and \refb{eexx2}
we use the relations
\be \label{esx3}
\sum_k e^{\bar s \beta_k} = \sum_{n=0}^\infty
{1\over n!} \bar s^n \sum_k \beta_k^n 
= \sum_{n=0}^\infty
{1\over n!} \bar s^n  Tr(\MM_1^n)\, ,
\ee
\be \label{esx3ab}
\sum_k e^{\bar s \wt\beta_k} = \sum_{n=0}^\infty
{1\over n!} \bar s^n \sum_k \wt\beta_k^n 
= \sum_{n=0}^\infty
{1\over n!} \bar s^n  Tr(\wt\MM_1^n)\, .
\ee
and
\be \label{esx3abc}
\sum_k e^{\bar s \wh\beta_k} = \sum_{n=0}^\infty
{1\over n!} \bar s^n \sum_k \wh\beta_k^n 
= \sum_{n=0}^\infty
{1\over n!} \bar s^n  Tr(\wh\MM_1^n)\, ,
\ee
Explicit computation gives
\ben\label{esx4}
Tr(\MM_1) &=&  0\nonumber \\
Tr(\MM_1^2) &=& 64 + 16 (l+1)^2 - 16 
\lambda^2 \nonumber \\
Tr(\MM_1^3) &=&  -144 (l+1)^2 - 144 
\lambda^2\nonumber \\
Tr(\MM_1^4) &=&  256 + 192 (l+1)^2 + 
80(l+1)^4 - 192 \lambda^2 
+ 32 (l+1)^2 \lambda^2 + 80 \lambda^4\, .
\een
\ben\label{esy1}
Tr(\wt\MM_1) &=&  -8\nonumber \\
Tr(\wt\MM_1^2) &=& 72 - 8 \lambda^2 \nonumber \\
Tr(\wt\MM_1^3) &=&  -152 - 120 \lambda^2\nonumber \\
Tr(\wt\MM_1^4) &=&  520 
- 112 \lambda^2 + 72 \lambda^4\, .
\een
\ben\label{esy1.9}
Tr(\wh\MM_1) &=&  -8\nonumber \\
Tr(\wh\MM_1^2) &=&   8 + 8 (l+1)^2\nonumber \\
Tr(\wh\MM_1^3) &=&   -8 - 24 (l+1)^2\nonumber \\
Tr(\wh\MM_1^4) &=&  8 + 48 (l+1)^2 + 8 (l+1)^4\, .
\een
Furthermore we also have the analogs of eqs.\refb{esa1}
and \refb{esa2}:
\ben\label{esa1a}
&& \int_0^\infty d\lambda \, \lambda\, \coth(\pi\lambda)\,
e^{-\bar s \lambda^2} \, \lambda^{2n}\cr
&=& {1\over 2} \bar s^{-1 - n} 
\Gamma(1+n) + 2 \sum_{m=0}^\infty \bar s^m 
{(2m+2n+1)!\over m!} \, (2\pi)^{-2(m+n+1)} \, (-1)^m\cr
&& \qquad \qquad \qquad \qquad \qquad \qquad
\zeta(2(m+n+1))\, .
\een
\ben\label{esa2a}
&& Im \int_0^{e^{i\kappa}\times \infty} 
d\wt\lambda \, \wt\lambda\, \cot(\pi\wt\lambda)\,
e^{-\bar s \wt\lambda^2} \, \wt\lambda^{2n}\cr
&=& {1\over 2} \bar s^{-1 - n} 
\Gamma(1+n) + 2 \sum_{m=0}^\infty \bar s^m 
{(2m+2n+1)!\over m!} \, (2\pi)^{-2(m+n+1)} (-1)^{n+1}\cr
&& \qquad \qquad \qquad \qquad \qquad \qquad
\zeta(2(m+n+1))\, .
\een
Using these relations we get the following 
order $s^0$ contributions
to various terms when expanded in a power series 
in $s$ around
$s=0$:
\ben \label{efin1}
\wt K^f_{(1)}(0;s) &:& {5\over 72\pi^2 a^4}
\nn
\wt K^f_{(2)}(0;s) &:& -{1\over 3\pi^2 a^4}
\nn
K^f_{(3)}(0;s) &:& - {1\over 3\pi^2 a^4}
\nn
K^f_{ghost}(0;s) &:& - {11\over 120\pi^2 a^4}
\, .
\een
Adding up all the contributions we get the net contribution to
$K(0;s)$ from the fermionic fields in the gravity multiplet:
\be \label{enetfer}
K^f(0;s) = -{31\over 45\pi^2 a^4}\, .
\ee

We now need to remove from this the zero mode
contribution.
Analysis of the zero modes in the fermionic sector requires
special care.
Among the $l=0$ modes in the first line of
\refb{eexx2} we have four vanishing $\wh\beta_k$ 
giving a net contribution of $-1/\pi^2 a^4$. 
Thus naively we must remove this contribution from
$K(0;s)$. However a detailed analysis shows that
although the matrix $\wh\MM^2$ describing the square of
the fermionic kinetic term has four eigenvectors with 
zero eigenvalues,
the matrix $\wh\MM$ has only a pair of eigenvectors
with zero eigenvalues.
These eigenvectors are
\ben \label{ezea1}
\psi^{(1)}_m &=& (i - \sigma_3\wh\Gamma^4 
- i\tau_3\wh\Gamma^5 +\sigma_3\tau_3 \wh\Gamma^4
\wh\Gamma^5)(a^{-1}
\Gamma_m+2\sigma_3 D_m)\chi\, , \nonumber \\
\psi^{(2)}_m &=& \sigma_3(i + \sigma_3\wh\Gamma^4 
+ i\tau_3\wh\Gamma^5 +\sigma_3\tau_3 \wh\Gamma^4
\wh\Gamma^5)(a^{-1}
\Gamma_m+2\sigma_3 D_m)\chi\, .
\een
The other two eigenvectors of $\wh \MM^2$ 
which are not zero modes of $\wh\MM$ are
\ben \label{ezea1new}
\xi^{(1)}_m &=& (i + \sigma_3\wh\Gamma^4 
- i\tau_3\wh\Gamma^5 -\sigma_3\tau_3 \wh\Gamma^4
\wh\Gamma^5)(a^{-1}
\Gamma_m+2\sigma_3 D_m)\chi\, , \nonumber \\
\xi^{(2)}_m &=& \sigma_3(i - \sigma_3\wh\Gamma^4 
+ i\tau_3\wh\Gamma^5 - \sigma_3\tau_3 \wh\Gamma^4
\wh\Gamma^5)(a^{-1}
\Gamma_m+2\sigma_3 D_m)\chi\, .
\een
The action of $\wh\MM$ on these modes are given by
\be \label{emaction}
\wh \MM \, \xi^{(1)}_m = - 2\, i \, a^{-1}\,
\psi^{(1)}_m, \quad
\wh \MM \, \xi^{(2)}_m = 2\, i \, a^{-1}\, \psi^{(2)}_m, \quad
\wh \MM \, \psi^{(1)}_m = 0, \quad
\wh \MM \, \psi^{(2)}_m = 0\, .
\ee
{}From this we conclude that the contribution of
only two of the four zero modes
of $\wh\MM^2$ will have to be removed from the
contribution to $K(0;s)$. 
This amounts to removing a factor of $-1/2\pi^2 a^4$
from $K(0;s)$.
Subtracting this from \refb{enetfer} we get the net
contribution to $K(0;s)$ from the non-zero modes of the
gravity multiplet fermions:
\be \label{enetferBB}
-{17\over 90\pi^2 a^4}\, .
\ee

For later use it will be useful to find the physical interpretation
of these zero modes. 
First we note that the zero modes satisfy the chirality projection
condition:
\be \label{echiral}
\sigma_3\tau_3\wh\Gamma^{4}\wh\Gamma^{5} \psi^{(k)}_m 
= i \psi^{(k)}_m,
\quad \sigma_3\tau_3\wh\Gamma^{4}\wh\Gamma^{5} 
\xi^{(k)}_m 
=- i \xi^{(k)}_m,
\quad \hbox{for $k=1,2$}\, .
\ee
Choosing $\chi=\chi^+_k(i)$ in  \refb{ezea1} and using
\refb{eimp1} and
the fact that $\chi$ has been chosen to satisfy
$\wh\Gamma^{45}\chi = i\chi$, we get
\ben \label{ezea2}
\psi^{(1)}_m &=& (i - \sigma_3\wh\Gamma^4 
-\wh\Gamma^4 -i\sigma_3)(\Gamma_m+2\sigma_3 D_m)\chi\, , \nonumber \\
\psi^{(2)}_m &=& \sigma_3(i + \sigma_3\wh\Gamma^4 
+\wh\Gamma^4 -i\sigma_3)(\Gamma_m+2\sigma_3 D_m)\chi\, .
\een
Thus we have $\psi^{(1)}_m = -\psi^{(2)}_m$, \i.e.\ they are
not independent. A similar analysis for $\chi=\eta^+_k(i)$
will give $\psi^{(1)}_m = \psi^{(2)}_m$ again showing that
they are not independent. This allows us to keep only
one of these modes, -- we shall take it
to be $\psi^{(1)}-i\psi^{(2)}$ in both cases.
Now one can show that
\ben \label{epsitrs2}
&& \psi^{(1)}_\mu-i\psi^{(2)}_{\mu}
=  D_\mu \eps +{1\over  \sqrt 2}\,
  \Gamma^\sigma\, 
\left(
\bar F^{(1)}_{\mu\sigma} \Gamma^{4}
+ \bar F^{(2)}_{\mu\sigma} \Gamma^{5}\right)\, 
\eps \, , \nn
&& \eps \equiv 2 (i - \sigma_3\wh\Gamma^4 
- i\tau_3\wh\Gamma^5 +\sigma_3\tau_3 \wh\Gamma^4
\wh\Gamma^5 
+\sigma_3 -i\wh\Gamma^4 
+\sigma_3\tau_3\wh\Gamma^5 -i \tau_3 \wh\Gamma^4
\wh\Gamma^5)
\sigma_3 \chi\, ,
\nn
&& \Gamma^{\rho\sigma} \, \left(
\bar F^{(1)}_{\rho\sigma} \Gamma^{4}
+ \bar F^{(2)}_{\rho\sigma} \Gamma^{5}\right)\eps
 = 0
\een
Comparing this with the 
supersymmetry
transformations laws for the gravitino and the dilatino 
in the convention of
\cite{1005.3044}
\ben \label{epsitrs}
\delta \psi_\mu &=& D_\mu \eps +{1\over 4\sqrt 2}\,
 \left( 4 \delta_\mu^{\rho} \Gamma^\sigma
- \Gamma_\mu \Gamma^{\rho\sigma}\right) 
\left(
\bar F^{(1)}_{\rho\sigma} \Gamma^{4}
+ \bar F^{(2)}_{\rho\sigma} \Gamma^{5}\right)
\eps + \cdots \nn
\delta\Lambda &=& -{1\over 4}\, \Gamma^{\rho\sigma}\,
\left(
\bar F^{(1)}_{\rho\sigma} \Gamma^{4}
+ \bar F^{(2)}_{\rho\sigma} \Gamma^{5}\right)
\eps\, 
\een
where $\eps$ is the supersymmetry transformation parameter
and $\cdots$ denotes terms which vanish in the near 
horizon background geometry, we see that 
$\psi^{(1)}_\mu-i\psi^{(2)}_\mu$ is associated with a
supersymmetry transformation
generated by the parameter $\eps$. However since
$\eps$ is obtained by the action of $\Gamma$ matrices on
$\chi^+(i)$ or $\eta^+(i)$, it is not normalizable.

\sectiono{Zero mode contribution} \label{szero}

Adding \refb{efgravnz} and \refb{enetferBB} we get
the net $s$-independent
contribution to $K(0;s)$ from all the non-zero modes:
\be \label{epx2}
-{101 \over 180\pi^2 a^4} -{17\over 90\pi^2 a^4}
 = - {3\over 4\pi^2 a^4}\, .
\ee
To this we must add the result of carrying out the zero
mode integration. This was described for the gauge
fields in appendix A of \cite{1005.3044}; we shall briefly
review the argument since it can also be generalized to
integration over the zero modes of the metric and
the gravitino fields.

Let $A_\mu$ be a vector field on $AdS_2\times S^2$ and $g_{\mu\nu}$
be the background metric of the form
$a^2 \, g^{(0)}_{\mu\nu}$
where $a$ is radius of curvature of 
$S^2$ and $AdS_2$ 
and $g^{(0)}_{\mu\nu}$
is independent of $a$. The path integral over 
$A_\mu$ is normalized such that
\be \label{eap2}
\int [DA_\mu] \exp\left[- \int d^4 x \, \sqrt{\det g} \, g^{\mu\nu} A_\mu A_\nu
\right] = 1\, ,
\ee
\i.e.\
\be \label{eap3}
\int [DA_\mu] \exp\left[- a^2 \int d^4 x \, \sqrt{\det g^{(0)}} \, 
g^{(0)\mu\nu} A_\mu A_\nu
\right] = 1\, .
\ee
{}From this we see that up to an $a$ independent 
normalization
constant, $[DA_\mu]$  actually corresponds to integration
with measure $\prod_{\mu,x} d(a A_\mu(x))$. 
On the other hand  the gauge field
zero modes are associated with
deformations produced by the gauge transformations
with non-normalizable parameters: $\delta A_\mu\propto
\p_\mu \Lambda(x)$ for some functions $\Lambda(x)$ with 
$a$-independent
integration range. Thus
the result of integration over the gauge field zero
modes can be found by first changing the integration over the
zero modes of
$(aA_\mu)$ to integration over $\Lambda$ and then
picking up the contribution from the Jacobian in this
change of variables. This gives a factor of $a$ from integration
over each zero mode of $A_\mu$.
Thus if there are $N$ zero modes then we shall get a
factor of $a^N$. Of course $N$ is infinite, but it
needs to be regularized by subtracting from it a term
proportional to the length of the boundary of $AdS_2$.
We shall now describe two equivalent ways of
computing $N$: one is a somewhat indirect but useful
method and the other is a more direct method, but involves a
little bit of additional computation.

First let us describe the indirect method.
For a non-zero mode, the path integral weighted by the exponential
of the action produces a factor of $\kappa_n^{-1/2}$ where $\kappa_n$
is the eigenvalue of the kinetic operator. Since $\kappa_n$ has the form
$b_n/a^2$ where $b_n$ is an $a$ independent constant, integration
over a non-zero mode produces a factor proportional to $a$. 
Including the zero mode contribution to $K(0,s)$ is equivalent
to counting the same factor of $a$ from integration over the
zero modes as well.
Thus when
we remove from the determinant
the contribution due to the zero modes, 
we remove a factor
of $a$ for each zero mode. However the analysis of the previous paragraph
showed that integration over the zero modes gives us back a factor of $a$.
Thus the net effect of integrating over the gauge 
field zero modes is to cancel the effect of the subtraction
of the zero mode contribution
from $K(0;s)$. In other words the net effect of integration
over the six gauge field zero modes amounts to effecively
adding a contribution of $6/8\pi^2 a^4$ to 
$K(0;s)$.\footnote{Note that this is not the actual 
modification of
the heat kernel, but represents the effective contribution to
be added to $K(0;s)$ that reproduces, via \refb{elogfor}, the
net contribution to the one loop determinant due to the
zero modes.} Using \refb{elogfor} we see that this
corresponds to a contribution 
of $-6\ln a$ in the entropy, \i.e.\ $-\ln a$ for each gauge
field.

Next we shall describe a direct method for evaluating the
zero mode contribution from the gauge fields which
does not make any reference to the result on integration
over the non-zero modes.  Let $f^{(\ell)}_m$ denote the
normalized zero mode wave functions of gauge fields
on
$AdS_2$ given in \refb{e24p}. Then the total number
of zero modes may be written as
\be \label{estx1}
\sum_{\ell\in \zzz, \ell \ne 0} \int d\theta d\eta 
\sqrt{\det g_{AdS_2}}\, 
g_{AdS_2}^{mn} f^{(\ell)*}_m f^{(\ell)}_n\, ,
\ee
where $g_{AdS_2}$ is the metric on $AdS_2$.
We now use the fact that $\sum_{\ell}
g_{AdS_2}^{mn} f^{(\ell)*}_m f^{(\ell)}_n$ must be
independent of the $AdS_2$ coordinates $(\eta,\theta)$ since
$AdS_2$ is a homogeneous space. Thus we can evaluate
this at $\eta=0$. In this case only $\ell=\pm 1$ modes
contribute, leading to the result $1/(2\pi a^2)$. Integrating
this over $AdS_2$ with a cut-off $\eta\le \eta_0$, 
we get the result
\be \label{egetthe}
{1\over 2\pi a^2} \, 2\pi a^2\, (\cosh\eta_0-1)\, .
\ee
The term proportional to $\cosh\eta_0$ can be interpreted as
a shift in the ground state energy. Thus we are left with an
effective contribution of $-1$.
{}From this we conclude that for every
gauge field the integration over the zero modes gives a
factor of $a^{-1}$ to $e^{S_{BH}}$, \i.e. 
$-\ln a$ to the black hole
entropy. 

The effect of integration over the zero modes of the 
fluctuations
$h_{\mu\nu}$ of the metric (including those of the
SU(2) gauge fields arising from the dimensional
reduction of the metric on $S^2$) 
can be found in a similar way,
with \refb{eap2}, \refb{eap3} replaced by
\be \label{ebp2}
\int [Dh_{\mu\nu}] \exp\left[- \int d^4 x \, \sqrt{\det g} \, 
g^{\mu\nu} g^{\rho\sigma} h_{\mu\rho} h_{\nu\sigma}
\right] = 1\, ,
\ee
\i.e.\
\be \label{ebp3}
\int [Dh_{\mu\nu}] \exp\left[- \int d^4 x \, 
\sqrt{\det g^{(0)}} \, 
g^{(0)\mu\nu} g^{(0)\rho\sigma}h_{\mu\rho} h_{\nu\sigma}
\right] = 1\, .
\ee
Thus the correctly normalized integration
measure, up to an $a$ independent constant, is 
$\prod_{x,(\mu\nu)} dh_{\mu\nu}(x)$.
We now note that the zero modes are associated
with diffeomorphisms with non-normalizable parameters:
$h_{\mu\nu}\propto D_\mu\xi_\nu + D_\nu\xi_\mu$, with
the diffeomorphism parameter $\xi^\mu(x)$
having $a$ independent integration range. Thus the $a$
dependence of the integral over the metric zero modes
can be found by finding the Jacobian from the change
of variables from $h_{\mu\nu}$ to $\xi^\mu$. Lowering
of the index of $\xi^\mu$ gives a factor of $a^2$, leading to
a factor of $a^2$ per zero mode. On the other hand following
the same logic as in the case of gauge fields we find that
the removal of the integration over the metric zero modes
from the heat kernel 
removes a factor of $a$ per zero mode from the integrand.
Thus the effect of integration over the metric zero
modes will be to add the
double of the contribution that one removes.
Since we had removed from $K(0;s)$ a contribution
of $3/8\pi^2 a^4 + 3/8\pi^2 a^4 = 6/8\pi^2 a^4$
(see eq.\refb{epx1}) we need to  add
a factor of $12/8\pi^2 a^4$.

Finally we turn to the fermion zero modes.\footnote{Naively
integration over the fermion zero modes will make the
integral vanish. However it was shown in \cite{0905.2686} using
localization techniques that the zeroes due to the
fermionic zero mode integrals cancel the infinities coming
from integration over the bosonic zero modes of the metric.}
The normalization of the zero modes is determined from
\be \label{ebp2fer}
\int [D\psi_{\mu}] [D\bar\psi_{\mu}]
\exp\left[- \int d^4 x \, \sqrt{\det g} \, 
g^{\mu\nu} \bar\psi_\mu \psi_{\nu}
\right] = 1\, ,
\ee
\i.e.\
\be \label{ebp3fer}
\int [D\psi_{\mu}] [D\bar\psi_{\mu}]
\exp\left[- a^{2} \int d^4 x \, \sqrt{\det g^{(0)}}
 \, 
g^{(0)\mu\nu}  \bar\psi_\mu \psi_{\nu}
\right] = 1\, ,
\ee
indicating that $a\psi_\mu$ and $a\bar\psi_\mu$
are the correctly normalized integration variables. 
As discussed in the previous section, the fermion
zero  modes are  associated with the asymptotic
supersymmetry transformations, with the
anti-commutator of a pair of supersymmetry
transformations generating a diffeomorphism with 
parameter $\bar\eps\Gamma^\mu\epsilon$.  Since
$\Gamma^\mu\sim a^{-1}$, and since 
$\bar\eps\Gamma^\mu\epsilon$ has $a$-independent
integration range, we see that the correctly
normalized $\epsilon$ is $\eps_0= 
a^{-1/2}\epsilon$ for which
the supersymmetry algebra
generated by  $\eps_0$ and $\xi^\mu$ does not involve
any $a$ dependence. Thus integration over each
$\psi_\mu$ zero mode is equivalent to integration over
$a\psi_\mu\sim a^{3/2}\, \eps_0$, producing a factor of $a^{-3/2}$.
On the other hand a non-zero mode of the fermion will
produce a factor of $a^{-1/2}$ after integration since
the kinetic operator of the fermion is of order $a^{-1}$.
Thus removing  a fermion zero mode contribution from
the heat kernel removes a factor of $a^{-1/2}$ for each zero 
mode. Thus
the effect of integration over the
fermion zero modes is to add back three times the amount
we remove from the heat kernel while removing the
fermion zero mode contribution. 
This gives a net
contribution of $-3/2\pi^2 a^4$  to the effective heat kernel.

Adding up the contribution from all the zero modes we
see that the net effect of integration over the zero modes
is to effectively add a factor of
\be \label{epx5}
{6\over 8\pi^2 a^4} +{12\over 8\pi^2 a^4}
- {3\over 2\pi^2 a^4} = {3\over 4\pi^2 a^4}\, ,
\ee
to $K(0;s)$. Note that the contribution from the graviton and the
gravitino zero modes cancel -- the final result $3/4\pi^2 a^4$ is 
the contribution of the six gauge fields in the gravity multiplet.

Adding \refb{epx5} to the contribution \refb{epx2} due to the
non-zero modes we get the net contribution to the
effective heat kernel  to be
\be \label{enet1ax}
-{3\over 4\pi^2 a^4} + {3\over 4\pi^2 a^4} =0
\, .
\ee
This is perfectly consistent with the
microscopic result \refb{en4}.

\sectiono{$\NN=8$ black holes} \label{n8}

In this section we shall briefly describe the
analysis of logarithmic corrections
to the entropy of 1/8 BPS black holes in $\NN=8$ 
supersymmetric string theories 
obtained by compactifying type
IIB string theory on $T^6$. For this we first note that
there is a consistent truncation of $\NN=8$ supergravity
to $\NN=4$ supergravity by projecting on to the 
$(-1)^{F_L}$ even states in which we set all the RR and
R-NS sector fields to zero. Using this embedding of the
$\NN=4$ supergravity into $\NN=8$ supergravity,
the quarter
BPS black hole in $\NN=4$ supergravity that
we have already analyzed can now be regarded as the
1/8  BPS black hole in the $\NN=8$ supergravity.
Since the projection on to the $\NN=4$ supergravity is a
consistent truncation it is guaranteed that at the quadratic
level, the fluctuations of the additional fields in the 
$(-1)^{F_L}$ odd sector
does not mix with the fluctuations of the fields in the 
$(-1)^{F_L}$ even sector. Thus the one loop effective
action  of full $\NN=8$ supergravity
receives the contribution already computed for
the $\NN=4$ black holes plus an additional contribution
from the determinant of the $(-1)^{F_L}$ odd fields.

We begin with the contribution due to the extra bosons.
There are sixteen gauge bosons, -- one 
from the ten dimensional
gauge field $A_\mu$ and
fifteen from the components $C_{mn\mu}$ of
the 3-form field with $m,n$ along $T^6$ and $\mu$ along
$AdS_2\times S^2$. There are also thirty two scalars,
-- six from the components $A_m$ of the ten dimensional
gauge field along $T^6$, twenty from the components
$C_{mnp}$ of the 3-form field along $T^6$ and six from
dualizing the components $C_{m\mu\nu}$ of the 3-form
field. These fields can be labelled as $\AAA^{(r)}_\mu$,
$\phi_{1r}$ and $\phi_{2r}$ with $1\le r\le 16$, 
and, in the Feynman gauge,
the quadratic terms in the
action in the near horizon background geometry takes the
form:
\ben \label{es81}
&& 
\int d^4 x \, \sqrt{\det g} \, 
\Bigg[ 
{1\over 2} \sum_{r=1}^{16}
\AAA_\mu^{(r)} (g^{\mu\nu}\square - R^{\mu\nu}) 
\AAA_\nu^{(r)}
 + {1\over 2} \sum_{r=1}^{16}
\left(\phi_{1r} \, \square \, \phi_{1r} + 
\phi_{2r} \, \square \, \phi_{2r}\right) \nonumber \\
&& \qquad \qquad 
+ \sum_{r=1}^8  \left( 2 a^{-1}  \phi_{2r}\,
\vareps^{\gamma\beta} 
\p_\gamma \AAA^{(r)}_\beta
-  \, a^{-2} \, \phi_{2r} \, \phi_{2r}\right) \nonumber \\
&& \qquad \qquad  +\sum_{r=9}^{16}
\left( -  2 i \, a^{-1} \, \phi_{1r} \, \vareps^{mn} 
\p_m \AAA^{(r)}_n +  
a^{-2} \, \phi_{1r} \, \phi_{1r} \right)
\bigg] \, .
\nonumber \\
\een
This has the same structure as the bosonic part of the matter
multiplet fields analyzed in \cite{1005.3044} except that
for $1\le r\le 8$ only the components of the gauge
fields along $S^2$ and the scalar fields $\phi_{2r}$
are affected by the background flux, while for
$9\le r\le 16$ only the components of the gauge fields
along $AdS_2$ and the scalar fields $\phi_{1r}$ are affected
by the flux.\footnote{This is not an accident but follows from
the following considerations. We could have gotten an
$\NN=4$ supergravity theory from the original
$\NN=8$ supergravity by projecting out all
fields which are odd under $\II_4$ where 
$\II_4$ represents the transformation that changes the
sign of the coordinates $x^6,\cdots x^9$. The eight
vectors $C_{mn\mu}$, $C_{45\mu}$ and $A_\mu$ 
with $6\le m,n\le 9$, $0\le\mu\le 3$
and the sixteen scalars $C_{mn4}$, $C_{mn5}$,
$A_4$, $A_5$ and the duals of $C_{4\mu\nu}$,
$C_{5\mu\nu}$ survive the projection. Four of the
gauge fields and the 16 scalars 
will form part of the bosonic sector the four matter
multiplets. 
For completing the matter multiplets we need eight more
scalars which will come from the components of the NSNS
2-form field and the metric along 6789 directions, but from
the analysis of \cite{1005.3044} we know that these describe
free scalars in $AdS_2\times S^2$ background. Thus the net
contribution from the eight vector and sixteen RR scalar 
fields to the heat kernel will be given
by that of four matter multiplets of $\NN=4$
supergravity minus eight free scalar
fields.
The four remaining gauge fields will
describe the four non-interacting vector fields of the
gravity multiplet, and their contribution to the heat
kernel will be given by that of four vector fields
in $AdS_2\times S^2$ as given in \refb{efourgauge}. 
For the RR fields which are odd
under $\II_4$ we can repeat the argument by using
projection by the operator $(-1)^{F_L}\times \II_4$, --
this will pick the complementary set.
Thus in total the contribution to the heat kernel will be 
given by that of the bosonic sector of eight matter
multiplets of the $\NN=4$ theory, plus that of eight free
vector fields minus that of sixteen free scalar fields
on $AdS_2\times S^2$. This is precisely \refb{es82}.}
Thus the analysis proceeds as in
\cite{1005.3044} and we find, after including the
contribution due to the ghosts, that the net contribution to
the heat kernel is given by
\be \label{es82}
8 \left[8 K^s_{AdS_2}(0;s) K^s_{S^2}(0;s) + {1\over 2\pi a^2}
\left\{K^s_{S^2}(0;s) - K^s_{AdS_2}(0;s)\right\}\right]\, .
\ee
The small $s$ expansion of this can be found by standard
methods described earlier and we get the $s$ independent
contribution to \refb{es82} to be
\be \label{es83}
{34\over 45\pi^2 a^4}\, .
\ee

Next we consider the contribution from the extra fermion
fields. These fields can be labelled by $\psi'_\mu$ ($0\le\mu
\le 3$), $\Lamb'$ and $\vp_r'$ ($4\le r\le 9$) where
for each $\mu$ and $r$, $\psi'_\mu$ and $\vp_r'$ are
16 component right handed Majorana-Weyl spinor of the
ten dimensional Lorentz group, and $\Lamb'$ is a
16 component left-handed Majorana-Weyl spinor of the
ten dimensional Lorentz group. Physically $\psi'_\mu$
and $\vp'_r$ are the four dimensional and internal
components of the ten dimensional gravitino arising in the
R-NS sector. 
In the
presence of the background field, the quadratic action of
these fermionic fields can be obtained by the dimensional
reduction of the ten dimensional action of type IIA
supergarvity. The result is:
\ben \label{es84}
&& -{1\over 2} \Bigg[\bar \psi'_\mu \Gamma^{\mu\nu\rho} D_\nu
\psi'_\rho + \bar \Lamb' \Gamma^\mu D_\mu \Lamb'
+ \sum_{r=4}^9\bar \vp'_r \Gamma^\mu D_\mu \vp'_r 
- {1\over 2} 
\bar \psi'_\mu \Gamma^\mu \Gamma^\nu 
D_\nu\Gamma^\rho
 \psi'_\rho \nonumber \\
 && - {1\over 2\sqrt 2} \Bigg\{
\left(- \bar\psi'_\rho
\Gamma^{\mu\nu}\Gamma^\rho +\sqrt 2
\bar\Lamb' \Gamma^{\mu\nu}
\right) (\vp'_4 \bar F^1_{\mu\nu} +\vp'_5 \bar F^2_{\mu\nu})
\nonumber \\
&& + 
(\bar\vp'_4 \bar F^1_{\mu\nu} +\bar \vp'_5 \bar 
F^2_{\mu\nu}) \left( -\Gamma^\rho \Gamma^{\mu\nu}
\psi'_\rho - \sqrt 2 \Gamma^{\mu\nu} \Lamb'
\right)\Bigg\}\Bigg]
 \, ,
\een
where the last term in the first line
is the gauge fixing term. This also
leads to ghosts which have the same action as given in
the last two terms in \refb{emghost}, except that the
new ghost fields $\tilde b'$, $\tilde c'$ and $\tilde e'$ have 
opposite ten dimensional chirality compared to the
superghosts
$\tilde b$, $\tilde c$ and $\tilde e$ respectively of $\NN=4$
supergravity.
Using \refb{efnon} we can express \refb{es84} as
\ben \label{es85}
&&-{1\over 2}\Bigg[ \bar \psi'_\mu \Gamma^{\mu\nu\rho} D_\nu
\psi'_\rho - {1\over 2} 
\bar \psi'_\mu \Gamma^\mu \Gamma^\nu 
D_\nu\Gamma^\rho
 \psi'_\rho
 + \bar \Lamb' \Gamma^\mu D_\mu \Lamb'
+\sum_{r=4}^9 \bar \vp'_r \Gamma^\mu D_\mu \vp'_r \nn
&& + {1\over 2a} \Bigg\{
\left(\bar\psi'_m
\Gamma^m  - \bar\psi'_\alpha \Gamma^\alpha 
+ \sqrt 2 \bar\Lamb'\right) \tau_3 \vp'_4 
+ i\left(-\bar\psi'_m
\Gamma^m  + \bar\psi'_\alpha \Gamma^\alpha 
+ \sqrt 2 \bar\Lamb'\right) \sigma_3 \vp'_5
\nonumber \\ &&
\qquad + 
\bar\vp'_4 \tau_3  \left( \Gamma^m
\psi'_m - \Gamma^\alpha \psi'_\alpha -
\sqrt 2  \Lamb'
\right)
+ i\bar\vp'_5 \sigma_3  \left( -\Gamma^m
\psi'_m + \Gamma^\alpha \psi'_\alpha -
\sqrt 2  \Lamb'
\right) \Bigg\}\Bigg]\, .\nonumber \\
\een
As in \refb{ezxa} we shall express this as
\be \label{es86}
-{1\over 2} \left[
\sum_{r=6}^9  \bar \vp'_r \Gamma^\mu D_\mu \vp'_r
+ \left(\bar 
\Lamb' \KK^{(1)}
+ \bar\psi^{\prime\alpha} \KK^{(2)}_\alpha + \bar \psi^{\prime m}
\KK^{(3)}_m + \bar \vp_4' \KK^{(4)} + \bar \vp_5' \KK^{(5)}
\right)\right]\, ,
\ee
where
\ben \label{es87}
\KK^{(1)} &=& (\not \hskip -4pt D_{S^2} +\sigma_3 \, \not \hskip -4pt D_{AdS_2})
\Lamb' 
+ {1\over \sqrt 2\, 
a} (\tau_3 \vp'_4 + i\sigma_3 \vp'_5)
\, , \nn
\KK^{(2)}_\alpha &=&  - {1\over 2} \Gamma^n 
(\not \hskip -4pt D_{S^2} +\sigma_3 \, \not \hskip -4pt D_
{AdS_2}) \Gamma_\alpha \psi_n' 
 - {1\over 2}\Gamma^\beta\left(\not \hskip -4pt D_{S^2} 
+\sigma_3 \, \not \hskip -4pt D_{AdS_2}\right) \Gamma_\alpha
\psi'_\beta -{1\over 2a} \Gamma_\alpha 
(\tau_3\vp'_4 - i\sigma_3\vp'_5)\nn
\KK^{(3)}_m &=& 
- {1\over 2} \Gamma^\beta (\not \hskip -4pt D_{S^2} 
+\sigma_3 \, \not \hskip -4pt D_{AdS_2}) \Gamma_m\psi_\beta'
- {1\over 2}\Gamma^n\left(\not \hskip -4pt D_{S^2} 
+\sigma_3 \, \not \hskip -4pt D_{AdS_2}\right) \Gamma_m
\psi_n'
+ {1\over 2a} \Gamma_m 
(\tau_3\vp'_4 - i\sigma_3\vp'_5)\, \nn
\KK^{(4)} &=& (\not \hskip -4pt D_{S^2} +\sigma_3 \, \not \hskip -4pt D_{AdS_2})\vp_4' +{1\over 2a}\tau_3
\left( \Gamma^m
\psi'_m - \Gamma^\alpha \psi'_\alpha -
\sqrt 2  \Lamb'
\right)
\nn 
\KK^{(5)} &=& (\not \hskip -4pt D_{S^2} +\sigma_3 \, \not \hskip -4pt D_{AdS_2})\vp_5' +{i\over 2a} \sigma_3
\left( -\Gamma^m
\psi'_m + \Gamma^\alpha \psi'_\alpha -
\sqrt 2  \Lamb'
\right)\, .
\een

The fields $\vp_6',\cdots \vp_9'$ represent free fermions in $AdS_2\times
S^2$ background, and the net contribution from these fields to the heat
kernel is given by\cite{1005.3044}
\ben \label{es88}
&& -{4\over \pi^2 a^4} \, \int_0^\infty d\lambda \lambda \coth(\pi\lambda)
e^{-\bar s\lambda^2} \, \sum_{l=0}^\infty (2l+2) e^{-\bar s (l+1)^2}\nn
&=& -{2\over \pi^2 a^4 \bar s^2} \left( 1 -{11\over 180}\bar s^2 + \OO(
\bar s^3)\right)\, .
\een
The overall normalization is fixed by noting that each of the
$\vp'_r$'s represent four Majorana fermions in four
dimensions. Thus altogether we have 16 Majorana or
equivalently eight Dirac fermions. The overall minus sign
is a reflection of the fact that the path integral over the fermions
gives the determinant of the kinetic operator instead of the
inverse of the determinant.

For computing the contribution from the other fields we expand them
as in \refb{ezx1}
\ben \label{es89}
\Lamb' &=& a_1 \chi + a_2 \sigma_3 \chi\nn
\psi_\alpha' &=& b_1 \Gamma_\alpha\chi + b_2 \sigma_3
\Gamma_\alpha\chi + b_3 D_\alpha\chi + b_4 \sigma_3  D_\alpha\chi \nn
\psi_m' &=& c_1 \Gamma_m\chi + c_2 \sigma_3
\Gamma_m\chi + c_3 \sigma_3 D_m\chi + c_4  
D_m\chi \nn
\vp'_4 &=& \tau_3(h_1 \chi + h_2 \sigma_3 \chi) \\
\vp'_5 &=& (g_1 \chi + g_2 \sigma_3 \chi)
\een
where $a_i$, $b_i$, $c_i$, $h_i$ and $g_i$ are grassman parameters
and $\chi$ is the product of an arbitrary spinor of the
$SO(6)$ Clifford algebra generated by $\wt\Gamma^4,
\cdots \wh \Gamma^9$, $\chi^+_{lm}$ (or
$\eta^+_{lm}$) defined in \refb{ed2} and 
$\chi^\pm_k(\lambda)$ (or $\eta^\pm_k(\lambda)$) 
defined in \refb{ed2a}. $\chi$ satisfies
\be \label{es90}
\not \hskip-4pt D_{S^2} \chi= i\zet_1 \, \chi,
\quad \not \hskip-4pt D_{AdS_2}\chi= i\zet_2 \, \chi,
\quad \zet_1 > 0 \, .
\ee
As in \refb{ezxb}, we expand 
$\KK^{(1)},\cdots \KK^{(5)}$ as
\ben \label{es89A}
\KK^{(1)} &=& A_1 \chi + A_2 \sigma_3 \chi\nn
\KK^{(2)}_\alpha &=& B_1 \Gamma_\alpha\chi + B_2 \sigma_3
\Gamma_\alpha\chi + B_3 D_\alpha\chi + B_4 \sigma_3  D_\alpha\chi \nn
\KK^{(3)}_m &=& C_1 \Gamma_m\chi + C_2 \sigma_3
\Gamma_m\chi + C_3 \sigma_3 D_m\chi + C_4  D_m\chi \nn
\KK^{(4)} &=& \tau_3(H_1 \chi + H_2 \sigma_3 \chi) \nn
\KK^{(5)} &=& (G_1 \chi + G_2 \sigma_3 \chi)
\een
Explicit computation yields
\ben \label{es91}
A_1 &=& i\zeta_1 a_1 + i\zeta_2 a_2 + {1\over \sqrt 2 a} 
h_1 + {i\over \sqrt 2 a} g_2\nn
A_2 &=& i\zeta_2 a_1 - i\zeta_1 a_2 
+ {1\over \sqrt 2 a} 
h_2 + {i\over \sqrt 2 a} g_1\nn
B_1 &=& -i\zeta_1 b_1 + {1\over 2}\zeta_1^2 b_3 
+ {1\over 2}\zeta_1 \zeta_2 b_4 
+ i\zeta_1 c_1 -{1\over 2}\zeta_1 \zeta_2 c_3 + 
{1\over 2}\left(\zeta_2^2 + {1\over a^2} \right) c_4
- {1\over  2 a} h_1 + {i\over 2a} g_2 \nn
B_2 &=& i\zeta_1 b_2 
+{1\over 2}\zet_1 \zet_2 b_3
-{1\over 2}\zeta_1^2 b_4 
+ i\zet_1 c_2- {1\over 2}\left(\zeta_2^2 
+ {1\over a^2} \right) c_3
-{1\over 2}\zet_1 \zet_2 c_4 +{1\over 2a} h_2
-{i\over 2a} g_1\nn
B_3 &=& i\zet_2 b_4  - 2c_1 - i\zet_2 c_3\nn
B_4 &=& i\zet_2 b_3 -2 c_2 -i\zet_2 c_4\nn
C_1 &=& -i\zet_2 b_2 + {1\over 2}\left(\zet_1^2 -{1\over a^2}
\right) b_3 +{1\over 2}\zet_1\zet_2 b_4 
-i\zet_2 c_2 -{1\over 2}\zet_1 \zet_2 c_3 +
{1\over 2}\zet_2^2 c_4
+ {1\over  2 a} h_1 - {i\over 2a} g_2 \nn
C_2 &=& i\zet_2 b_1 -{1\over 2}\zet_1\zet_2 b_3 
+{1\over 2}\left( \zet_1^2 -{1\over a^2}\right) b_4
-i\zet_2 c_1 +{1\over 2}\zet_2^2 c_3 +{1\over 2} \zet_1
\zet_2 c_4+{1\over 2a} h_2
-{i\over 2a} g_1\nn
C_3 &=& 2 b_2 + i\zet_1 b_4 - i\zet_1 c_3\nn
C_4 &=& -2 b_1 -i\zet_1 b_3 + i\zet_1 c_4\nn
H_1 &=& i\zeta_1 h_1 - i\zeta_2 h_2 
-{1\over \sqrt 2 a}  a_1 -{1\over a} b_1 -{i\over 2 a}
\zet_1 b_3 + {1\over a} c_1 +{i\over 2a} \zet_2 c_3\nn
H_2 &=& -i\zeta_2 h_1 - i\zeta_1 h_2 
-{1\over \sqrt 2 a}  a_2 +{1\over a} b_2 +{i\over 2 a}
\zet_1 b_4 + {1\over a} c_2 +{i\over 2a} \zet_2 c_4\nn
G_1 &=& i\zeta_1 g_1 + i\zeta_2 g_2 
-{i\over \sqrt 2 a}  a_2 -{i\over a} b_2 +{1\over 2 a}
\zet_1 b_4 - {i\over a} c_2 +{1\over 2a} \zet_2 c_4\nn
G_2 &=& i\zeta_2 g_1 - i\zeta_1 g_2 
-{i\over \sqrt 2 a}  a_1 +{i\over a} b_1 -{1\over 2 a}
\zet_1 b_3 - {i\over a} c_1 +{1\over 2a} \zet_2 c_3\, .
\een
We can express this as
\be \label{es92}
\pmatrix{\vec A\cr \vec B\cr \vec C\cr \vec H\cr \vec G} 
= \MM \, \pmatrix{\vec a\cr
\vec b\cr \vec c\cr \vec h\cr \vec g}\, ,
\ee
$\MM$ being a $14\times 14$ matrix. 
Let us also introduce a matrix $\MM_1$ through
\be \label{esx1A}
\MM^2 = - (\zet_1^2 + \zet_2^2) I_{14} + 
a^{-2} \MM_1 \, ,
\ee
where $I_{14}$ denotes the $14\times 14$ identity matrix,
and denote
by $\beta_k$ for $1\le k\le 14$ the 14 eigenvalues of the
matrix $\MM_1$. 
Then following the logic leading to \refb{esx2} one can show that
the contribution to the heat kernel
from the fermionic modes for $|\zet_1| >1$, \i.e.\ $l>0$, 
will be given by
\be \label{esx2A}
K^f_{(1)}(0;s) = -{1\over 2\pi^2 a^4}\,
\sum_{l=1}^\infty (2l+2) \int_0^\infty d\lambda \lambda
\coth(\pi\lambda) e^{-\bar s (l+1)^2 -\bar s\lambda^2}
\sum_{k=1}^{14} e^{\bar s \beta_k}\, ,
\ee
where $l$ and $\lambda$ are related to $\zeta_1$ and $\zeta_2$ via
\be \label{esx1.5A}
|\zet_1| = (l+1)/a, \qquad \zet_2 = \lambda/a\, .
\ee
The overall normalization is fixed by noting that $\psi_\mu'$,
$\vp_r'$ for $r=4,5$ and $\Lamb'$ altogether has degrees
of freedom equal to that of $(4+2+1)\times 4=28$ Majorana
fermions or equivalently 14 Dirac fermions.

The contribution from the $|\zet_1|=1/a$,
\i.e.\ $l=0$ term has to be
evaluated separately following the same logic that lead to
\refb{esy2}. We choose the coefficients $b_3$ and $b_4$ to be
zero and replace
in  \refb{es91}
the expressions for $B_k$ by that of
$B_k + {i\over 2a} B_{k+2}$ for $k=1,2$. This leads to a
12$\times$12 matrix $\wt\MM$. We now
define 
a matrix $\wt\MM_1$ through
\be \label{esx1.9A}
\wt\MM^2 = -(a^{-2}+\zet_2^2) I_{12} + a^{-2} \wt\MM_1 \, ,
\ee
where $I_{12}$ denotes the $12\times 12$ identity matrix.
If $\wt\beta_k$'s are the eigenvalues of $\wt\MM_1$
then the contribution from the $l=0$ modes to the
heat kernel may be expressed as
\be \label{esy2A}
K^f_{(2)}(0;s) = -{1\over \pi^2 a^4}\,
\int_0^\infty d\lambda \lambda
\coth(\pi\lambda) e^{-\bar s  -\bar s\lambda^2}
\sum_{k=1}^{12} e^{\bar s \wt\beta_k}\, .
\ee
We can combine \refb{esx2A} and \refb{esy2A} to write
\be \label{esy3A}
K^f_{(1)}(0;s) +  K^f_{(2)}(0;s) = \wt
K^f_{(1)}(0;s) +  \wt K^f_{(2)}(0;s)\, ,
\ee
where
\ben \label{esy4aA}
\wt K^f_{(1)}(0;s) &=&  -{1\over 2\pi^2 a^4}\,
\sum_{l=0}^\infty (2l+2) \int_0^\infty d\lambda \lambda
\coth(\pi\lambda) e^{-\bar s (l+1)^2 -\bar s\lambda^2}
\sum_{k=1}^{14} e^{\bar s \beta_k}\nn
&=& -{1\over \pi^2 a^4} \, Im \,
\int_0^{e^{i\kappa}\times \infty} 
d\wt\lambda \, \wt\lambda \, 
\cot(\pi\wt\lambda)\, \int_0^\infty d\lambda \lambda
\coth(\pi\lambda) e^{-\bar s\wt\lambda^2-\bar s\lambda^2}
\sum_{k=1}^{14} e^{\bar s \beta_k|_{l+1 \to \wt\lambda}}\nn
\een
\be \label{esy4bA}
\wt K^f_{(2)}(0;s) = -{1\over \pi^2 a^4}\,
\int_0^\infty d\lambda \lambda
\coth(\pi\lambda) e^{-\bar s  -\bar s\lambda^2}
\left[\sum_{k=1}^{12} e^{\bar s \wt\beta_k}
- \sum_{k=1}^{14} e^{\bar s \beta_k|_{l=0}}\right]\, .
\ee

We also need to compute the contribution 
due to the discrete
modes described in \refb{eadd1}. 
For this we set the fields $\Lamb'$, $\psi_\alpha'$,  $\vp_4'$
and $\vp_5'$ to
0, and expand $\psi_m'$ as in \refb{es89} with $c_{k+2}=2c_ka$
for $k=1,2$, with $\zet_2=i/a$,
$|\zet_1|\ge 1/a$, \i.e.\ $l\ge 0$.
It can be seen that with this choice
$A_i$, $B_i$, $H_i$, $G_i$ computed from 
\refb{es91} vanish and we have $C_{k+2}=2C_ka$ for $1\le k \le 2$. 
Thus we can express
these relations as
\be \label{eexx1A}
\pmatrix{C_1\cr C_2} = \wh\MM \, 
\pmatrix{c_1\cr c_2} \, ,
\ee
for some $2\times 2$ matrix $\wh\MM$. If $\wh\beta_k$
denote the eigenvalues of 
$a^2\{\wh \MM^2 + (\zet_1^2
-a^{-2}) I_2 \}$ then the 
contribution to $K(0;s)$ from these modes is given by
\ben \label{eexx2A}
 K^f_{(3)}(0;s) &=& -{1\over 2\pi^2 a^4}\, 
\sum_{l=0}^\infty (2l+2) e^{\bar s-\bar s(l+1)^2} \sum_{k=1}^2
e^{\bar s \wh\beta_k}\nn
&=& -{1\over \pi^2 a^4} \, Im \,
\int_0^{e^{i\kappa}\times \infty} 
d\wt\lambda \, \wt\lambda \, 
\cot(\pi\wt\lambda)\, e^{\bar s-\bar s \wt\lambda^2} 
\sum_{k=1}^2
e^{\bar s \wh\beta_k|_{l+1\to\wt\lambda}}\nn\, ,
\een
Explicit computation using \refb{es91} gives
$\wh\beta_1=\wh\beta_2=-1$. Hence we have
\be \label{ekf3fin}
 K^f_{(3)}(0;s) = -{2\over \pi^2 a^4} \, Im \,
\int_0^{e^{i\kappa}\times \infty} 
d\wt\lambda \, \wt\lambda \, 
\cot(\pi\wt\lambda)\, e^{-\bar s \wt\lambda^2} 
\, .
\ee

Finally the three sets of bosonic ghosts 
$\wt b'$, $\wt c'$ and $\wt e'$ associated with
gauge fixing of local supersymmetry, 
each of which gives rise
to four Majorana fermions in four dimensions, 
contributes
\ben \label{efghx1A}
 K^f_{ghost} &=& {3\over \pi^2 a^4}\,
\sum_{l=0}^\infty (2l+2) \int_0^\infty d\lambda \lambda
\coth(\pi\lambda) e^{-\bar s (l+1)^2 -\bar s\lambda^2}
\nn
&=& {6\over \pi^2 a^4} \, Im \,
\int_0^{e^{i\kappa}\times \infty} 
d\wt\lambda \, \wt\lambda \, 
\cot(\pi\wt\lambda)\, \int_0^\infty d\lambda \lambda
\coth(\pi\lambda) e^{-\bar s\wt\lambda^2-\bar s\lambda^2}\, ,
\een
to
$K(0;s)$.

To evaluate the right hand sides of  \refb{esy4aA} and
\refb{esy4bA}
we use the relations
\be \label{esx3A}
\sum_k e^{\bar s \beta_k} = \sum_{n=0}^\infty
{1\over n!} \bar s^n \sum_k \beta_k^n 
= \sum_{n=0}^\infty
{1\over n!} \bar s^n  Tr(\MM_1^n)\, ,
\ee
\be \label{esx3abA}
\sum_k e^{\bar s \wt\beta_k} = \sum_{n=0}^\infty
{1\over n!} \bar s^n \sum_k \wt\beta_k^n 
= \sum_{n=0}^\infty
{1\over n!} \bar s^n  Tr(\wt\MM_1^n)\, .
\ee
Explicit computation gives
\ben\label{esx4A}
Tr(\MM_1) &=&  0\nonumber \\
Tr(\MM_1^2) &=& -8 + 16 (l+1)^2 - 16 \lambda^2 \nonumber \\
Tr(\MM_1^3) &=& -6 (l+1)^2 - 6 \lambda^2 \nonumber \\
Tr(\MM_1^4) &=& 8 - 16 (l+1)^2 + 40 (l+1)^4 + 
16 \lambda^2 - 48 (l+1)^2 \lambda^2 + 40 \lambda^4 \, .
\een
\ben\label{esy1A}
Tr(\wt\MM_1) &=& -2 \nonumber \\
Tr(\wt\MM_1^2) &=& 6 - 16 \lambda^2 \nonumber \\
Tr(\wt\MM_1^3) &=&  -8 - 6 \lambda^2\nonumber \\
Tr(\wt\MM_1^4) &=&  30 - 32 \lambda^2 
+ 40 \lambda^4\, .
\een

Following the procedure of \S\ref{sferm} we can now
carry out the small $s$ expansion of the heat kernels.
We get the following
contribution to the order $s^0$ term in the small $s$
expansion of various terms:
\ben \label{efin1A}
\wt K^f_{(1)}(0;s) &:& -{43\over 360 \pi^2 a^4}
\nn
\wt K^f_{(2)}(0;s) &:& {1\over 6\pi^2 a^4}
\nn
 K^f_{(3)}(0;s) &:& {1\over 6\pi^2 a^4}
\nn
 K^f_{ghost}(0;s) &:& - {11\over 120\pi^2 a^4}
\, .
\een
Adding up all the contributions in eq.\refb{efin1A}
and the contribution from \refb{es88}
we get the net contribution to
$K(0;s)$ from the extra fermionic fields of $\NN=8$
supergravity:
\be \label{enetferA}
K^f(0;s) = { 11 \over  45 \pi^2 a^4}\, .
\ee

Adding \refb{enetferA} to the bosonic contribution
\refb{es83} we get the
net contribution to the order $s^0$ terms in the heat
kernel from all the extra fields appearing in $\NN=8$
supergravity:
\be \label{eabfin}
{1\over \pi^2 a^4}\, .
\ee

It is also easy to see that the only zero modes
among these extra fields arise from the gauge fields.
In particular there are no fermion zero modes since both
the $\wh\beta_k$'s in \refb{eexx2A} take the value
$-1$ for $l=0$.
Now we have already seen that for the gauge fields
the integration over the zero modes gives us back the same
result that we remove from the heat kernel. Thus removing the
zero mode contribution of the sixteen gauge fields
from the heat kernel and then including
the contribution due to the zero mode integrals
does not give any net contribution, and 
\refb{eabfin} represents the net extra contribution to the
heat kernel from the extra fields of $\NN=8$ supergravity.
Since for the $\NN=4$ supergravity the net $s$
independent contribution to the effective heat kernel
vanished, \refb{eabfin} represents the net contribution in
$\NN=8$ supergravity.
According
to \refb{elogfor} this gives a logarithmic
correction to the black hole entropy of the form:
\be \label{enetlog}
-4\ln a^2 = -2\ln \Delta\, .
\ee
This is 
in perfect
agreement with the microscopic answer
\refb{en8}. 

For identifying separately the contributions from the zero 
modes and the non-zero modes we note that the $\NN=8$
supergravity has 28 gauge fields whose zero mode contribution
to the entropy is $-28\ln a=-7\ln\Delta$. This represents the
net zero mode contribution since the contribution from the
graviton and the gravitino zero modes cancel. The rest of
the contribution $5\ln\Delta$ comes from non-zero modes.

\sectiono{Half BPS black holes in STU model} \label{sstu}

Our analysis also gives the result for logarithmic corrections to the
entropy of half BPS black holes in the 
STU model\cite{9508064,9901117}
which has been studied recently 
in \cite{0711.1971,0810.1233,0905.4115} in the context of
black hole entropy. 
The STU model is
constructed by beginning with type IIA string theory on $T^4\times T^2$
and taking an orbifold of this theory with a $\ZZZ_2\times \ZZZ_2$
group. The first $\ZZZ_2$ acts as $(-1)^{F_L}$ times half a unit of
shift along
one of the circles of $T^2$ and the second $\ZZZ_2$ acts as $\II_4$
times a shift along the second circle of $T^2$ where $\II_4$
denotes changing the sign of all the coordinates of $T^4$. 
If we label the
two circles of $T^2$ by $x^4$ and $x^5$ then the black hole solution
described at the beginning of
\S\ref{squad} survives the orbifold projection and hence
continues to describe a black hole solution in this theory. The first
$\ZZZ_2$ projection removes from the spectrum all the masless
RR and R-NS sector states and hence the low energy theory is
an $\NN=4$ supergravity theory, -- with a structure identical to that
of heterotic string theory on $T^6$ except that the sixteen matter
multiplet fields associated
with the dimensional reduction of ten dimensional $E_8\times E_8$
gauge fields are absent.
The action of the second $\ZZZ_2$ orbifold projection breaks
the $\NN=4$ supersymmetry to $\NN=2$. Under this a matter
multiplet of $\NN=4$ supergravity decomposes into a vector
multiplet and a hypermultiplet, and we need to examine which
components of the fields survive the projection.
Similarly the gravity multiplet fields of the $\NN=4$
supergravity decompose into different supermultiplets of
$\NN=2$ supergravity, and only some of these survive
the orbifold projection.

For later use it is useful to note that
in the fermionic sector the orbifold operation
projects onto modes which are even under the action of
$\Gamma^{6789}$ accompanied by
$(x^6,\cdots x^9)\to (-x^6,\cdots -x^9)$. 
Using the ten dimensional chirality
of $\Lambda$ and the ten dimensional gravitino field
$\psi_M$ ($0\le M\le 9$), this condition translates
to
\be\label{eorbproj}
\sigma_3\tau_3\wh\Gamma^{4}\wh\Gamma^{5} \psi_\mu
=i\psi_\mu\, , \quad \sigma_3\tau_3\wh\Gamma^{4}
\wh\Gamma^{5} \psi_{4,5}
=i\psi_{4,5}\, , \quad \sigma_3\tau_3\wh\Gamma^{4}
\wh\Gamma^{5} \psi_{6,7,8,9}
=-i\psi_{6,7,8,9}, \quad
\sigma_3\tau_3\wh\Gamma^{4}\wh\Gamma^{5} \Lambda
=-i\Lambda\, ,
\ee
together with similar projection on the ghost fields.

Let $G_{MN}$ and $B_{MN}$ be the ten dimensional
metric and NSNS 2-form fields.
We begin with the 
 two matter multiplet fields of
$\NN=4$ supergravity whose vector
fields come from $G_{4\mu}-B_{4\mu}$ and $G_{5\mu}-B_{5\mu}$.
Their scalar partners are $G_{44}$, $G_{45}$, $G_{55}$, $B_{45}$,
$G_{4m} - B_{4m}$  and $G_{5m}-B_{5m}$ for
$6\le m\le 9$. Under the orbifold
projection the two vector fields as well as the scalars
 $G_{44}$, $G_{45}$, $G_{55}$, $B_{45}$ survive, but the
 rest of the scalars are projected out. The
 surviving fields belong
 to two vector multiplets
 of $\NN=2$ supersymmetry. The contribution to the
 heat kernel from these scalar and vector fields 
 and the ghosts associated with the vector fields
 can be read out from
 the results of \cite{1005.3044}. The vector couples to the two scalars
 due to the presence of the background flux and the net contribution to the
 heat kernel from the bosonic fields (including the ghosts) is given by 
 $4K^s(0;s)$. 
  There are zero modes of the gauge fields whose contribution
 needs to be removed from this and then added separately, 
 but as we have seen before, this does not change the
 result.
 
 The fermionic components of these two
 matter multiplets come from the
 components $\psi_4$ and $\psi_5$ of the ten
 dimensional gravitino. As was shown in \cite{1005.3044},
 acting on these fermions, the kinetic operator takes the form:
 \be \label{erefkin}
\not \hskip -4pt D_{S^2} + \sigma_3 \not \hskip -4pt D_{AdS_2}
- {i\over 2} \, a^{-1} \, \wh\Gamma^5\, \tau_3 
- {1\over 2}  a^{-1} \,\sigma_3\, \wh\Gamma^4
 \, .
\ee
It follows from \refb{eorbproj} that acting on the fields
$\psi_{4,5}$ the last two terms in \refb{erefkin}
cancel and the kinetic operator reduces to 
$\not \hskip -4pt D_{S^2} + \sigma_3 \not \hskip -4pt D_{AdS_2}$,
\i.e.\ that of a free fermion in $AdS_2\times S^2$. The 
heat kernel of this is given by 1/8 of the
contribution shown in \refb{es88}. 
Adding this to the bosonic
contribution $4 K^s(0;s)$ given in \refb{e10} we get the net contribution
to the heat kernel from each of the vector multiplets to be:
\be \label{ekhyper}
K^{vector}(0;s) = {1\over 180\pi^2 a^4} + {11\over 720\pi^2 a^4}
+\cdots
= {1\over 48\pi^2 a^4}+\cdots \, .
\ee
This corresponds to a correction of $-{1\over 24}\ln\Delta$ per vector
multiplet, \i.e.\ a total of $-{1\over 12}\ln\Delta$ to the
black hole entropy from the two vector multiplets coming from the two
matter multiplets of $\NN=4$ supergravity.

Next we turn to the 
four matter multiplet fields 
of $\NN=4$ supergravity whose vector 
fields come 
from $G_{m\mu}- B_{m\mu}$ where $m$ is along $T^4$ and
$\mu$ is along the non-compact direction. Their scalar
components are $G_{mn}$, $B_{mn}$, $G_{m4}-B_{m4}$
and $G_{m5}-B_{m5}$. Under the orbifold projection the
scalars $G_{mn}$, $B_{mn}$ survive but the vector fields as
well as the scalars $G_{m4}-B_{m4}$
and $G_{m5}-B_{m5}$ are projected out. This corresponds to
removing the vector multiplets and keeping the hypermltiplet
fields. Since the net contribution to $K(0;s)$ from a 
hypermultiplet and a vector multiplet vanishes, we could
directly conclude that the hypermultiplet contribution
to $K(0;s)$ will be negative of the contribution
\refb{ekhyper} from the vector multiplet. However it is
instructive to carry out the computation directly.
For each hypermultiplet we have four scalars without any
coupling to the background gauge fields, and their 
contribution to the heat kernel is given by 
$4 K^s(0;s)$. 
In the fermionic sector we have the fields $\psi_6,\cdots \psi_9$
subject to the orbifold projection \refb{eorbproj}.
This makes the contribution from the last two terms in
\refb{erefkin} identical, and we can express the
operator as $\not \hskip -4pt D_{S^2} + \sigma_3 \not \hskip -4pt D_{AdS_2}
- {i} \, a^{-1} \, \wh\Gamma^5\, \tau_3$.
We need to compute its determinant on the subspace
of states subject to the projection \refb{eorbproj}.
Now note that since $\sigma_3\tau_3\wh\Gamma^4$ 
anti-commutes with
the projection operator it takes a state satisfying the
orbifold projection to a state satisfying the opposite
projection and vice versa. Since 
it also anti-commutes with the kinetic operator
$\not \hskip -4pt D_{S^2} + \sigma_3 
\not \hskip -4pt D_{AdS_2}
- {i} \, a^{-1} \, \wh\Gamma^5\, \tau_3$, the action
of $\sigma_3\tau_3\wh\Gamma^4$ changes the eigenvalue
of the kinetic operator. Thus we see that 
the matrix representing the kinetic operator
in the subspace satisfying opposite projection is 
just the negative of the kinetic operator acting on
the subspace
satisfying the correct projection. Thus we could
evaluate the determinant ignoring the projection condition
and then take the square root of the modulus
of the determinant. We now note that in the
unprojected space the operator 
$\not \hskip -4pt D_{S^2} 
- {i} \, a^{-1} \, \wh\Gamma^5\, \tau_3$
anti-commutes with the operator 
$\sigma_3 \not \hskip -4pt D_{AdS_2}$. 
Thus the squares of the
eigenvalues of 
\hbox{$\not \hskip -4pt D_{S^2} 
+ \sigma_3 \not \hskip -4pt D_{AdS_2}
- {i} \, a^{-1} \, \wh\Gamma^5\, \tau_3$} will be given by the
sum of the squares of the eigenvalues of
$\not \hskip -4pt D_{S^2} 
- {i} \, a^{-1} \, \wh\Gamma^5\, \tau_3$ and
$\not 
\hskip -4pt D_{AdS_2}$. Of these $\not 
\hskip -4pt D_{AdS_2}$ has eigenvalues $\pm i
a^{-1}\lambda
$. On the other hand
since $\not \hskip -4pt D_{S^2}$ has eigenvalues
$\pm ia^{-1}(l+1)$ and $\wh\Gamma^5\, \tau_3$ has 
eigenvalues
$\pm 1$, and they act on independent spaces,
the eigenvalues of $\not \hskip -4pt D_{S^2} 
- {i} \, a^{-1} \, \wh\Gamma^5\, \tau_3$ are given by
$\pm i a^{-1}(l+1\pm 1)$ 
with $l=0,1,\cdots \infty$. This gives the net
contribution to $K(0;s)$ from the fermionic 
components of the hypermultiplet to be
\be \label{ehyperfermi}
-{1\over 2\pi^2 a^4} Im
\int_0^{e^{i\kappa}\times\infty}
d\wt\lambda \, \wt\lambda \, \coth\pi\wt\lambda \, 
\int_0^\infty d\lambda\, \lambda\,
\coth\pi\lambda \, 
e^{-s\lambda^2 -s\wt\lambda^2} \, 
\left[e^{-s - 2s\wt\lambda} + e^{-s+2s\wt\lambda}\right]\, .
\ee
We can evaluate this by expanding the term in the
square bracket in a power series in $s$, or by shifting the 
sum over $l$ as in \cite{1005.3044}.
Both ways give the same result and adding this to the
scalar contribution $4K^s(0;s)$ we get
the contribution to the heat kernel from each hypermultiplet fields
to be
\be \label{ekvector}
K^{hyper}(0;s) = -{1\over 48\pi^2 a^4}+\cdots \, .
\ee
This corresponds to a correction of ${1\over 24}\ln\Delta$ per
hypermultiplet, \i.e.\ a total of ${1\over 6}\ln\Delta$ to the
black hole entropy from the four hypermultiplets coming from the
matter multiplets of $\NN=4$ supergravity.

Finally we have to compute the contribution from the fields
 which survive from the gravity multiplet of $\NN=4$ supergravity.
 In the bosonic sector the four gauge fields $G_{m\mu}+B_{m\mu}$
 for $m$ along $T^4$ are projected out but all other fields
 survive. Thus we need to remove the contribution given by
 \refb{efourgauge} together with a contribution of 
 $-8 K^s_{AdS_2\times S^2}(0;s)$
 representing the contribution of the eight ghost fields associated
 with these four gauge fields. 
 The small $s$ expansion of this is given by\cite{1005.3044}:
 \be \label{esmall1}
 {1\over \pi^2 a^4 \bar s^2} \left(1 +{16\over 45}
\bar s^2 - {1\over 2} - {1\over 90} \bar s^2 +\cdots \right)
={1\over \pi^2 a^4 \bar s^2} \left({1\over 2}
+{31\over 90} \bar s^2 +\cdots\right)\, ,
\ee
where the $-{1/2} -\bar s^2/90$ is the 
contribution due to the ghosts. Of these $4/8\pi^2 a^4$ can be
identified as the contribution due to the gauge field zero
modes. Thus the net non-zero mode contribution from
these four gauge fields is $31/90\pi^2 a^4 - 1/2\pi^2a^4
=-7/45\pi^2 a^4$, -- this needs to be removed from the
non-zero mode contribution 
\refb{efgravnz} from the gravity multiplet of full $\NN=4$
supergravity. Thus the net $s$-independent 
contribution to the heat kernel 
from  the  bosonic non-zero modes of $\NN=4$ gravity multiplet
which survive the orbifold projection is given by:
\be \label{estuboson}
-{101\over 180 \pi^2 a^4} + {7\over 45\pi^2 a^4}
= -{73 \over 180 \pi^2 a^4}\, .
\ee

The fermionic components of the $\NN=4$ gravity multiplet are
given by $\psi_\mu$ and $\Lambda$,
but we need to work in the subspace of these fermions which
satisfy the conditions \refb{eorbproj}. This 
requires us to impose the following restriction on the various
coefficients appearing in \S\ref{sferm}:
\ben \label{eorbrestr}
&& a_4 = i\, a_1, \quad a_3 = -i\, a_2, \quad 
 a_4' = i\, a_1', \quad a_3' = -i\, a_2', \nn &&
b_4 = -i\, b_1, \quad b_3 = i\, b_2, \quad b_8 = - i\, b_5, \quad
b_7 = i \, b_6\nn &&
b'_4 = -i\, b'_1, \quad b'_3 = i\, b'_2, \quad b'_8 = - i\, b'_5, \quad
b'_7 = i \, b'_6\nn &&
c_4 = -i\, c_1, \quad c_3 = i\, c_2, \quad c_8 = - i\, c_5, \quad
c_7 = i \, c_6\nn &&
c'_4 = -i\, c'_1, \quad c'_3 = i\, c'_2, \quad c'_8 = - i\, c'_5, \quad
c'_7 = i \, c'_6\, .
\een
Furthermore after the action of the kinetic operator on the
fields the result will be
a fermion of opposite chirality and hence the coefficients $A_i,A_i',
B_i, B_i', C_i, C_i'$ are no longer all independent. This allows us
to remove half of these coefficients and keep 
$A_i, A_i'$ for $i=1,2$ and $B_i,B_i',C_i, C_i'$ for $i=1,2,5,6$
as the independent constants labelling the state obtained by the action
the kinetic operator on the fields. This essentially halves the dimensions
of all the matrices $\MM_1$, $\wt\MM_1$  and $\wh\MM_1$
appearing in \S\ref{sferm}. The rest of the analysis proceeds
exactly as in \S\ref{sferm},
and we find the following results for the traces of the various matrices:
\ben\label{esx4n=2}
Tr(\MM_1) &=&  16\nonumber \\
Tr(\MM_1^2) &=& 64 - 32 (l+1)^2 - 32 
\lambda^2 \nonumber \\
Tr(\MM_1^3) &=&  256 -192 (l+1)^2 - 192 
\lambda^2\nonumber \\
Tr(\MM_1^4) &=&  1024 -1024 (l+1)^2 + 
128(l+1)^4 - 1024 \lambda^2 
+256 (l+1)^2 \lambda^2 + 128 \lambda^4\, . \nn
\een
\ben\label{esy1n=2}
Tr(\wt\MM_1) &=&  8\nonumber \\
Tr(\wt\MM_1^2) &=& 16-16 \lambda^2 \nonumber \\
Tr(\wt\MM_1^3) &=&  32-96 \lambda^2\nonumber \\
Tr(\wt\MM_1^4) &=&  64
- 384 \lambda^2 + 64 \lambda^4\, .
\een
\ben\label{esy1.9n=2}
Tr(\wh\MM_1) &=&  0\nonumber \\
Tr(\wh\MM_1^2) &=&   0 \nonumber \\
Tr(\wh\MM_1^3) &=&   0 \nonumber \\
Tr(\wh\MM_1^4) &=&  0 \, .
\een
This in turn gives the following order $\bar s^0$ terms in the small
$\bar s$ expansion of various parts of the fermionic
heat kernel:
\ben \label{efin1n=2}
\wt K^f_{(1)}(0;s) &:& {11\over 144\pi^2 a^4}
\nn
\wt K^f_{(2)}(0;s) &:& -{5\over 12\pi^2 a^4}
\nn
K^f_{(3)}(0;s) &:& - {5\over 12\pi^2 a^4}
\nn
K^f_{ghost}(0;s) &:& - {11\over 240\pi^2 a^4}
\, .
\een
Adding up all the contributions and subtracting the zero mode 
contribution $-1/2\pi^2 a^4$ we get the net contribution to
$K(0;s)$ from the non-zero modes of the 
surviving fermionic fields in the $\NN=4$ gravity multiplet after the
orbifold projection:
\be \label{enetfern=2}
K^f(0;s) = -{109\over 360\pi^2 a^4} +\cdots\, .
\ee
Adding this to \refb{estuboson} we get a net contribution of
$-17/24\pi^2 a^4$ from the non-zero modes. On the other hand the
zero modes of two gauge fields, the metric and the gravitino gives
a net contribution of
\be \label{enetzeron=2}
{2\over 8\pi^2 a^4} +{12\over 8\pi^2 a^4} - {3\over 2\pi^2 a^4} 
= {1\over 4\pi^2 a^4}\, ,
\ee
to the effective heat kernel.
Adding this to the sum of \refb{estuboson} and \refb{enetfern=2} we
get 
\be \label{enetn=2grav}
-{11\over 24\pi^2 a^4}\, ,
\ee
leading to a correction of ${11\over 12}\ln \Delta$ to the entropy.
Adding this to the contribution of ${1\over 6}\ln\Delta$ from
the four hypermultiplets and $-{1\over 12} \ln \Delta$ from the 
two vector multiplets we arrive at a net correction of
\be \label{en=2fin}
\ln \Delta\, ,
\ee
to the entropy of a half BPS black hole in the STU model.

We can identify separately the zero mode and the non-zero
mode contributions by noting that the four gauge field zero
modes give a contribution of $-\ln\Delta$ and the
contributions from the metric and the gravitino zero
modes cancel. The rest of the
contribution $2\ln\Delta$ comes from the non-zero
modes.

Finally we note that the analysis of this section can be extended to
any $\NN=2$ supergravity theory whose low energy effective action
can be obtained by a consistent truncation of the $\NN=4$
supergravity action in which two of the six
vector fields of the gravity
multiplet survive. In that case we can consider a black hole solution
whose electric and magnetic charges are carried by these vector fields
and the analysis of logarithmic corrections proceed in an identical
manner. The FHSV model of \cite{9505162} is another 
example of such a model.

\bigskip

{\bf Acknowledgement:} We would like to thank Atish
Dabholkar, Justin David, Frederik Denef, Joao Gomes, 
Rajesh Gopakumar, Dileep Jatkar and
Sameer Murthy
for useful discussions. 
The work of R.K.G. 
is part of research programme of FOM, which is financially supported by the Netherlands Organization for Scientific Research (NWO).
The work of I.M. was supported in part by
the project 11-R\&D-HRI-5.02-0304. The work of A.S. was
supported in part by the J. C. Bose fellowship of 
the Department of Science and Technology, India and the 
project 11-R\&D-HRI-5.02-0304.


\begin{thebibliography}{99}

\bibitem{9307038}
  R.~M.~Wald,
  ``Black hole entropy in the Noether charge,''
  Phys.\ Rev.\ D {\bf 48}, 3427 (1993)
  [arXiv:gr-qc/9307038].

\bibitem{9312023}
  T.~Jacobson, G.~Kang and R.~C.~Myers,
  ``On Black Hole Entropy,''
  Phys.\ Rev.\ D {\bf 49}, 6587 (1994)
  [arXiv:gr-qc/9312023].

\bibitem{9403028}
  V.~Iyer and R.~M.~Wald,
  ``Some properties of Noether 
  charge and a proposal for dynamical black hole
  entropy,''
  Phys.\ Rev.\ D {\bf 50}, 846 (1994)
  [arXiv:gr-qc/9403028].

\bibitem{9502009}
  T.~Jacobson, G.~Kang and R.~C.~Myers,
  ``Black hole entropy in higher curvature gravity,''
  arXiv:gr-qc/9502009.


\bibitem{0506177}
  A.~Sen,
  ``Black hole entropy function and the attractor mechanism in higher
  derivative gravity,''
  JHEP {\bf 0509}, 038 (2005)
  [arXiv:hep-th/0506177].

\bibitem{0508042}
  A.~Sen,
  ``Entropy function for heterotic black holes,''
  JHEP {\bf 0603}, 008 (2006)
  [arXiv:hep-th/0508042].

\bibitem{0809.3304}
  A.~Sen,
  ``Quantum Entropy Function from AdS(2)/CFT(1) Correspondence,''
  Int.\ J.\ Mod.\ Phys.\  A {\bf 24}, 4225 (2009)
  [arXiv:0809.3304 [hep-th]].

\bibitem{9607026}
R.~Dijkgraaf, E.~P.~Verlinde and H.~L.~Verlinde,
``Counting dyons in N = 4 string theory,''
Nucl.\ Phys.\ B {\bf 484}, 543 (1997)
[arXiv:hep-th/9607026].

\bibitem{0412287}
  G.~Lopes Cardoso, B.~de Wit, J.~Kappeli and T.~Mohaupt,
  ``Asymptotic degeneracy of dyonic N = 4 string states and black hole
  JHEP {\bf 0412}, 075 (2004)
  [arXiv:hep-th/0412287].

\bibitem{0505094}
D.~Shih, A.~Strominger and X.~Yin,
``Recounting dyons in N = 4 string theory,''
arXiv:hep-th/0505094.

\bibitem{0506249}
D.~Gaiotto,
``Re-recounting dyons in N = 4 string theory,''
arXiv:hep-th/0506249.

\bibitem{0508174}
  D.~Shih and X.~Yin,
  ``Exact black hole degeneracies and the topological string,''
  JHEP {\bf 0604}, 034 (2006)
  [arXiv:hep-th/0508174].

\bibitem{0510147}
  D.~P.~Jatkar and A.~Sen,
  ``Dyon spectrum in CHL models,''
  JHEP {\bf 0604}, 018 (2006)
  [arXiv:hep-th/0510147].

\bibitem{0602254}
  J.~R.~David, D.~P.~Jatkar and A.~Sen,
  ``Product representation of dyon partition function in CHL models,''
  JHEP {\bf 0606}, 064 (2006)
  [arXiv:hep-th/0602254].

\bibitem{0603066}
  A.~Dabholkar and S.~Nampuri,
  ``Spectrum of dyons and black holes in
  CHL orbifolds using Borcherds lift,''
  arXiv:hep-th/0603066.

\bibitem{0605210}
  J.~R.~David and A.~Sen,
  ``CHL dyons and statistical entropy function from D1-D5 system,''
  arXiv:hep-th/0605210.

\bibitem{0607155}
  J.~R.~David, D.~P.~Jatkar and A.~Sen,
  ``Dyon spectrum in N = 4 supersymmetric type II string theories,''
  arXiv:hep-th/0607155.

\bibitem{0609109}
  J.~R.~David, D.~P.~Jatkar and A.~Sen,
  ``Dyon spectrum in generic N = 4 supersymmetric Z(N) orbifolds,''
  arXiv:hep-th/0609109.

\bibitem{0612011}
  A.~Dabholkar and D.~Gaiotto,
  ``Spectrum of CHL dyons from genus-two partition function,''
  arXiv:hep-th/0612011.


\bibitem{0708.1270}
  A.~Sen,
  ``Black Hole Entropy Function,
Attractors and Precision Counting of
  Microstates,''
Gen.\ Rel.\ Grav.\  {\bf 40}, 2249 (2008)
  [arXiv:0708.1270 [hep-th]].


\bibitem{0802.0544}
  S.~Banerjee, A.~Sen and Y.~K.~Srivastava,
  ``Generalities of Quarter BPS Dyon
Partition Function and Dyons of Torsion
  Two,''
  arXiv:0802.0544 [hep-th].

\bibitem{0802.1556}
  S.~Banerjee, A.~Sen and Y.~K.~Srivastava,
  ``Partition Functions of Torsion $>1$ Dyons in Heterotic
String Theory on $T^6$,''
  arXiv:0802.1556 [hep-th].

\bibitem{0803.2692}
  A.~Dabholkar, J.~Gomes and S.~Murthy,
  ``Counting all dyons in N =4 string theory,''
  arXiv:0803.2692 [hep-th].
  
\bibitem{0903.1477}
  A.~Sen,
  ``Arithmetic of Quantum Entropy Function,''
  JHEP {\bf 0908}, 068 (2009)
  [arXiv:0903.1477 [hep-th]].
 
 \bibitem{1009.3226}
  A.~Dabholkar, J.~Gomes, S.~Murthy, A.~Sen,
  ``Supersymmetric Index from Black Hole Entropy,''
  JHEP {\bf 1104}, 034 (2011).
  [arXiv:1009.3226 [hep-th]].
 
\bibitem{0908.0039}
  A.~Sen,
  ``Arithmetic of N=8 Black Holes,''
  JHEP {\bf 1002}, 090 (2010)
  [arXiv:0908.0039 [hep-th]].

\bibitem{9507090}
  M.~Cvetic and D.~Youm,
  ``Dyonic BPS saturated black holes of heterotic string on a six torus,''
  Phys.\ Rev.\  D {\bf 53}, 584 (1996)
  [arXiv:hep-th/9507090].

\bibitem{9512031}
M.~Cvetic and A.~A.~Tseytlin,
``Solitonic strings and BPS saturated dyonic black holes,''
  Phys.\ Rev.\  D {\bf 53}, 5619 (1996)
  [Erratum-ibid.\  D {\bf 55}, 3907 (1997)]
  [arXiv:hep-th/9512031].

\bibitem{9407001}
  S.~N.~Solodukhin,
  ``The Conical singularity and quantum corrections to entropy of black hole,''
  Phys.\ Rev.\  D {\bf 51}, 609 (1995)
  [arXiv:hep-th/9407001].

\bibitem{9408068}
  S.~N.~Solodukhin,
  ``On 'Nongeometric' contribution 
  to the entropy of black hole due to quantum
  corrections,''
  Phys.\ Rev.\  D {\bf 51}, 618 (1995)
  [arXiv:hep-th/9408068].


\bibitem{9412161}
  D.~V.~Fursaev,
  ``Temperature And Entropy Of A 
  Quantum Black Hole And Conformal Anomaly,''
  Phys.\ Rev.\  D {\bf 51}, 5352 (1995)
  [arXiv:hep-th/9412161].
  
\bibitem{9604118}
  R.~B.~Mann and S.~N.~Solodukhin,
  ``Conical geometry and quantum entropy of a charged Kerr black hole,''
  Phys.\ Rev.\  D {\bf 54}, 3932 (1996)
  [arXiv:hep-th/9604118].

\bibitem{9709064}
  R.~B.~Mann and S.~N.~Solodukhin,
  ``Universality of quantum entropy for extreme black holes,''
  Nucl.\ Phys.\  B {\bf 523}, 293 (1998)
  [arXiv:hep-th/9709064].

\bibitem{0002040}
  R.~K.~Kaul and P.~Majumdar,
  ``Logarithmic correction to the Bekenstein-Hawking entropy,''
  Phys.\ Rev.\ Lett.\  {\bf 84}, 5255 (2000)
  [arXiv:gr-qc/0002040].

\bibitem{0005017}
  S.~Carlip,
  ``Logarithmic corrections to black hole entropy from the Cardy formula,''
  Class.\ Quant.\ Grav.\  {\bf 17}, 4175 (2000)
  [arXiv:gr-qc/0005017].

\bibitem{0104010}
  T.~R.~Govindarajan, R.~K.~Kaul and V.~Suneeta,
  ``Logarithmic correction to the 
  Bekenstein-Hawking entropy of the BTZ  black
  hole,''
  Class.\ Quant.\ Grav.\  {\bf 18}, 2877 (2001)
  [arXiv:gr-qc/0104010].

\bibitem{0112041}
  K.~S.~Gupta, S.~Sen,
  ``Further evidence for the conformal structure of a Schwarzschild black hole in an algebraic approach,''
  Phys.\ Lett.\  {\bf B526}, 121-126 (2002).
  [hep-th/0112041].


\bibitem{0406044}
  A.~J.~M.~Medved,
  ``A comment on black hole entropy or why Nature abhors a logarithm,''
  Class.\ Quant.\ Grav.\  {\bf 22}, 133 (2005)
  [arXiv:gr-qc/0406044].

\bibitem{0409024}
  D.~N.~Page,
  ``Hawking radiation and black hole thermodynamics,''
  New J.\ Phys.\  {\bf 7}, 203 (2005)
  [arXiv:hep-th/0409024].


\bibitem{0805.2220}
  R.~Banerjee and B.~R.~Majhi,
  ``Quantum Tunneling Beyond Semiclassical Approximation,''
  JHEP {\bf 0806}, 095 (2008)
  [arXiv:0805.2220 [hep-th]].

\bibitem{0808.3688}
  R.~Banerjee and B.~R.~Majhi,
  ``Quantum Tunneling, Trace Anomaly and Effective Metric,''
  Phys.\ Lett.\  B {\bf 674}, 218 (2009)
  [arXiv:0808.3688 [hep-th]].

\bibitem{0809.1508}
  B.~R.~Majhi,
  ``Fermion Tunneling Beyond Semiclassical Approximation,''
  Phys.\ Rev.\  D {\bf 79}, 044005 (2009)
  [arXiv:0809.1508 [hep-th]].

\bibitem{0911.4379}
  R.~G.~Cai, L.~M.~Cao and N.~Ohta,
  ``Black Holes in Gravity with 
  Conformal Anomaly and Logarithmic Term in Black
  Hole Entropy,''
  arXiv:0911.4379 [hep-th].


\bibitem{1003.1083}
  R.~Aros, D.~E.~Diaz and A.~Montecinos,
  ``Logarithmic correction to BH entropy as Noether charge,''
  arXiv:1003.1083 [hep-th].

\bibitem{1008.4314}
  S.~N.~Solodukhin,
  ``Entanglement entropy of round spheres,''
  Phys.\ Lett.\  {\bf B693}, 605-608 (2010).
  [arXiv:1008.4314 [hep-th]].


\bibitem{1005.3044}
  S.~Banerjee, R.~K.~Gupta, A.~Sen,
  ``Logarithmic Corrections to Extremal Black Hole Entropy from Quantum Entropy Function,''  
  [arXiv:1005.3044 [hep-th]].

\bibitem{0804.1773}
  S.~Giombi, A.~Maloney and X.~Yin,
  ``One-loop Partition Functions of 3D Gravity,''
  JHEP {\bf 0808}, 007 (2008)
  [arXiv:0804.1773 [hep-th]].
  
\bibitem{0911.5085}
  J.~R.~David, M.~R.~Gaberdiel and R.~Gopakumar,
  ``The Heat Kernel on $AdS_3$ and its Applications,''
  arXiv:0911.5085 [hep-th].

\bibitem{1009.6087}
  M.~R.~Gaberdiel, R.~Gopakumar, A.~Saha,
  ``Quantum $W$-symmetry in $AdS_3$,''
  JHEP {\bf 1102}, 004 (2011).
  [arXiv:1009.6087 [hep-th]].
  
\bibitem{gilkey}
P.~B.~Gilkey, 
``Invariance theory, the heat equation and the Atiyah-Singer index theorem,''
Publish or Perish Inc., USA (1984).


\bibitem{0306138}
  D.~V.~Vassilevich,
  ``Heat kernel expansion: User's manual,''
  Phys.\ Rept.\  {\bf 388}, 279 (2003)
  [arXiv:hep-th/0306138].

\bibitem{0805.0095}
  A.~Sen,
  ``Entropy Function and AdS(2) / CFT(1) Correspondence,''
  JHEP {\bf 0811}, 075 (2008).
  [arXiv:0805.0095 [hep-th]].

\bibitem{0608021}
  C.~Beasley, D.~Gaiotto, M.~Guica, L.~Huang, 
  A.~Strominger and X.~Yin,
  ``Why Z(BH) = |Z(top)|**2,''
  arXiv:hep-th/0608021.

\bibitem{0905.2686}
  N.~Banerjee, S.~Banerjee, R.~Gupta, I.~Mandal and A.~Sen,
  ``Supersymmetry, Localization and Quantum Entropy Function,''
  arXiv:0905.2686 [hep-th].

\bibitem{1012.0265}
  A.~Dabholkar, J.~Gomes, S.~Murthy,
  [arXiv:1012.0265 [hep-th]].
  
\bibitem{dabappear}
  A.~Dabholkar, J.~Gomes, S.~Murthy, to appear.

\bibitem{9508064}
  A.~Sen, C.~Vafa,
  ``Dual pairs of type II string compactification,''
  Nucl.\ Phys.\  {\bf B455}, 165-187 (1995).
  [hep-th/9508064].
  
  \bibitem{9901117}
  A.~Gregori, C.~Kounnas, P.~M.~Petropoulos,
  ``Nonperturbative triality in heterotic and type II N=2 strings,''
  Nucl.\ Phys.\  {\bf B553}, 108-132 (1999).
  [hep-th/9901117].

\bibitem{0711.1971}
  J.~R.~David,
  ``On the dyon partition function in N=2 theories,''
  JHEP {\bf 0802}, 025 (2008).
  [arXiv:0711.1971 [hep-th]].

\bibitem{0405146}
  H.~Ooguri, A.~Strominger, C.~Vafa,
  ``Black hole attractors and the topological string,''
  Phys.\ Rev.\  {\bf D70}, 106007 (2004).
  [hep-th/0405146].
  
\bibitem{0808.2627}
  G.~L.~Cardoso, B.~de Wit, S.~Mahapatra,
 ``Subleading and non-holomorphic corrections to N=2 BPS black hole entropy,''
  JHEP {\bf 0902}, 006 (2009).
  [arXiv:0808.2627 [hep-th]].
  
\bibitem{0702146}
  F.~Denef, G.~W.~Moore,
 ``Split states, entropy enigmas, holes and halos,''
  [hep-th/0702146 [HEP-TH]].

\bibitem{christ-duff1}
  S.~M.~Christensen and M.~J.~Duff,
  ``New Gravitational Index Theorems And Supertheorems,''
  Nucl.\ Phys.\  B {\bf 154}, 301 (1979).

\bibitem{christ-duff2}
  S.~M.~Christensen and M.~J.~Duff,
  ``Quantizing Gravity With A Cosmological Constant,''
  Nucl.\ Phys.\  B {\bf 170}, 480 (1980).

\bibitem{duffnieu}
  M.~J.~Duff and P.~van Nieuwenhuizen,
  ``Quantum Inequivalence Of Different Field Representations,''
  Phys.\ Lett.\  B {\bf 94}, 179 (1980).


\bibitem{duffroc}
  S.~M.~Christensen, M.~J.~Duff, G.~W.~Gibbons and M.~Rocek,
  ``Vanishing One Loop Beta Function In Gauged N $>$ 4 Supergravity,''
  Phys.\ Rev.\ Lett.\  {\bf 45}, 161 (1980).


\bibitem{campo}
  R.~Camporesi,
  ``Harmonic analysis and propagators on homogeneous spaces,''
  Phys.\ Rept.\  {\bf 196} (1990) 1.

\bibitem{camhig1}
  R.~Camporesi and A.~Higuchi,
  ``Spectral functions and zeta functions in hyperbolic spaces,''
  J.\ Math.\ Phys.\  {\bf 35}, 4217 (1994).

\bibitem{campo2}
  R.~Camporesi,
  ``The Spinor heat kernel in maximally symmetric spaces,''
  Commun.\ Math.\ Phys.\  {\bf 148} (1992) 283.

\bibitem{camhig2}
  R.~Camporesi and A.~Higuchi,
  ``Arbitrary spin effective potentials in anti-de Sitter space-time,''
  Phys.\ Rev.\  D {\bf 47}, 3339 (1993).

\bibitem{9505009}
  R.~Camporesi and A.~Higuchi,
  ``On The Eigen Functions Of 
  The Dirac Operator On Spheres And Real Hyperbolic
  Spaces,''
  J.\ Geom.\ Phys.\  {\bf 20}, 1 (1996)
  [arXiv:gr-qc/9505009].


\bibitem{0810.1233}
  G.~L.~Cardoso, J.~R.~David, B.~de Wit, S.~Mahapatra,
  ``The Mixed black hole partition function for the STU model,''
  JHEP {\bf 0812}, 086 (2008).
  [arXiv:0810.1233 [hep-th]].
  
\bibitem{0905.4115}
  J.~R.~David,
  ``On walls of marginal stability in N=2 string theories,''
  JHEP {\bf 0908}, 054 (2009).
  [arXiv:0905.4115 [hep-th]].

\bibitem{9505162}
  S.~Ferrara, J.~A.~Harvey, A.~Strominger, C.~Vafa,
  ``Second quantized mirror symmetry,''
  Phys.\ Lett.\  {\bf B361}, 59-65 (1995).
  [hep-th/9505162].

\end{thebibliography}
\end{document}